\documentclass[a4paper,oneside,11pt]{article}
\pdfoutput=1
\usepackage{cprform}
\usepackage{amsmath,amstext,amsfonts,amsbsy,amssymb,amscd,slashed}
\usepackage{pdflscape,color,svg,epsfig,cite,hyperref,bbm,multirow}
\usepackage{graphicx}
\usepackage{tabularx}
\usepackage{booktabs}
\usepackage{floatpag}

\usepackage[makeroom]{cancel}
\usepackage{blkarray}
\usepackage{stackengine}

\usepackage{tikz}
\usetikzlibrary{arrows,shapes}
\usetikzlibrary{trees}
\usetikzlibrary{matrix,arrows}                          
\usetikzlibrary{positioning}                            
\usetikzlibrary{calc,through}                           
\usetikzlibrary{decorations.pathreplacing}  
\usetikzlibrary{decorations.pathmorphing}       
\usetikzlibrary{decorations.markings}
\tikzset{
    vector/.style={decorate, decoration={snake}, draw},
        provector/.style={decorate, decoration={snake,amplitude=2.5pt}, draw},
        antivector/.style={decorate, decoration={snake,amplitude=-2.5pt}, draw},
    fermion/.style={draw=black, postaction={decorate},
        decoration={markings,mark=at position .55 with {\arrow[draw=black]{>}}}},
    fermionbar/.style={draw=black, postaction={decorate},
        decoration={markings,mark=at position .55 with {\arrow[draw=black]{<}}}},
    fermionnoarrow/.style={draw=black},
    gluon/.style={decorate, draw=black,
        decoration={coil,amplitude=4pt, segment length=5pt}},                           
    scalar/.style={dashed,draw=black, postaction={decorate},
        decoration={markings,mark=at position .55 with {\arrow[draw=black]{>}}}},
    scalarbar/.style={dashed,draw=black, postaction={decorate},
        decoration={markings,mark=at position .55 with {\arrow[draw=black]{<}}}},
    scalarnoarrow/.style={dashed,draw=black},
    electron/.style={draw=black, postaction={decorate},
        decoration={markings,mark=at position .55 with {\arrow[draw=black]{>}}}},
        bigvector/.style={decorate, decoration={snake,amplitude=4pt}, draw},
}

\newcommand{\ba}{\begin{array}}
\newcommand{\ea}{\end{array}}

\newcommand{\req}[1]{Eq.~(\ref{#1})}

\newcommand{\ret}[1]{Table~\ref{#1}}

\newcommand{\Dslash}{\relax{\kern+.25em / \kern-.70em D}}

\newcommand{\fm}{{\rm fm}}
\newcommand{\MeV}{{\rm MeV}}

\newcommand{\Real}{\relax{\mathsf{\Gamma\kern-.35em R}}}
\newcommand{\Int}{\relax{\mathsf{Z\kern-.40em Z}}}

\newcommand{\ihalf}{{\scriptstyle{{i\over 2}}}}

\newcommand{\NF}{N_\mathrm{\scriptstyle f}}

\newcommand{\MSbar}{{\overline{\rm MS}}}

\newcommand{\gbar}{\kern1pt\overline{\kern-1pt g\kern-0pt}\kern1pt}
\newcommand{\mbar}{\kern2pt\overline{\kern-1pt m\kern-1pt}\kern1pt}
\newcommand{\obar}[1]{\kern3pt\overline{\kern-2pt #1\kern-0pt}\kern1pt}

\newcommand{\mcrit}{m_{\rm cr}}

\newcommand{\ZP}{Z_{\rm\scriptscriptstyle P}}

\newcommand{\ZA}{Z_{\rm\scriptscriptstyle A}}

\newcommand{\Oa}{\mbox{O}(a)}
\newcommand{\Oasq}{\mbox{O}(a^2)}

\newcommand{\icsw}{c_{\rm sw}}

\newcommand{\abar}{\kern1pt\overline{\kern-1pt a\kern-0.5pt}\kern1pt}

\newcommand{\cO}{{\cal O}}

\newcolumntype{C}{>{$}c<{$}}
\newcolumntype{R}{>{$}r<{$}}
\newcolumntype{L}{>{$}l<{$}}

\usepackage[normalem]{ulem}
\newcommand\xoutpars[1]{\let\helpcmd\xout\parhelp#1\par\relax\relax}
\newcommand\soutpars[1]{\let\helpcmd\sout\parhelp#1\par\relax\relax}
\long\def\parhelp#1\par#2\relax{%
  \helpcmd{#1}\ifx\relax#2\else\par\parhelp#2\relax\fi%
}

\begin{document}


\begin{titlepage}


\vspace*{-30truemm}
\begin{flushright}
IFT-UAM/CSIC-23-114\\[10pt]

\end{flushright}
\vspace{15truemm}


\centerline{\bigbf Hadronic physics from a Wilson fermion mixed-action approach:}
\centerline{\bigbf Charm quark mass and $D_{(s)}$ meson decay constants}
\vskip 10 true mm
\begin{center}
\includegraphics[width=0.16\textwidth]{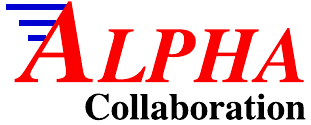}\\

\end{center}

\vskip -2 true mm
\centerline{\elfrm 
Andrea~Bussone$^{a}$,
Alessandro~Conigli$^{b,c}$,
Julien~Frison$^{d}$,
Gregorio~Herdo\'{\i}za$^{b,c}$,
}
\centerline{\elfrm 
Carlos~Pena$^{b,c}$,
David~Preti$^e$,
Alejandro S\'aez$^{b,c}$,
Javier~Ugarrio$^{b,c}$}
\vskip 4 true mm
\centerline{\tenit $^a$ Humboldt Universit\"at zu Berlin, Institut f\"ur Physik and IRIS Adlershof}
\centerline{\tenit Zum Gro{\ss}en Windkanal 6, 12489 Berlin, Germany}
\vskip 3 true mm
\centerline{\tenit $^b$ Instituto de F\'{\i}sica Te\'orica UAM-CSIC} 
\centerline{\tenit c/~Nicol\'as Cabrera 13-15, Universidad Aut\'onoma de Madrid}
\centerline{\tenit Cantoblanco E-28049 Madrid, Spain}
\vskip 3 true mm
\centerline{\tenit $^c$ Dpto. de F\'{\i}sica Te\'orica, Universidad Aut\'onoma de Madrid}
\centerline{\tenit Cantoblanco E-28049 Madrid, Spain}
\vskip 3 true mm
\centerline{\tenit $^d$ John von Neumann-Institut f\"ur Computing NIC, Deutsches Elektronen-Synchrotron DESY}
\centerline{\tenit Platanenallee 6, 15738 Zeuthen, Germany}
\vskip 3 true mm
\centerline{\tenit $^e$ INFN, Sezione di Torino}
\centerline{\tenit Via Pietro Giuria 1, I-10125 Turin, Italy}
\vskip 15 true mm


\noindent{\tenbf Abstract:}
We present our first set of results for charm physics,
using the mixed-action setup introduced in a companion
paper~\cite{MA1}. Maximally twisted Wilson valence fermions are used
on a sea of non-perturbatively $O(a)$-improved Wilson fermions, made
up by CLS $\NF=2+1$ ensembles. Our charm-sector observables are free
from $O(am_c)$ discretisation effects, without need of tuning any
improvement coefficient, and show continuum-limit scaling properties consistent with 
leading cutoff effects of $O(a^2)$. We consider a subset
of CLS ensembles -- including four values of the lattice spacing and
pion masses down to 200\,MeV -- allowing to take the continuum limit
and extrapolate to the physical pion mass. A number of techniques are
incorporated in the analysis in order to estimate the systematic
uncertainties of our results for the charm quark mass and the
$D_{(s)}$-meson decay constants.
This first study of observables in the charm sector, where the emphasis has been on the control of the methodology, demonstrates the potential of our setup to achieve high-precision results.

\vspace{10truemm}

\eject
\end{titlepage}

\cleardoublepage

\section{Introduction}

Heavy flavour physics is a key frontline in the endeavour to test the limits of
the Standard Model, and look for new fundamental physics. Ever-increasing precision
for fundamental parameters such as quark masses and Cabibbo-Kobayashi-Maskawa
matrix elements, as well as for weak matrix elements that control the low-energy
hadronic contribution to weak decay amplitudes, is necessary to keep pace
with experimental developments.

First-principles, systematically improvable computations performed in Lattice QCD
--- possibly, beyond a certain precision threshold, with QED corrections ---
are of course the basic source of input. When dealing with heavy quark physics, however,
lattice computations face a non-trivial multiscale problem. Since computations involve
both an ultraviolet cutoff --- the inverse lattice spacing $a^{-1}$ --- and an infrared
cutoff --- the inverse size $L^{-1}$ of the finite box computations are performed in ---
all physical scales should best fit comfortably between the cutoffs, lest control on their
removal is compromised. A standard criterion for finite-volume effects to be sufficiently
suppressed in typical hadronic quantities involves a constraint on the box size
$m_\pi L \gtrsim 4$; for the typical range of pion masses explored, which nowadays
routinely reaches the physical point, this implies box sizes in the 3~to~7~fm ballpark. 
Having $m_c \ll a^{-1}$, and especially, $m_b \ll a^{-1}$, then requires values of the
lattice size $L/a$ that are close to or simply beyond current computational capabilities.
This problem is much worsened by the extra difficulty to approach the very fine lattice
spacings needed to accommodate heavy quark masses: the computational cost of typical
simulations scales as $\sim a^{-7}$~\cite{Luscher:2011qa}, and for $a \lesssim 0.05~\fm$ the algorithmic
problem of topology freezing sets in, which in practice impedes simulations long enough
to control statistical uncertainties reliably~\cite{Schaefer:2010hu}.

Facing these problems requires a specific toolset for heavy quark physics on the lattice,
that, in particular, relies on input from effective theories to try and control the
ultraviolet cutoff dependence: Symanzik effective theory~\cite{Symanzik:1983dc,Symanzik:1983gh,Luscher:1984xn,Luscher:1996sc} to understand and
suppress the leading cutoff effects, heavy quark effective theory input to guide the
construction of lattice actions or the extraction of physics,\footnote{See, e.g., App.~A.1.3
of~\cite{FlavourLatticeAveragingGroup:2019iem} for a summary of existing approaches.} etc.
The resulting sophisticated frameworks often rely on assumptions about the systematic impact
of the use of effective theory, and/or require the determination of ancillary quantities
such as $O(a)$ improvement coefficients. A full overview of lattice techniques and results
for heavy quark physics can be obtained from the latest FLAG review~\cite{FlavourLatticeAveragingGroupFLAG:2021npn}.
A main theme underpinning all studies in the charm and, especially, the B sector is that
having results from a variety of approaches is essential to gain confidence on the
systematic uncertainties affecting hadronic observables relevant for flavour physics.

The main motivation of the mixed-action setup used in this work, and fully discussed in~\cite{MA1},
is to devise an optimal framework for heavy quark physics that bypasses many of the difficulties
mentioned above. The first ingredient is the use of CLS $\NF=2+1$ ensembles obtained
with non-perturbatively $O(a)$ improved Wilson fermions~\cite{Bruno:2014jqa} and
open boundary conditions for the gauge field~\cite{Luscher:2011kk,Luscher:2012av},
which allows to enter the realm of very
fine lattice spacings while keeping control on statistical uncertainties. The second ingredient is
to compute heavy quark observables by means of a valence twisted-mass Wilson
setup~\cite{Frezzotti:2000nk,Pena:2004gb},
which leads to automatic $O(a)$ improvement~\cite{Frezzotti:2003ni}. Working with a mixed action of course
leads to new requirements, such a precise matching between the valence and sea sectors,
and a careful analysis of the relative cutoff effects. This is discussed in the companion
paper~\cite{MA1}. Here we will focus on illustrating how the technique is able to obtain
precise, reliable results for basic observables in the hadronic sector.
Progress report of this long-term project have been given in~\cite{Herdoiza:2017bcc,Ugarrio:2018ghf,Bussone:2018ljj,Bussone:2019mlt,Conigli:2021rqu,Bussone:2022wjb,Saez:2022ptd,Conigli:2022kgq}.

In particular, we will focus on determining the value of the charm quark mass, and of
the leptonic decay constants of the $D$ and $D_s$ mesons. Our results are based on a
subset of the available CLS ensembles which allow us to illustrate the properties of the setup.
We also emphasise our development of a variant of the existing
techniques to assess systematic uncertainties in lattice observables based on information
criteria~\cite{Jay:2020jkz,Frison:2023lwb} applied to appropriate goodness-of-fit 
estimators~\cite{Bruno:2022mfy}. Still, despite the fact that our current results use a subset of the CLS ensembles, they are already at a point where they have competing precision in the
context of the state-of-the-art determination of these quantities that enter current FLAG averages~\cite{McNeile:2010ji,Davies:2010ip,FermilabLattice:2011njy,Na:2012iu,ETM:2013jap,Chakraborty:2014aca,EuropeanTwistedMass:2014osg,Alexandrou:2014sha,Yang:2014sea,Carrasco:2014poa,Nakayama:2016atf,Bazavov:2017lyh,Boyle:2017jwu,FermilabLattice:2018est,Balasubramamian:2019wgx,Petreczky:2019ozv,Hatton:2020qhk,ExtendedTwistedMass:2021gbo,Heitger:2021apz}.
Results with a larger set of CLS ensembles, including finer lattice spacings
and physical pion mass ensembles, will be the object of upcoming publications.

Let us conclude this introduction by describing the organisation of the paper. 
Sec.~\ref{sec:mixed_action_setup} summarises the main aspects of our mixed-action
approach, discussed at length in~\cite{MA1}. Sec.~\ref{sec:charm_basics} deals with
our approach to matching the scale of our partially-quenched charm quark, and numerical
aspects of computations in the charm sector. Secs.~\ref{sec:mc} and~\ref{sec:fDs}
discuss our determination of the charm mass and decay constants, respectively.
Finally, Sec.~\ref{sec:conc} contains our conclusions and outlook.

\section{Mixed-action setup}
\label{sec:mixed_action_setup}

In this section we review the basic features of our setup, with an emphasis on their implications
for heavy quark physics.
We refer the reader to~\cite{MA1} for a fully detailed discussion of our approach.

\subsection{Generalities}

All our results stem from a mixed-action setup.
In the sea sector we employ a tree-level improved gauge action~\cite{Luscher:1984xn,Luscher:1985zq},
and a non-perturbatively $\Oa$-improved Wilson fermion action~\cite{Sheikholeslami:1985ij}.
This has indeed been used in the generation of the CLS $\NF=2+1$ ensembles~\cite{Bruno:2014jqa,Bruno:2016plf,Mohler:2017wnb,Mohler:2020txx}
that we employ.
In the valence sector, on the other hand, we use a fully-twisted tmQCD~\cite{Frezzotti:2000nk}
fermion action.
Both actions include the same massless Wilson-Dirac operator~\cite{Wilson:1974sk,Sheikholeslami:1985ij}
\begin{gather}
\label{eq:Dphys}
D=\frac{1}{2}\gamma_\mu(\nabla_\mu^*+\nabla_\mu)-
\frac{a}{2}\nabla_\mu^*\nabla_\mu 
+\frac{i}{4}\,a\icsw\sigma_{\mu\nu}\hat F_{\mu\nu},
\end{gather}
where $\nabla_\mu$ and $\nabla_\mu^*$ are, respectively, the forward and backward covariant derivatives,
$\sigma_{\mu\nu}=\ihalf[\gamma_\mu,\gamma_\nu]$, and $\hat F_{\mu\nu}$ is
the clover-leaf definition of the field strength tensor as spelled out in~\cite{Luscher:1996sc}.
The mass term in the sea has the form
\begin{gather}
\label{eq:msea}
\bar\psi \mathbf{m}^{(s)} \psi\,,
\end{gather}
while the tmQCD action is obtained by adding a mass term of the form~\cite{Frezzotti:2000nk,Pena:2004gb}\footnote{While, other, versions of the valence sector \`a la Osterwalder-Seiler~\cite{Frezzotti:2004wz}
can be used without substantial changes to the discussion below, in this work the form
in \req{eq:mval} will suffice to extract all the relevant physics, and we will therefore stick to it
for definiteness.}
\begin{gather}
\label{eq:mval}
i \bar\psi \boldsymbol{\mu} \gamma_5 \psi
+\mcrit\bar\psi\psi
\,; \qquad
\boldsymbol{\mu} = {\rm diag}(\mu_u,-\mu_d,-\mu_s,\mu_c)\,,
\end{gather}
where $\mcrit$ is the standard Wilson critical mass, and the signs have been set
so that the values of the twisted masses $\mu_{\rm f}$ are implied to be positive.
We will always work in the isospin limit, where the up and down quark masses take
the same values both in the sea and in the valence (i.e., $m_u^{(s)}=m_d^{(s)}$ and
$\mu_u=\mu_d \equiv \mu_l$).
The procedure to fully define the mixed action involves the matching between Wilson and
tmQCD valence actions, and a specific prescription to define the critical mass used in our setup.
To that purpose, for any given ensemble we first tune $\mu_l$, $\mu_s$ and $\mcrit$ such that the quantities $\phi_2 $ and $\phi_4$ -- depending on pion and kaon masses, as defined in Eq.~(\ref{eq:phi2_and_phi4}) -- coincide for sea and valence actions, while imposing that the ($u$,$d$) standard PCAC quark mass
-- including all known $\Oa$-improvement counterterms -- vanishes in the valence sector.
This ensures equivalent physics and sets the twist angle to $\pi/2$,
ensemble by ensemble.

\subsection{Properties of the twisted valence sector}

The most interesting property of this setup for the purpose of the results presented
in this paper is that it results in automatic
$\Oa$ improvement of observables extracted from valence correlation functions~\cite{Frezzotti:2003ni},
up to terms proportional to the trace of the subtracted sea quark mass matrix,
$a{\rm tr}\{\mathbf{m}^{(s)}_{\rm q}\}$~\cite{MA1}.
Since the latter only involves up, down, and strange quarks, the value of the
trace in lattice units is of $\cO(10^{-2})$ on our ensembles.
Furthermore, these terms arise from loop effects, and their coefficient is thus formally
of perturbative order $\alpha_{\rm\scriptscriptstyle s}^2$.
Given our typical statistical uncertainties, the natural size of these $a{\rm tr}\{\mathbf{m}^{(s)}_{\rm q}\}$ lattice artefacts  therefore amounts to a subdominant contribution. This property can be furthermore verified a posteriori by inspecting the scaling of observables towards the continuum limit.
This is very important for heavy quark observables, since we are then assured that
the leading cutoff effects associated to a quark of mass $\mu_h$
are of order $(a\mu_h)^2$, without need of fine-tuning
improvement coefficients to ensure the elimination of linear effects, as would be
the case in the standard $\Oa$ improved setup.
Note that automatic $\Oa$ improvement holds even in the absence of the clover term
in the valence fermion action; we have however kept it for a number of reasons.
First, it simplifies the matching between sea and valence, since the regularisations
coincide in the chiral limit.
Secondly, for the same reason, it allows to use non-perturbative renormalisation constants
determined with standard methods --- e.g., to obtain renormalised quark masses~\cite{Campos:2018ahf}.
Finally, it has been observed that keeping the clover term leads to a better control
over the $\Oasq$ flavour-breaking effects induced by the twisted mass term, thus improving
the overall scaling of the setup~\cite{Frezzotti:2005gi,Dimopoulos:2009es}.
A second, more generic benefit is that the use of a twisted mass regularisation
implies multiplicative renormalisation of (twisted) quark masses, and the possibility
to determine decay constants without need of finite normalisation factors such as $\ZA$.
This is a result of the explicit chiral symmetry breaking pattern at full twist,
which leaves exactly conserved axial currents.
In the twisted quark field basis implicitly assumed when writing our valence mass terms,
the relevant on-shell ($x \neq 0$) Ward-Takahashi identity reads
\begin{equation}
	\langle \partial_\mu^* \tilde{V}^{qr}_\mu(x)\, O(0) \rangle = 
	i(\mu_q + \mu_r) \langle P^{qr}(x)\, O(0)\rangle,
	\label{eq:vector_ward_id}
\end{equation}
where $\partial_\mu^*$ is the backward lattice derivative;
$O$ is any gauge-invariant local operator;
$\mu_{q,r}$ are the {\em Lagrangian} twisted masses for the corresponding flavours $q,r$,
that are here assumed to carry different signs in the twisted mass matrix $\boldsymbol{\mu}$ of
\req{eq:mval};\footnote{With our conventions, this applies to any of the pairs
$(u,d)$, $(u,s)$, $(d,c)$ and $(s,c)$.}
$P^{qr}=\bar\psi_q\gamma_5\psi_r$ is a non-singlet pseudoscalar density;
and $\tilde{V}^{qr}_\mu$ is the point-split vector current\footnote{This is
indeed the physical axial current, chirally rotated by the relation between
physical and twisted quark variables --- see, e.g., \cite{Frezzotti:2000nk,Shindler:2007vp}.}
\begin{equation}
	\label{eq:awi}
	\tilde{V}_\mu^{qr}(x) = \frac{1}{2}
	\bigg[
	\overline{\psi}^q(x) (\gamma_\mu-1) U_\mu(x)\psi^r(x+a\hat{\mu}) 
	+ 
	\overline{\psi}^q(x+a\hat{\mu}) (\gamma_\mu+1) U_\mu^\dagger(x)\psi^r(x)
	\bigg].
\end{equation}
Since the current is exactly conserved, there are two important consequences.
First, current and Lagrangian quark masses coincide, and renormalise with
$Z_\mu = \ZP^{-1}$.\footnote{It can be separately proven that renormalisation
is indeed multiplicative.}
Second, meson decay constants can be extracted from a two-point function of the pseudoscalar
density, by setting $O=P^{rq}$ in \req{eq:vector_ward_id} and using the fact that the l.h.s.
of the Ward identity is exactly normalised.
These will be the basis of our determinations of the charm quark mass in Sec.~\ref{sec:charm_basics}
and of $f_{D_{(s)}}$ in Sec.~\ref{sec:fDs}.

\subsection{Ensembles and line of constant physics}

CLS ensembles have been generated along three different lines of constant physics.
Our results are based on a subset of the ensembles generated at (approximately)
constant value of ${\rm tr}\{\mathbf{m}^{(s)}_{\rm q}\}$, which we list
in~\ret{tab:CLS}.
In order to define a precise line of constant physics, we use the quantities
\begin{gather}
\phi_2 \equiv 8t_0m_\pi^2\,, \qquad
\phi_4 \equiv 8t_0\left(\frac{1}{2}m_\pi^2+m_K^2\right)\,,
\label{eq:phi2_and_phi4}
\end{gather}
where $t_0$ is the gradient flow scale introduced in~\cite{Luscher:2010iy}, and whose
value in physical units has already been determined using CLS ensembles 
in~\cite{Bruno:2016plf,RQCD:2022xux,Strassberger:2021tsu,MA1}.
A renormalised line of constant physics can thus be fixed by setting $\phi_4$
constant and equal to its physical value; extraction of the physics will then
proceed by a combined continuum-chiral limit fit that hits the physical value
of $\phi_2$.
The condition that $\phi_4$ is constant is well approximated by keeping
${\rm tr}\{\mathbf{m}^{(s)}\}$
fixed, since it is proportional to $\phi_4$ at leading order in the effective
chiral description of QCD dynamics.
Small deviations from the correct value of $\phi_4$ in each ensemble can be
corrected by means of the mass shifting prescription introduced in~\cite{Bruno:2016plf},
and incorporated into the fitting procedure --- see~\cite{MA1} for technical
details.
Our renormalised chiral trajectory is ultimately set at
$\phi_4^{\mathrm{phys}}=\phi_4^{\mathrm{isoQCD}}=1.101(7)(5)$, where the second error
quoted is the systematic uncertainty coming from our Bayesian model averaging (see below),
and the first error comprises the statistical uncertainty, the one associated to chiral-continuum
extrapolations, and those related to input parameters --- improvement coefficients, renormalisation
constants, and the input pion and kaon masses.
The values of the latter employed to fix $\phi_4$ are those in the QCD isospin symmetric limit (isoQCD)  given by \cite{FlavourLatticeAveragingGroupFLAG:2021npn}
\begin{eqnarray}
	m_\pi^{\mathrm{isoQCD}} &=& 134.9768(5) \ \mathrm{MeV},
	\\
	m_K^{\mathrm{isoQCD}} &=& 497.611(13) \ \mathrm{MeV}.
\end{eqnarray}
In the remainder of this paper we will use the superscript ``phys'' for quantities defined
in the isoQCD prescription for the continuum theory, as fixed above.

In this work we employ our  determination of the physical scale from the gradient flow scale $t_0$. To set the scale, we use the following combination of pion and kaon decay constants
\begin{equation}
	\sqrt{8t_0}f_{\pi K} = \sqrt{8t_0}\bigg(
	\frac{2}{3}f_K + \frac{1}{3}f_\pi
	\bigg).
\end{equation}
At NLO in $SU(3)$ $\chi$PT, this quantity remains constant up to logarithmic terms. From the chiral-continuum extrapolated value of $\sqrt{8t_0}f_{\pi K}$ we eventually extract the flow scale $t_0$ in physical units by using as physical inputs the isoQCD values for $f_\pi$ and $f_K$. Specifically, we use \cite{FlavourLatticeAveragingGroupFLAG:2021npn}
\begin{eqnarray}
	f_\pi^{\mathrm{isoQCD}} &=& 130.56(13) \ \mathrm{MeV},
	\\
	f_K^{\mathrm{isoQCD}} &=& 157.2(5) \ \mathrm{MeV}.
\end{eqnarray}
The full details of our scale setting procedure through a combination of the $O(a)$-improved Wilson results with the ones from the valence Wilson Twisted Mass regularisation can be found in \cite{MA1}.
The resulting value of $t_0$ we will use to convert our results to physical units is
\begin{gather}
\sqrt{t_0^{\mathrm{phys}}}=0.1445(5)(3) \ \fm\,,
\label{eq:t0phys}
\end{gather}
where the uncertainty is split in the same way as described above for $\phi_4^{\mathrm{phys}}$.

\begin{table}
	\begin{center}
			\begin{tabular}{c  c r r l  l  c  c  c}
				\toprule
				\text{Id} & $\beta $ & $L/a$ & $T/a$ & $\kappa_l$ & $\kappa_s$ & $m_\pi$ [MeV] & $m_K$ [MeV] & $m_\pi L $\\
					\multicolumn{5}{c}{}\\[-1em]
				\hline 
				\multicolumn{5}{c}{}\\[-1em]
				H101 & 3.40 & 32 & 96 & 0.13675962 &  0.13675962 &  416 & 416 & 5.8  \\
				\multicolumn{5}{c}{}\\[-1em]
				H102 &   & 32 & 96 & 0.136865 &  0.136549339 & 352 & 437 & 4.9 \\
				\multicolumn{5}{c}{}\\[-1em]
				H105 &   & 32 & 96 & 0.136970 &  0.13634079 & 277 & 462 & 3.9 \\
				\multicolumn{5}{c}{}\\[-1em]
				\hline 
				\multicolumn{5}{c}{}\\[-1em]
				H400 & 3.46 & 32 & 96 & 0.13688848  &  0.13688848  & 415 & 415 & 5.1  \\
				\multicolumn{5}{c}{}\\[-1em]
				\hline
				\multicolumn{5}{c}{}\\[-1em]
				N202 & 3.55 & 48 & 128 & 0.137000  &  0.137000 & 412 & 412 & 6.4  \\
				\multicolumn{5}{c}{}\\[-1em]
				N203 &  & 48 & 128 &  0.137080  &  0.136840284   & 346 & 442 & 5.4  \\
				\multicolumn{5}{c}{}\\[-1em]
				N200 &  & 48 & 128 &   0.137140  &  0.13672086  & 284 & 463 & 4.4  \\
				\multicolumn{5}{c}{}\\[-1em]
				D200 &  & 64 & 128 &   0.137200  &  0.136601748    & 200 & 480 & 4.2  \\
				\multicolumn{5}{c}{}\\[-1em]
				\hline
				\multicolumn{5}{c}{}\\[-1em]
				N300 & 3.70 & 48 & 128 &  0.137000  &  0.136601748   & 419 & 419 & 5.1 \\
				\multicolumn{5}{c}{}\\[-1em]
				J303 &  & 64 & 196 &   0.137123  &  0.1367546608  &   257 & 474 & 4.1  \\
				\multicolumn{5}{c}{}\\[-1em]
				\toprule
			\end{tabular}
			\caption{List of CLS $\NF=2+1$ ensembles used in the present study. $L/a$ and $T/a$ refer to the spatial and temporal extent respectively of the lattice. The values $\kappa_l$ and $\kappa_s$ refer to the hopping parameters of the light and strange quark masses in the sea sector. Approximate values of the pion mass $m_\pi$, the kaon mass $m_K$, and of the product $m_\pi L$ are provided in the last three columns.}
			\label{tab:CLS}
	\end{center}
\end{table}

\section{Charm correlators and scale setting}
\label{sec:charm_basics}

In this section we discuss the technical details behind the computation of physical observables in the charm region from our mixed action setup. We  introduce the GEVP setup used to extract meson masses and matrix elements throughout this work and explain our strategy to match the charm quark mass to its physical value.

\subsection{Computation of correlation functions}

To extract physical observables we have measured two-point correlation functions at zero momentum on CLS gauge configurations listed in Table \ref{tab:CLS}. Fermionic two-point correlators have the form
\begin{equation}
\label{eq:f2p}
	f^{q,r}(x_0,y_0) = \frac{a^6}{L^3}\sum_{\vec{x}, \vec{y}}
	\langle O_\Gamma^{q,r}(x_0,\vec{x})  O_{\Gamma'}^{r,q}(y_0,\vec{y})\rangle, 
\end{equation}
where $y_0$  and $x_0$ are, respectively, the source and sink time coordinates;
$q$ and $r$ are flavour indices; and a trace over spin and colour is implicit.
$O_\Gamma^{q,r}$ are quark bilinear operators defined as 
\begin{equation}
	O_\Gamma^{q,r}(x) = \overline{\psi}^q(x) \Gamma \psi^r(x),
        \label{eq:biloperators}
\end{equation}
where $\Gamma$ is a spin matrix. The operator content will be denoted by subscripts in
straightforward notation --- we will refer to $f_{\rm\scriptscriptstyle PP}$ when
$\Gamma=\Gamma'=\gamma_5$, $f_{\rm\scriptscriptstyle AP}$ when $\Gamma=\gamma_0\gamma_5$ and
$\Gamma'=\gamma_5$, and so on.

In all computations in this work we have fixed the time position of the source at $y_0=T/2$, to maximise the distance from the boundaries:
when dealing with heavy-light and heavy-heavy flavour content in the operators $O_\Gamma^{q,r}$ in Eq.~(\ref{eq:biloperators}), we observe that
the region in which the signal for the considered two-point function is accessible
lies entirely within the lattice bulk, and that the boundary effects are strongly suppressed.
Ten time-diluted $U(1)$ stochastic sources are employed in the computation of the quark propagators in each gauge field configuration.
Moreover, the numerical inversion of the quark propagator in the charm region is 
performed using distance preconditioning techniques~\cite{deDivitiis:2010ya,Collins:2017iud}, in order to
reduce signal deterioration and enhance accuracy at large  Euclidean times.
Error analysis and propagation are based on the Gamma method of~\cite{Wolff:2003sm} and
automatic differentiation, as implemented in the \texttt{ADerrors} package~\cite{Ramos:2018vgu}. 

Light and strange propagators are computed at the values of $m_{\mathrm{cr}}$, $\mu_l$ and $\mu_s$
determined to ensure maximal twist and pion and kaon masses matched to the sea
(see Section \ref{sec:mixed_action_setup}). We note that this is a independent set of computations of the  propagators with respect to those employed in the matching procedure~\cite{MA1}, where a grid of values for the mass parameters is employed to accurately interpolate to the
matching point. Moreover, this grid was also employed to compute the mass corrections to the renormalised 
chiral trajectory \cite{MA1}.  Heavy propagators are computed at three different values of the twisted 
mass $\mu_c^{(i)}$ around the physical charm region (save for one ensemble where only two masses
have been used), so that observables are interpolated at the physical 
value of the charm quark mass. In Table \ref{tab:tm_mass_values}  we specify the twisted mass values
and the critical hopping parameter $\tilde{\kappa}_{\mathrm{cr}}$ used to impose the maximal twist condition for each ensemble used in the analysis.

\begin{table}
	\begin{center}
		\begin{tabular}{c  c  c  c  c  c  c c }
			\toprule
			\text{Id} & $\beta $ & $\tilde{\kappa}_{\mathrm{cr}}$ & $a\mu_l$ & $a\mu_s$  & $a\mu_c^{(1)}$  & $a\mu_c^{(2)}$ & $a\mu_c^{(3)}$\\
			\multicolumn{5}{c}{}\\[-1em]
			\hline 
			\multicolumn{5}{c}{}\\[-1em]
			H101 & 3.40 & 0.137277 & 0.006592  & 0.006592 & 0.237975 & 0.250500 & 0.263025   \\
			\multicolumn{5}{c}{}\\[-1em]
			H102 &  & 0.137291& 0.004711 & 0.010090 & 0.228285 & 0.240300 & 0.252315   \\
			\multicolumn{5}{c}{}\\[-1em]
			H105 &   & 0.137319 & 0.002958 & 0.013690 &0.230108 &  0.242219 & 0.254330 \\
			\multicolumn{5}{c}{}\\[-1em]
			\hline 
			\multicolumn{5}{c}{}\\[-1em]
			H400 & 3.46 & 0.137292 & 0.006006 & 0.006006 & 0.204155 & 0.214900 & 0.225645   \\
			\multicolumn{5}{c}{}\\[-1em]
			\hline
			\multicolumn{5}{c}{}\\[-1em]
			N202 & 3.55 &  0.137298 & 0.005160 & 0.005160 & 0.167105 & 0.175900 &  0.184695 \\
			\multicolumn{5}{c}{}\\[-1em]
			N203 &  & 0.137307 & 0.003609 & 0.010770 & 0.172805  & 0.181900 & 0.190995   \\
			\multicolumn{5}{c}{}\\[-1em]
			N200 &  & 0.137310  & 0.002403 & 0.008432 & 0.173375 & 0.182500 & 0.191625  \\
			\multicolumn{5}{c}{}\\[-1em]
			D200 &  & 0.137316 & 0.001227 & 0.013170 & 0.172900 & 0.191100 &  --   \\
			\multicolumn{5}{c}{}\\[-1em]
			\hline
			\multicolumn{5}{c}{}\\[-1em]
			N300 & 3.70 & 0.137207 & 0.004060 & 0.004060 & 0.130910 & 0.137800 & 0.144690 \\
			\multicolumn{5}{c}{}\\[-1em]
			J303 &  & 0.137212 & 0.001610 & 0.009570 & 0.133000 & 0.140000 & 0.147000  \\
			\multicolumn{5}{c}{}\\[-1em]
			\toprule
		\end{tabular}
		\caption{List of run parameters for each ensemble in Table \ref{tab:CLS}. The critical value of the hopping parameter required to set the valence sector to maximal twist~\cite{MA1} is denoted by $\tilde{\kappa}_{\mathrm{cr}}$. The values of $a\mu_l$ and $a\mu_s$ are the light and strange bare twisted quark masses, in lattice units, that match the corresponding sea quark masses~\cite{MA1}. Finally, the last three columns contain the three values of heavy bare twisted quark masses in the charm region. In the case of the D200 ensemble two values that straddle the charm point were considered.}
		\label{tab:tm_mass_values}
	\end{center}
\end{table}

\subsection{Extraction of meson masses}
 
In our analysis meson masses are employed to fix the renormalised line of constant physics and match the quark masses to some target physical value.  Light and strange quark masses are matched between the sea and valence sectors using $\phi_2$ and $\phi_4$ in Eq.~(\ref{eq:phi2_and_phi4}), whereas for the partially quenched charm quark we use different combinations of mesons masses matched to their physical values, as explained in Section  \ref{subsec:matching_charm}.
 
The ground state meson masses are extracted from a generalised eigenvalue problem (GEVP) variational method defined as 
  \begin{equation}\label{eq:gevp_sec3}
 	C(t) v_n(t,t_{\mathrm{ref}}) = \lambda_n(t,t_{\mathrm{ref}}) C(t_{\mathrm{ref}})v_n(t,t_{\mathrm{ref}}) \qquad n=0,\ldots,N-1 , \quad t>t_{\mathrm{ref}},
 \end{equation}
where $C(t)$ is a matrix of Euclidean correlation functions of the form
in \req{eq:f2p}, such that the indices $i,j$ in $C_{ij}(t)$ correspond to different
choices of $\Gamma,\Gamma'$ and source/sink location, and $t=x_0-y_0$.
This leads to the spectral expansion
 \begin{gather}
 \begin{split}
 	C_{ij}(t) &= \sum_{n=0}^{\infty} e^{-E_n t}\varphi_{ni} \varphi_{nj}^*, \quad i,j=0,\ldots, N-1 \,;\\
 \varphi_{ni} &\equiv \langle0|O_i|n\rangle.
\end{split}
\end{gather}
Here $N$ denotes the matrix dimension, and we have assumed non-degenerate energy levels.
 The GEVP  is solved in the regime  $t_{\mathrm{ref}} \geq t/2$, where a better control over excited state contributions is achieved \cite{Blossier:2009kd}.  The matrix $C(t)$ in our setup is built from pseudoscalar 
 two-point functions $f_{\rm\scriptscriptstyle PP}$ shifted in time as
 \begin{equation}
 	C_{\rm\scriptscriptstyle P}(t) = \bigg[\begin{matrix}
 		f_{\rm\scriptscriptstyle PP}(t)  &  f_{\rm\scriptscriptstyle PP}(t+\tau)
 		\\
 		f_{\rm\scriptscriptstyle PP}(t+\tau)  & f_{\rm\scriptscriptstyle PP}(t+2\tau)
 	\end{matrix}\bigg] \,,
 	\label{eq:gevp_matrix}
 \end{equation}
 where $\tau$ is the value of the time shift. Several values of the time shift have been tested, and we observe a mild dependence on small values of $\tau$ for the extraction of eigenvalues and eigenvectors. We refer to Appendix \ref{app:gevp} for a detailed discussion of our setup, together with sanity checks on the GEVP. In what follows we set $\tau=3a$.
  
\begin{figure}
  	\centering
  	\includegraphics[scale=0.5]{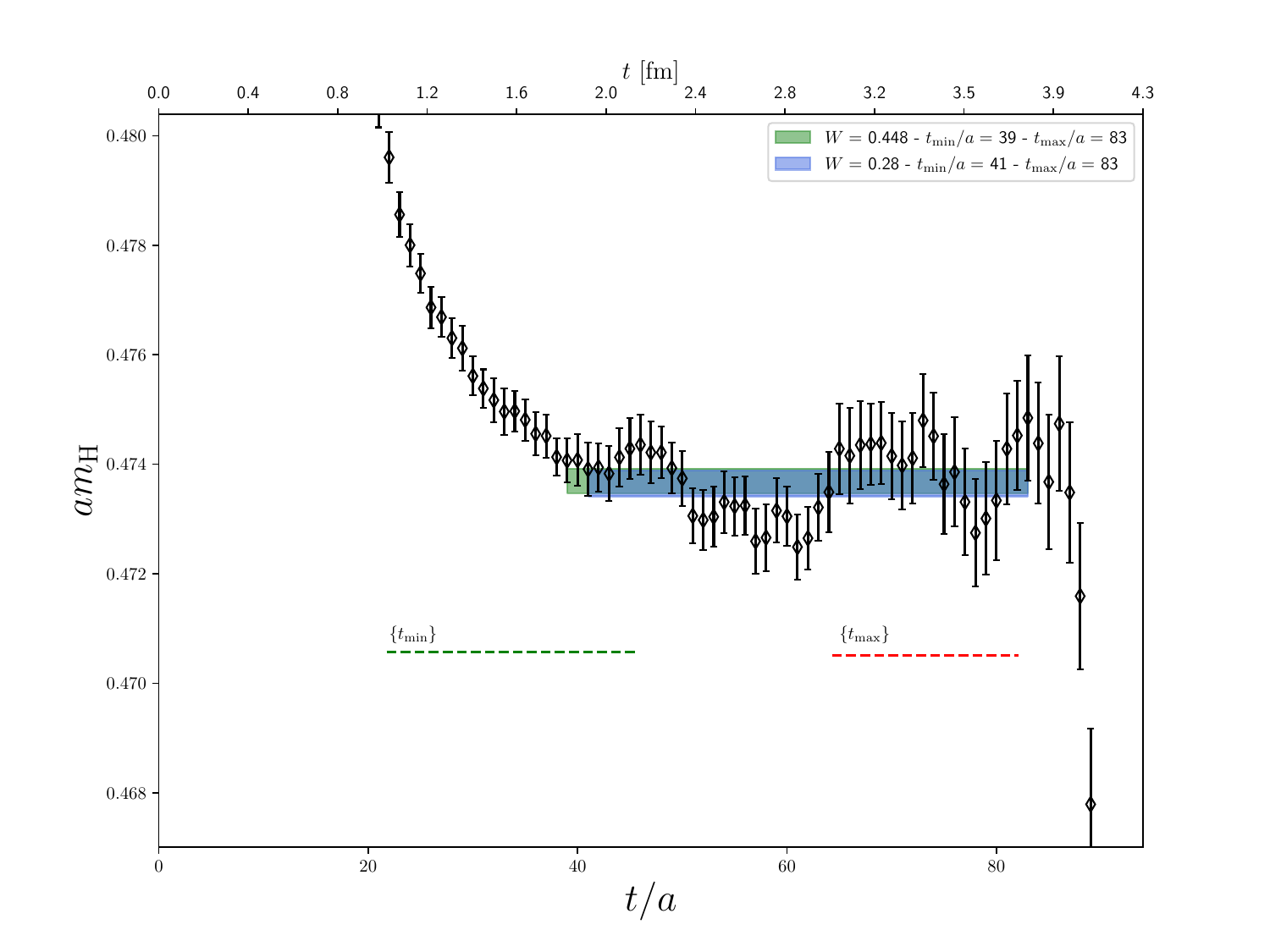}
  	\includegraphics[scale=0.5]{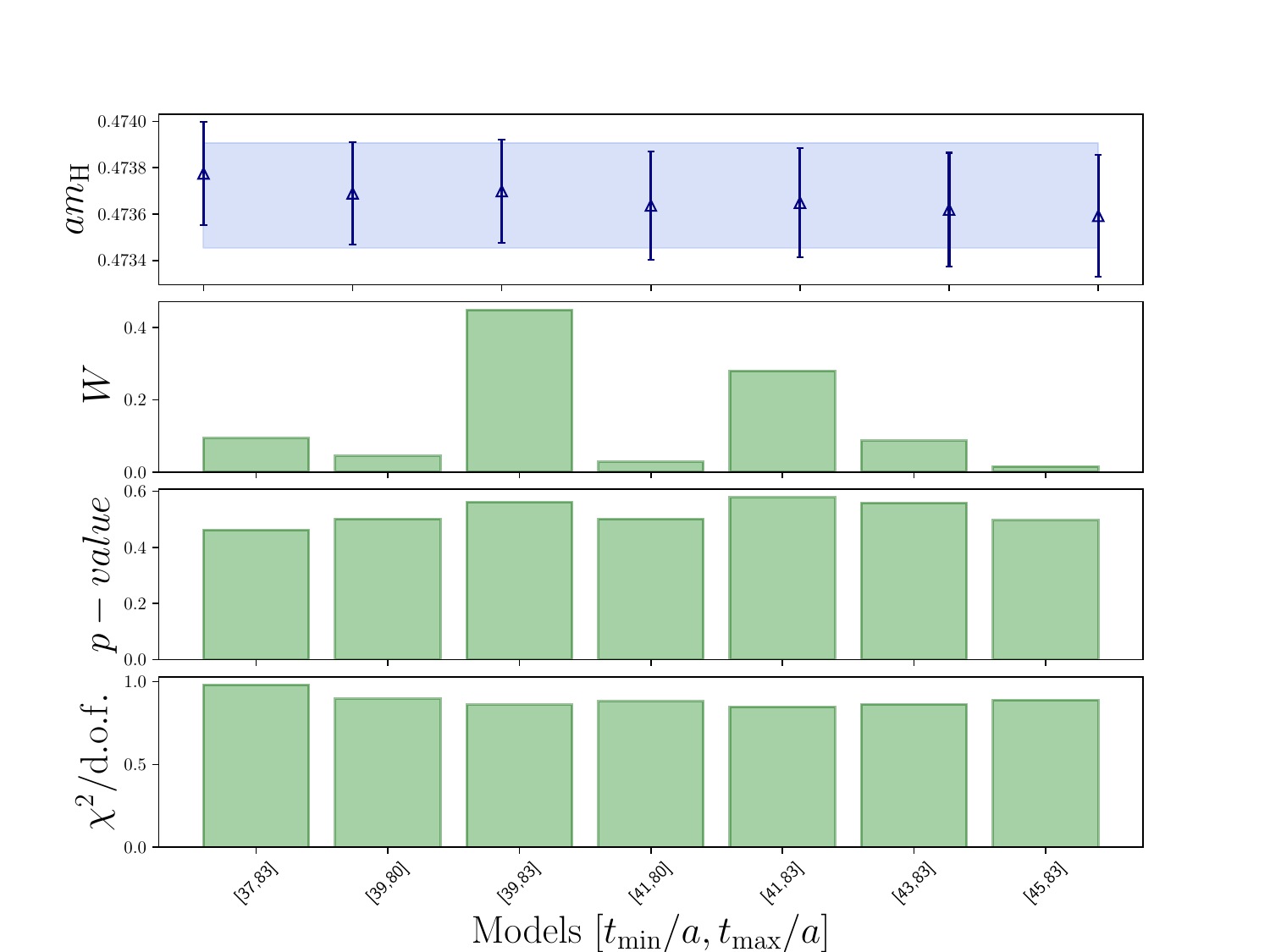}
  	\caption{Illustration of the extraction of the ground-state mass after applying a GEVP analysis, illustrated for the ensemble J303. \textit{Top}: Heavy-light pseudoscalar meson mass plateau showing the  two fit intervals with higher weights $W$ contributing to the model average. We also indicate the range of variations allowed for the interval in Euclidean time where the plateau is taken. \textit{Bottom}: Summary of determinations of $am_{\mathrm H}$
when considering variations over the fit intervals $[t_{\rm min}/a, t_{\rm max}/a]$ together with  the corresponding normalised weights $W$ based on Takeuchi's Information
Criterion (TIC), p-values and $\chi^2/{\rm d.o.f.}$. In the upper panel, the shaded blue band  corresponds to the model average result.} 
  	\label{fig:meff_plateau} 
  \end{figure}

The ground state meson mass is extracted from the eigenvalues of the GEVP using Eq.~(\ref{eq:eff_en_gevp}). 
In order to assess the systematic effects and correctly identify the plateau region, we perform several  uncorrelated $\chi^2$ fits to a constant, 
by varying the time ranges of the fitting interval. Correlated fits are impractical due to the fact that sample covariance matrices 
display very small modes and thus have ill-behaved inverses.
However, as the data is correlated, the uncorrelated $\chi^2$ is not a suitable quantity
to assess the goodness-of-fit;
we therefore quantify the latter with the expectation value of $\chi^2$,
denoted $\chi^2_{\mathrm{exp}}$, and the corresponding p-value, as introduced
in~\cite{Bruno:2022mfy}. Through this procedure we assign a weight to each fit based on 
the $\chi^2$ minimisation, and we eventually extract our ground state masses by means of the model 
averaging procedure described in Appendix~\ref{app:TIC}.
An example of a GEVP plateau for the heavy-light pseudoscalar mass together with a summary of the model average procedure for an ensemble used in the analysis is shown in Figure \ref{fig:meff_plateau}. 
  
\subsection{Matching of the charm quark mass}
\label{subsec:matching_charm}
In Section \ref{sec:mixed_action_setup} we recalled the matching of the light sector worked out in~\cite{MA1}, which ensures that physical observables involving only light and strange quarks computed in the valence and sea sectors coincide up to cutoff effects, so that
unitarity is recovered in the continuum limit.
A similar procedure is needed for the charm quark, designed to ensure that its physical value
is obtained upon taking the continuum limit and performing chiral fits.
Since the charm is partially quenched this matching procedure involves observables with only valence charm quark propagators.
In order to establish a connection with the physical point, we require that some charm-like
observable $\mathcal{O}_c$ matches its physical value.
In this paper we studied three different charm scale settings based on three choices of
$\mathcal{O}_c$, all in terms of meson masses; we will denote the latter as $m_H^{(i)},~i=1,2,3$,
and often express them in units of $\sqrt{8t_0}$ as $\phi_H^{(i)} = \sqrt{8t_0}m_H^{(i)}$.

The first possibility, corresponding to $\phi_H^{(1)}$, consists in using the flavour average meson mass combination
\begin{equation}
        m_H^{(1)} = m_{\overline{H}} \equiv \frac{2}{3} m_H + \frac{1}{3}m_{H_s},
        \label{eq:fl_av_matching}
\end{equation}
built from heavy-light $H$ and heavy-strange $H_s$ pseudoscalar meson masses with heavy-quark masses in neighbourhood of the charm.
Since we require the considered CLS ensembles to hold a constant value of the flavour average combination of pion and kaon masses -- denoted as $\phi_4$ in Eq.~(\ref{eq:phi2_and_phi4}) --
  we also expect the flavour average combination  $\phi_H^{(1)}$ to remain fairly constant along the chiral trajectory. The physical value of $m_H^{(1), \mathrm{phys}}$ is obtained by setting $m_{H_{(s)}}$ to the following prescription for the isoQCD values of $D_{(s)}$ meson masses,
  \begin{equation}
    m_D^{\mathrm{isoQCD}} = 1867.1 \pm 2.6  \ \mathrm{MeV}, \qquad m_{D_s}^{\mathrm{isoQCD}} = 1967.1 \pm 1.3 \ \mathrm{MeV}.
    \label{eq:DsisoQCDinputs}
  \end{equation}
  The uncertainties in these isoQCD values are chosen to cover the deviation with respect to the experimental values~\cite{ParticleDataGroup:2022pth} of the $D^{\pm}$ and $D_s^{\pm}$ meson masses, $m_{D^\pm}^{\mathrm{exp}} = 1869.66(5) \ \mathrm{MeV}$ and $m_{D_s^\pm}^{\mathrm{exp}} = 1968.35(7) \ \mathrm{MeV}$, respectively. We observe that the larger uncertainty in the isoQCD inputs of the $D$ and $D_s$ meson masses in Eq.~(\ref{eq:DsisoQCDinputs}) --- as compared to the corresponding experimental values --- does not induce a significant increase in the uncertainties of our target results. The input values in Eq.(~\ref{eq:DsisoQCDinputs}) lead to the following flavour averaged meson mass,
\begin{equation}
         m_H^{(1), \mathrm{phys}} = m_{\overline{D}} = 1900.4(1.8) \ \mathrm{MeV}\,.
\end{equation}

\begin{figure}[!t]
   	\centering
   	\includegraphics[scale=0.52]{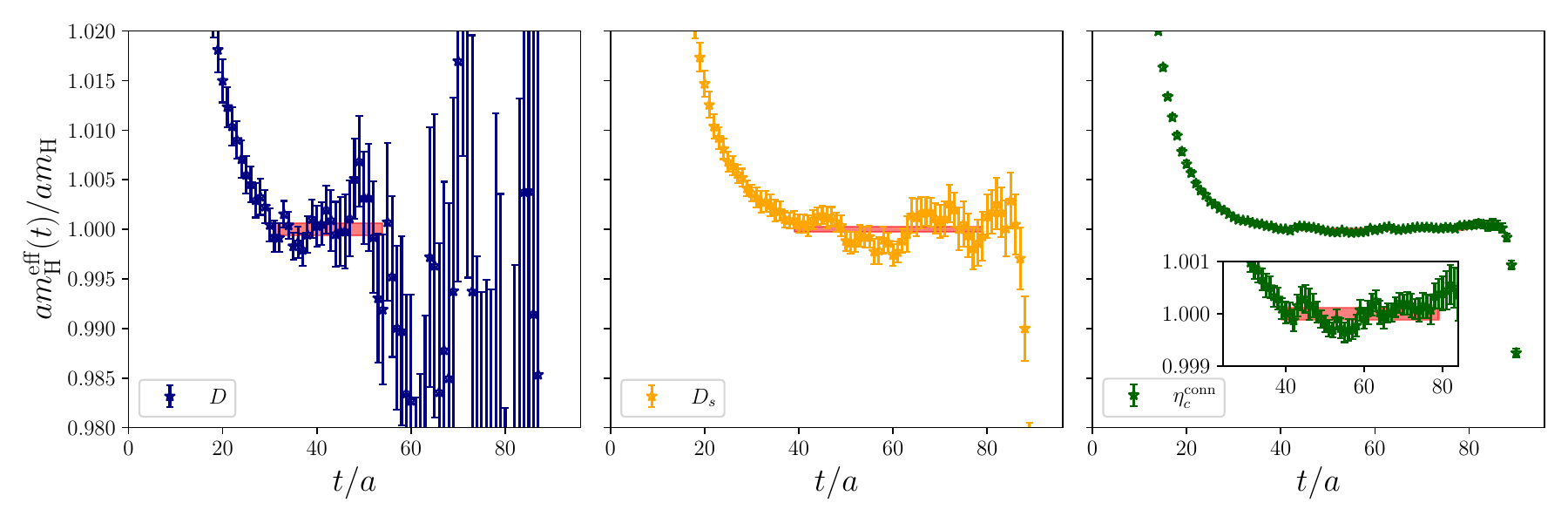}
   	\caption{Illustration of the effective meson masses involved in the matching procedure to the physical charm scale for the ensemble J303. We show three cases where the effective mass of the pseudoscalar meson $H$ is that of the $D$ (\textit{left}), $D_s$ (\textit{center}) and $\eta_c^{\mathrm{conn}}$  (\textit{right}), normalised by the central value of the corresponding plateau averaged mass.  The horizontal red bands show the results of the highest weight fit contributing to the model average procedure and the corresponding plateau interval. We observe the expected increase of the statistical uncertainties at large time separations when increasing the mass-difference among the quarks propagators of the pseudoscalar two-point correlators. 
}
   	\label{fig:corr_comparison}
   \end{figure}

\begin{figure}[!t]
 	\centering
 	\includegraphics[scale=0.3]{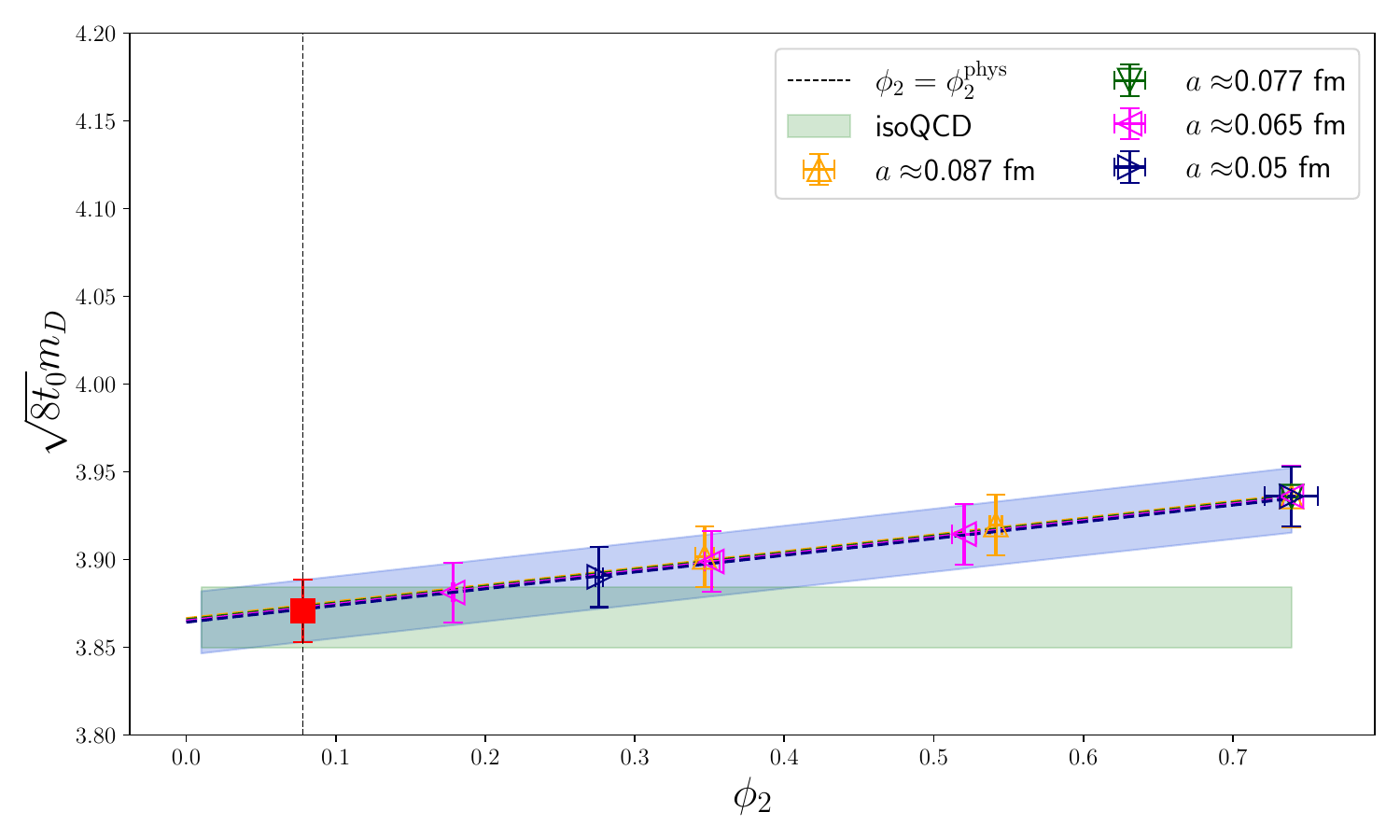}
 	\includegraphics[scale=0.3]{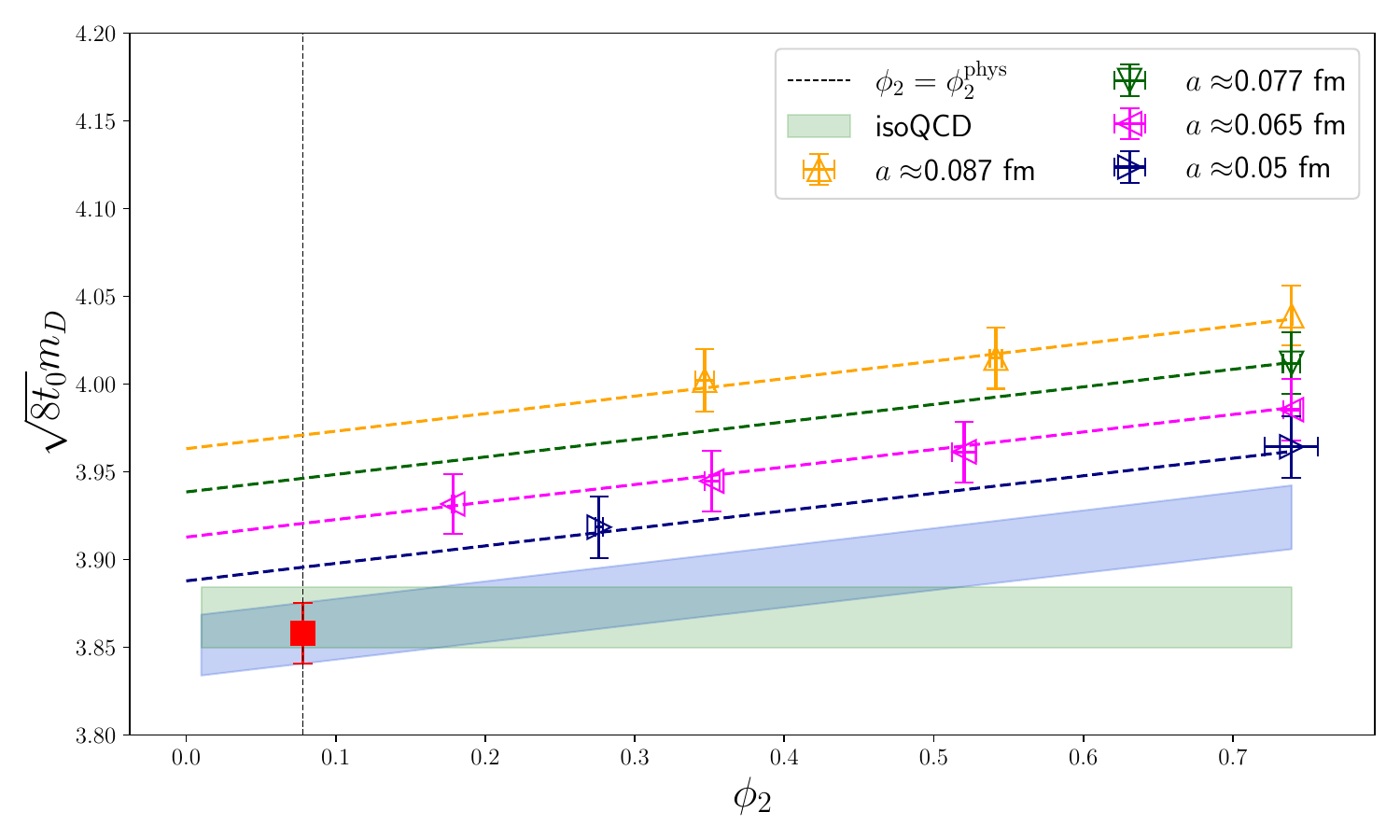}
 	\\
 	\includegraphics[scale=0.3]{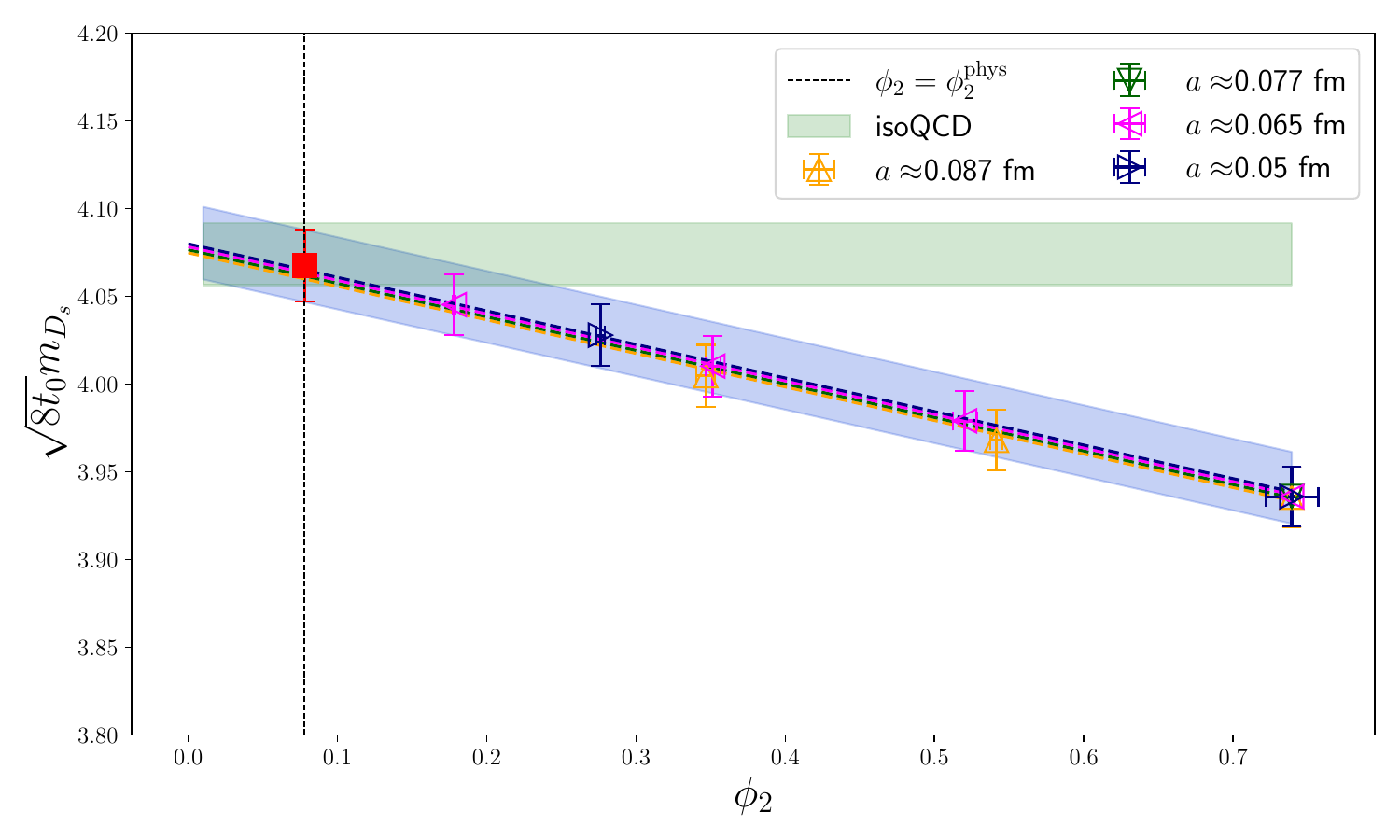}
 	\includegraphics[scale=0.3]{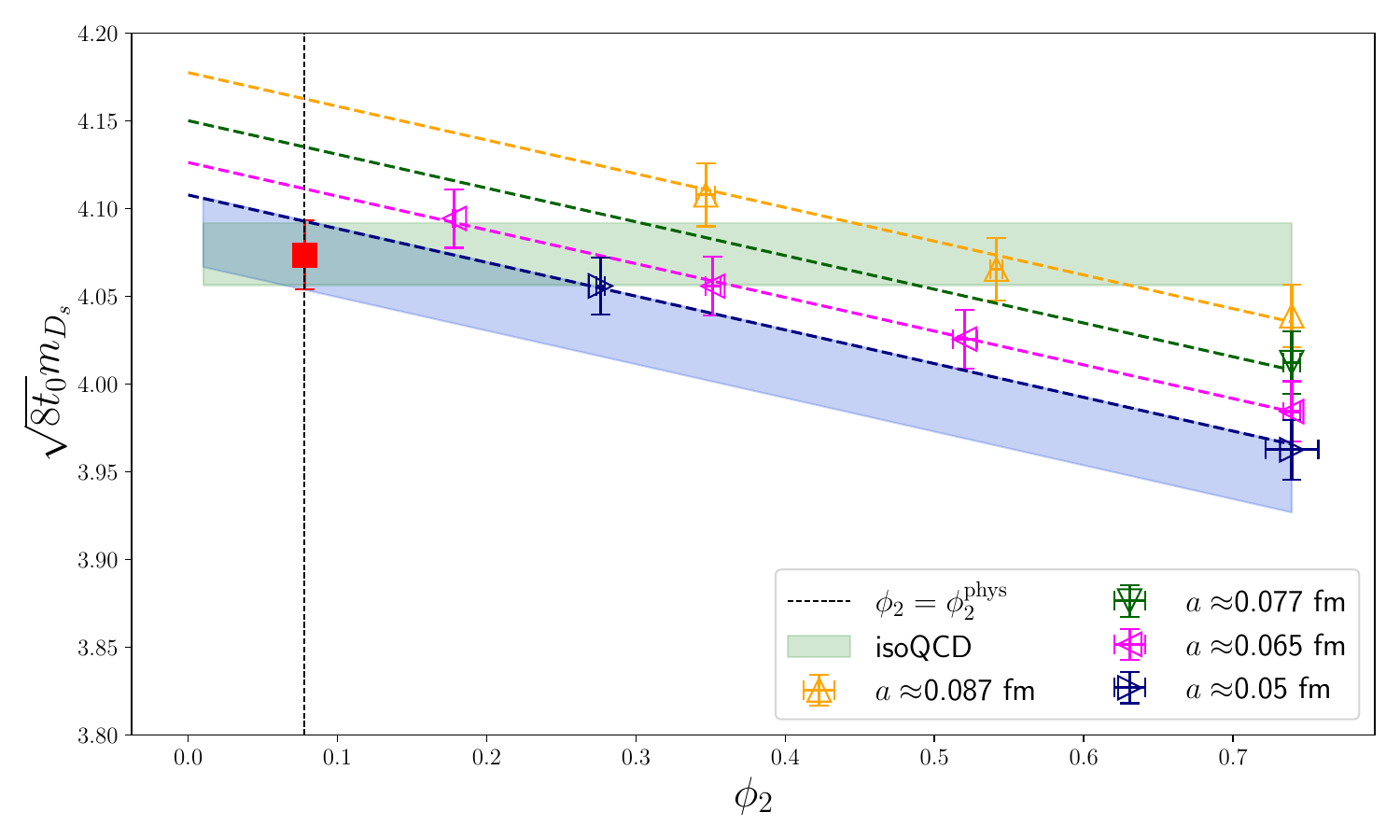}
 	\caption{Consistency checks of our charm matching strategy. We show the chiral extrapolation to the physical point of
	the ${D_{(s)}}$ meson mass in units of $\sqrt{8t_0}$ using all the ensembles listed in Table \ref{tab:CLS}.
	The left panels use the flavour-averaged mass combination, $m_H^{(1)}= m_{\overline{H}}$, while those on the right
	use the mass-degenerate pseudoscalar meson mass, $m_H^{(2)}=m_{\eta_h}^{\mathrm{conn}}$. The empty symbols correspond to the $D_{(s)}$ meson masses determined on a given ensemble,
	while the red square symbols show the extrapolated values at the physical point. Dashed lines
	show the fit forms projected to each individual lattice spacing, and the blue shaded bands
	are a projection to the continuum limit on the chiral plane. Data points are projected to the physical point $\phi_H^{(i),\mathrm{phys}}$. Finally, the green horizontal band shows
	the isoQCD input values for the corresponding masses in Eq.~(\ref{eq:DsisoQCDinputs}), in units of $\sqrt{8t_0}$.
        }
 	\label{fig:ds_matching} 
 \end{figure}

Our second strategy, corresponding to $\phi_H^{(2)}$, is to
consider the mass-degenerate pseudoscalar meson mass $m_{\eta_h}^{\mathrm{conn}}$ extracted from
the quark-connected two-point correlation function made of heavy quark
propagators with a mass in the neighbourhood of the charm mass,
\begin{equation}
  m_H^{(2)} = m_{\eta_h}^{\mathrm{conn}}\,.
                  \label{eq:etac_matching}
\end{equation}
The physical value for this mass, $m_H^{(2), \mathrm{phys}}$,  is set from the experimental
value of the $\eta_c$ meson mass~\cite{ParticleDataGroup:2022pth},
$m_{\eta_c}^{\mathrm{exp}} = 2983.9(4)\,\MeV$, from which a
correction of about 6\,\MeV, with 100\% error, is subtracted to account
for the absence of quark-disconnected diagrams and QED effects~\cite{deForcrand:2004ia, Donald:2012ga,Colquhoun:2015oha,Hatton:2020qhk,Colquhoun:2023zbc}. Specifically, we employ, 
\begin{equation}
  m_H^{(2), \mathrm{phys}} = m_{\eta_c}^{\mathrm{conn}} = 2978(6) \ \mathrm{MeV}\,.
\end{equation}
One potential advantage of this choice of matching observable is that
the overall precision of the $\eta_c^{\mathrm{conn}}$ meson mass is substantially better than the one
for heavy-light meson masses, as it does not suffer from the increase in noise-to-signal
ratio with Euclidean time; this is illustrated in Figure \ref{fig:corr_comparison},
where we show the $D$, $D_s$ and $\eta_c^{\mathrm{conn}}$ pseudoscalar correlators for a one specific ensemble.
Finally, as a third matching quantity we also tested the spin-flavour averaged mass combination
 \begin{equation}
 	m_H^{(3)} = m_{\overline{H}^*} = \frac{1}{12} \left(
 	2m_H + m_{H_s} + 6 m_{H^*} + 3 m_{H_s^*}
 	\right),
 	\label{eq:spin_flavour_av}
 \end{equation}
which involves a combination of heavy-light pseudoscalar  $m_{H_{(s)}}$ and vector  $m_{H_{(s)}^*}$ meson masses in the charm region, and
is motivated by heavy-quark symmetry. However, we observe that chiral-continuum fits coming from the spin flavour-averaged matching condition
lead to worse $\chi^2$ values, and as a result their weights are highly suppressed by  our model average prescription. We interpret this finding as a reflection of relatively poor control of heavy-light vector states, whose masses are extracted with significantly larger errors than those of heavy-light pseudoscalar states. In the rest of the discussion we will therefore focus on the results coming from the other two matching conditions.
Any of these matching conditions can in principle be imposed ensemble by ensemble,
even away from the physical point.
However, by doing so we would as a result build in the charm quark mass a dependence on the value of
the reference scale $t_0^{\mathrm{phys}}$, as well as $O(a^2)$ effects coming from the
specific choice of $\mathcal{O}_c$.
To avoid this, we have opted instead for setting the physical charm quark mass
jointly with the chiral-continuum
extrapolation, in a similar way as the one we employ to hit the physical point in the
light and strange sector.
What this means in practice is that the charm quark mass dependence of any given observable
$\cO$ is parameterised as $\mathcal{O}(a, \phi_2, \phi_H^{(i)})$, and we perform a global
fit to obtain its physical value $\mathcal{O}(0, \phi_2^{\mathrm{phys}}, \phi_H^{(i),\mathrm{phys}})$.
This will be the procedure applied below in the determination of the physical value of
the charm quark mass and of the decay constants $f_D$ and $f_{D_s}$.
Note that, as a consequence of our matching procedure and of working on a line of constant
physics where $\phi_4$ is kept constant, it is non-trivial that by adopting any of our 
matching procedures the mass of any particular meson reaches its physical value in the
chiral-continuum limit;
checking that it does is therefore a test of the robustness of our procedure.
As an illustration, we show in Fig.~\ref{fig:ds_matching} how the physical values
of the $D$ and $D_s$ meson masses arise when the charm scale is matched through either
$m_{\overline D}$ or $m_{\eta_c}^{\mathrm{conn}}$.
In either case we show results for the specific model of the lattice spacing, charm mass and pion mass
dependence of the form
\begin{equation}
\sqrt{8t_0}\, m_{D_{(s)}}(a, \phi_2, \phi_H^{(i)}) = p_0 + p_1\phi_2 +  p_2 \phi_H^{(i)} + c_1\frac{a^2}{8t_0},
\end{equation}
where $i=1,2$ according to the notation introduced above and where $c_1$ and $p_j$, $j=1,2,3$, stand for the fit parameters.
Note that the agreement is excellent, in spite of the different implications of the
two setups for the specific case of $m_{D_{(s)}}$; for instance, when $m_{\overline D}$ is used
for the matching cutoff effects are very small by construction, while the use of
$m_{\eta_c}^{\mathrm{conn}}$ leads to sizeable cutoff effects which are however
very well described by an $\mathcal{O}(a^2)$ term.

\section{Determination of the charm quark mass}
\label{sec:mc}

\subsection{Renormalised charm quark masses}

In Sec.~\ref{sec:mixed_action_setup} we have summed up the argument why renormalised quark
masses can be easily retrieved from bare Lagrangian twisted masses.
In our mixed-action setup, as discussed in detail in~\cite{MA1}, the resulting $O(a)$-improved expression for the renormalised  charm mass $\mbar_c(\mu)$ reads
\begin{equation}
	\mbar_c(\mu) = 
	 \ZP^{-1}(g_0^2, a\mu) [1 + a\overline{b}_\mu(g_0^2) \rm{tr}\{\mathbf{m}^{(s)}\}]\,\mu_c\,,
	\label{eq:renormalised_charm_mass}
\end{equation}
where $\ZP(g_0^2, a\mu)$ is a suitably defined renormalisation constant for the non-singlet
pseudoscalar density at renormalisation scale $\mu$.
As we have already discussed, the improvement term $\propto\rm{tr}\{\mathbf{m}^{(s)}\}$
can be neglected in practice, so $O(a)$-improved renormalised quark masses can be obtained
by just applying the renormalisation constants to the exactly known Lagrangian masses.
In this work we will use the non-perturbative values of $\ZP$ computed in~\cite{Campos:2018ahf}
in the Schr\"odinger Functional scheme, at a fixed renormalisation scale
$\mu_{\rm\scriptscriptstyle had} = 233(8)~\MeV$ and for the range of values of $g_0^2$ covered
by CLS.
It will be used to obtain renormalised quark masses for each of our ensembles, that can then
be used to determine the value of the charm quark mass in the continuum and at physical kinematics.
Contact with other renormalisation schemes can then be made by computing the renormalisation
group invariant (RGI) quark mass $M_c^{\mathrm{RGI}}$, using the continuum (flavour-independent)
ratio also computed in~\cite{Campos:2018ahf}
\begin{equation}
	\frac{M}{\overline{m}(\mu_{\mathrm{had}})} = 0.9148(88)\,.
	\label{eq:rgi_running_factor}
\end{equation}
Values of renormalised masses in, say, the $\MSbar$ scheme can then be obtained by
using the perturbative value of $\frac{\overline{m}(\mu)}{M}$ at any convenient scale $\mu$.
%


\subsection{Charm quark mass chiral-continuum fits}
\label{subsec:mc_chiral_continuum}

Having determined the  renormalised charm quark masses in the Schr\"odinger Functional scheme at the hadronic renormalisation scale $\mu_{\mathrm{had}}$
\begin{equation}
\mbar_c(\mu_{\rm\scriptscriptstyle had}) \equiv \mu_c^{\rm\scriptscriptstyle R}\,,
\end{equation}
for all the ensembles listed in Table \ref{tab:CLS}, we now describe our 
strategy to obtain results in the continuum limit and at the physical point,
following the approach outlined in Sec.~\ref{sec:charm_basics}. The matching procedure of the light
and strange sectors is already devised so that the physical value of the kaon mass is recovered
at $\phi_2 = \phi_2^{\mathrm{phys}}$, where the physical value of $\phi_2$ is computed
with the isospin-symmetric values of the pion mass quoted 
in~\cite{FlavourLatticeAveragingGroupFLAG:2021npn}, and the physical scale $t_0^{\mathrm{phys}}$
is the one determined in~\cite{MA1}. The charm scale is matched through the two different
prescriptions described in Sec.~\ref{sec:charm_basics}. All quantities entering the fit
are made dimensionless through the appropriate power of the factor $\sqrt{8t_0}$,
and physical units for the final result are restored by using our value for $t_0^{\mathrm{phys}}$.

\begin{figure}[!htb]
 	\centering
 	\includegraphics[scale=0.50]{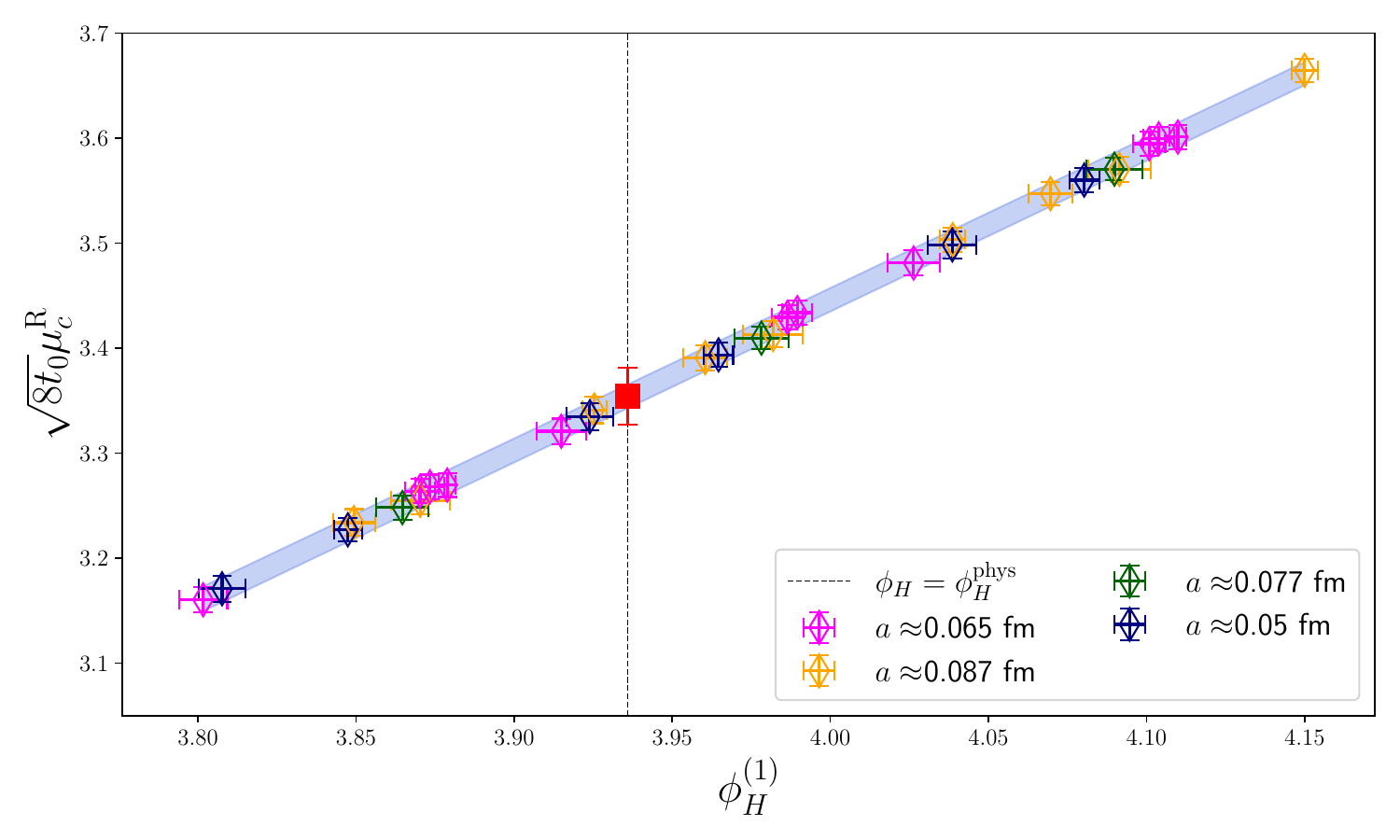}
 	\includegraphics[scale=0.50]{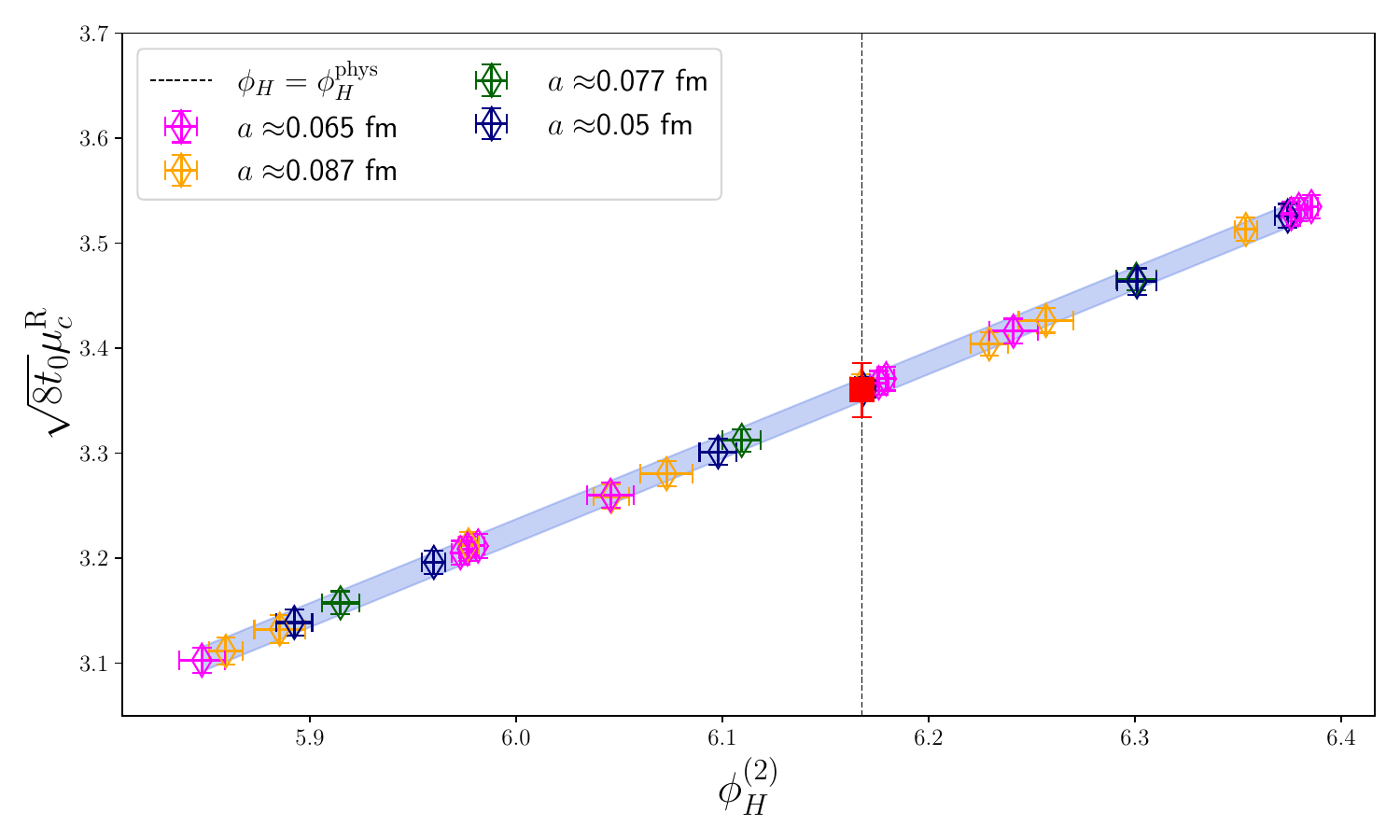}
 	\caption{ Heavy mass dependence of the renormalised charm quark mass $\mu_c^{\rm\scriptscriptstyle R}$ in units of $\sqrt{8t_0}$ for the fits with larger weights according to the TIC criteria. \textit{Top}: Results shown for the flavour-averaged matching condition $\phi_{H}^{(1)} = \sqrt{8t_0} m_{\overline{H}}$. \textit{Bottom}: Results shown for the $\eta_h^{\mathrm{conn}}$ matching condition $\phi_{H}^{(2)} = \sqrt{8t_0} m_{\eta_h}^{\mathrm{conn}}$. Dependencies other than $\phi_H^{(i)}$ in the chiral-continuum extrapolation have been projected to the physical point. The red square symbols indicate the continuum results at the physical value $\phi_H^{\mathrm{phys}}$. We observe a linear dependence of the charm quark mass on the different matching conditions used in this work. }
 	\label{fig:mc_mh_dependence}
 \end{figure}

We parameterise the continuum dependence of the renormalised charm quark mass on $\phi_2$
and any of the $\phi_H^{(i)}$ with the functional form
\begin{equation}
	\sqrt{8t_0}\, \mu_c^{\rm\scriptscriptstyle R}(a=0, \phi_2, \phi_H) = p_0 + p_1\phi_2 + p_2\phi_H\,.
	\label{eq:mc_continuum_parameterization}
\end{equation}
Based on the heavy quark effective theory expansion~\cite{Georgi:1990um} at lowest order,
we expect a linear dependence of the charmed meson masses as a function of the the charm quark 
mass, hence the latter term in the ansatz. This assumption is supported by our data that show indeed a 
linear behaviour in the charmed meson masses, as illustrated in Figure \ref{fig:mc_mh_dependence}. Note that this form is used only to describe the dependence
within a short interval in mass values, and interpolate the charm scale from points close by. When considering the pion dependence of the charm quark mass, we assume that the  leading order contributions exhibit a linear behaviour in $\phi_2$. With the current set of ensembles employed in this work we do not observe any deviations from the leading order term in the pion mass dependence.

Regarding the lattice spacing dependence of the charm quark mass, we assume the leading cutoff effects to 
be $O(a^2)$, as discussed above. Corrections of odd order in $a$ are generically expected to be highly
suppressed at maximal twist, by way of the extension of the argument for automatic $O(a)$
improvement; we thus include $a^4$ terms to account for deviations from linear behaviour
in $a^2$. Finally, we allow for terms proportional to $m_\pi^2$ and to various powers of the charm
mass. The generic ansatz to parameterise lattice spacing dependence thus take the following form
\begin{equation}
	c_{\mu_c}(a, \phi_2, \phi_H) = \frac{a^2}{8t_0} \big(
	c_1 + c_2\phi_2 + c_3 \phi_H^2
	\big)
	+
	\frac{a^4}{(8t_0)^2}\big(
	c_4 + c_5\phi_H^2 + c_6 \phi_H^4
	\big).
	\label{eq:lattice_spacing_dependence}
\end{equation} 

In order to estimate the systematic effects arising from the model variation, we consider all the possible 
combinations where some of the $c_i$ coefficients vanish, save for $c_1$ which is always kept.
Furthermore, following~\cite{Heitger:2021apz}, we allow for cutoff effects to enter either linearly or 
non-linearly, viz.,
  \begin{eqnarray} 	\label{eq:tot_model}
 	\sqrt{8t_0}\mu_c^{R,\text{linear}}(a, \phi_2,\phi_H) &=&
 	\sqrt{8t_0}\mu_c^{R,\text{cont}}(0, \phi_2,\phi_H) + c_{\mu_c}(a, \phi_2,\phi_H),
 	\\
 	\sqrt{8t_0}\mu_c^{R,\text{non-lin}}(a, \phi_2,\phi_H) &=& 
 	\sqrt{8t_0}\mu_c^{R,\text{cont}}(0, \phi_2,\phi_H) \big(1+ c_{\mu_c}(a, \phi_2,\phi_H)\big). \nonumber
 \end{eqnarray}
We thus end up with a total of 64 functional forms for each of the two charm matching conditions,
i.e., a total of 128 models.
Fit parameters are estimated minimising an uncorrelated
$\chi^2$ where, however, the covariance between the independent variables and the data is taken into account. As previously discussed, the goodness-of-fit of fit can still be obtained in this case from the measurement of the $\chi^2_{\mathrm{exp}}$ and the associated p-value. The TIC result for each model is then fed into the model averaging procedure summarised in App.~\ref{app:TIC},
which finally allows to quote a systematic uncertainty that reflects the fluctuations
engendered by the variety of fit ansaetze.

In Table \ref{tab:mc_results_all_matching} we report the results for $\mu_c^{\rm\scriptscriptstyle R}$
in units of $\sqrt{8t_0}$ obtained with each of the two matching conditions independently,
as well as for the combined model average.  

\begin{figure}[!htb]
	\centering
	\includegraphics[scale=0.42]{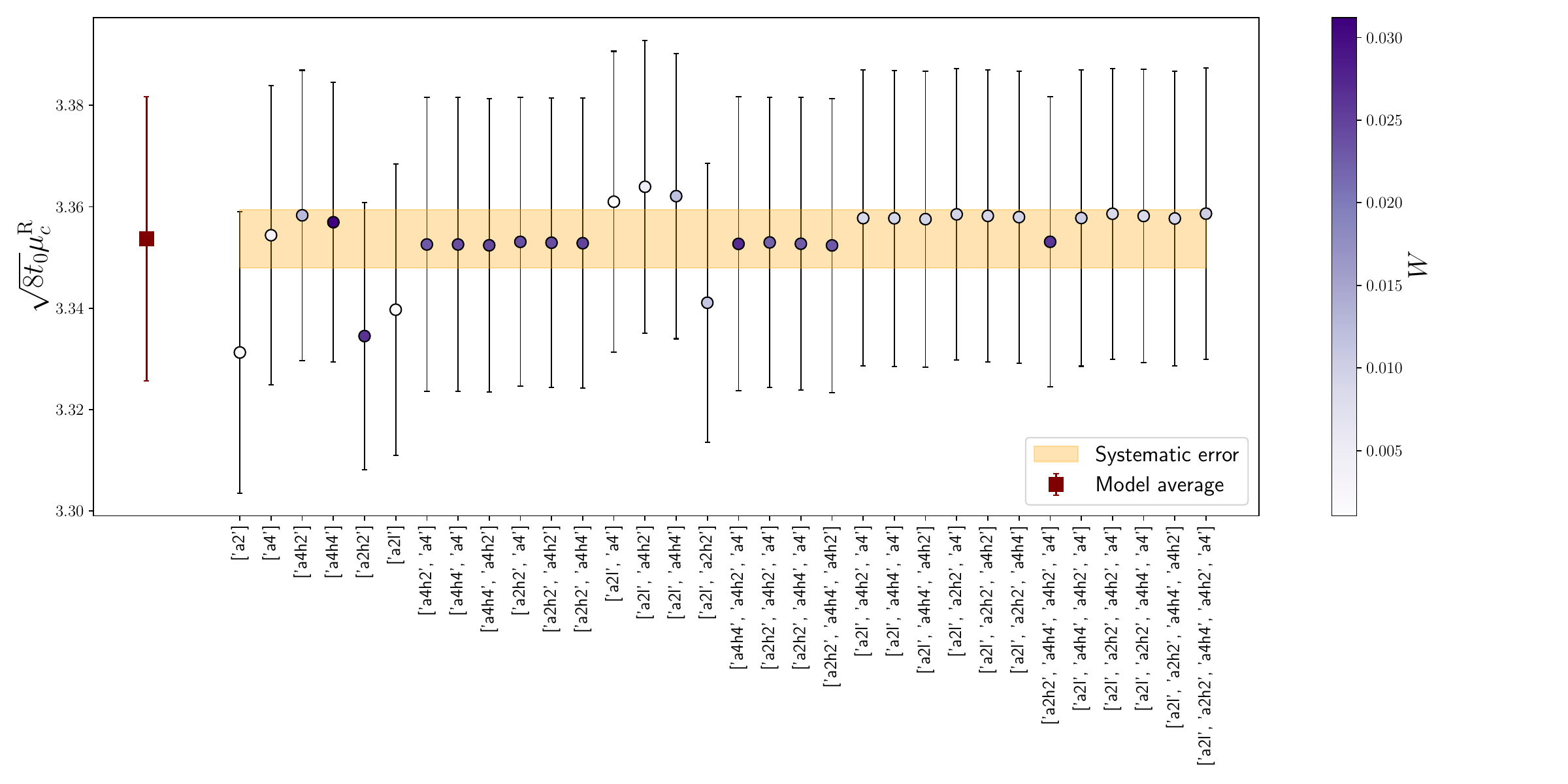}
	\caption{Model average procedure for the renormalised charm quark mass $\mu_c^{\rm\scriptscriptstyle R}$ in units of $\sqrt{8t_0}$. We collect a subset of the best results according to the TIC procedure, coming from different models, for the flavour-averaged  matching condition $\phi_H^{(1)}$.   The opacity of each circle data point reflects the associated normalised weights $W$ as given from the TIC. The yellow shaded band represents the systematic error computed with Eq.~(\ref{eq:weighted_variance}), while the left-most red square symbol corresponds to the result extracted from the model average procedure. The labels of the 32 models specified in the horizontal axis are related to the terms appearing in Eq.~(\ref{eq:lattice_spacing_dependence}) -- characterising the lattice spacing dependence -- in the following way: \texttt{`a2'} corresponds to the term depending on the fit parameter $c_1$. Similarly, \texttt{`a2l', `a2h2', `a4', `a4h2', `a4h4'} refer to $c_2,\dots, c_6$, respectively. Given that the parameter $c_1$ is included in all the models, the associated label is not explicitly specified for all cases appearing in the horizontal axis.}
	\label{fig:mc_model_av_summary}
\end{figure}

\begin{figure}[!htb]
	\centering
	\includegraphics[scale=0.65]{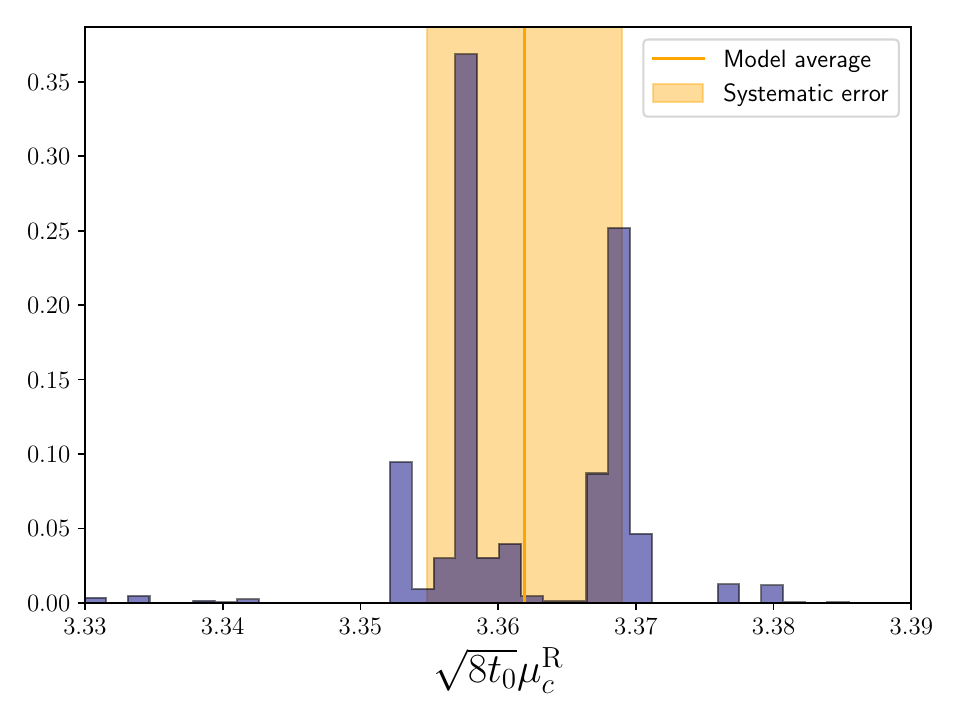}
	\caption{Weighted histogram illustrating the model average procedure for $\sqrt{8t_0}\,\mu_c^{\rm\scriptscriptstyle R}$. The result from each of the 128 models -- including both matching conditions $\phi_H^{(1)}$ and $\phi_H^{(2)}$ -- parameterising the lattice spacing dependence is weighted by its normalised weight $W$ based on the TIC. The vertical line represents the central value from the model average, while the vertical band shows the corresponding estimate of the systematic error.            }
	\label{fig:mc_histogram}
\end{figure}

In Figure~\ref{fig:mc_model_av_summary} we summarise the model average procedure, showing some of the best  
fit results coming from the functional forms defined in Eq.~(\ref{eq:tot_model}) for the two
 matching 
conditions studied in this work. Each circle corresponds to a result coming from  a particular model, and 
the opacity is associated to its weight determined from our Takeuchi's Information Criterion (TIC)  as explained in App.~\ref{app:TIC}.  We observe that for both
matching conditions the majority of the models with relevant weights nicely agree, and as a result the 
systematic error is subleading with respect to the statistical uncertainty. Figure~\ref{fig:mc_histogram}  shows a weighted histogram of our results 
coming from different fits. We observe that models cluster mainly around two values, which are adequately
covered by our quoted systematic uncertainty.

\begin{table}[t!]
	\begin{center}	
		\begin{tabular}{c ||  c c  c  }
			\hline
			 &  $\phi_{H}^{(1)}$ & $\phi_{H}^{(2)} $  &   \text{combined} \\ [0.5ex]
			\hline\hline
			$\sqrt{8t_0}\mu_c^{\rm\scriptscriptstyle R}$ & 3.354(28)(6) & 3.363(27)(6)  &   3.361(26)(7)   
		\end{tabular}
		\caption{Results of the model average for the renormalised charm quark mass  in units of $\sqrt{8t_0}$ based on the two
		 charm quark mass matching conditions --- $\phi_H^{(1)}$ denotes the flavour-averaged matching 
		 condition in \req{eq:fl_av_matching} and  $\phi_H^{(2)}$ the $\eta_h^{\mathrm{conn}}$ matching prescription in 
		 \req{eq:etac_matching}. The last column reports the combined result from these two matching procedures according to our model average prescription. The first error is 
		 statistical, while the second is the systematic uncertainty arising from the model variation.
                }
		\label{tab:mc_results_all_matching}
	\end{center}
\end{table}

\begin{figure}[!htb]
	\centering 
	\includegraphics[scale=0.6]{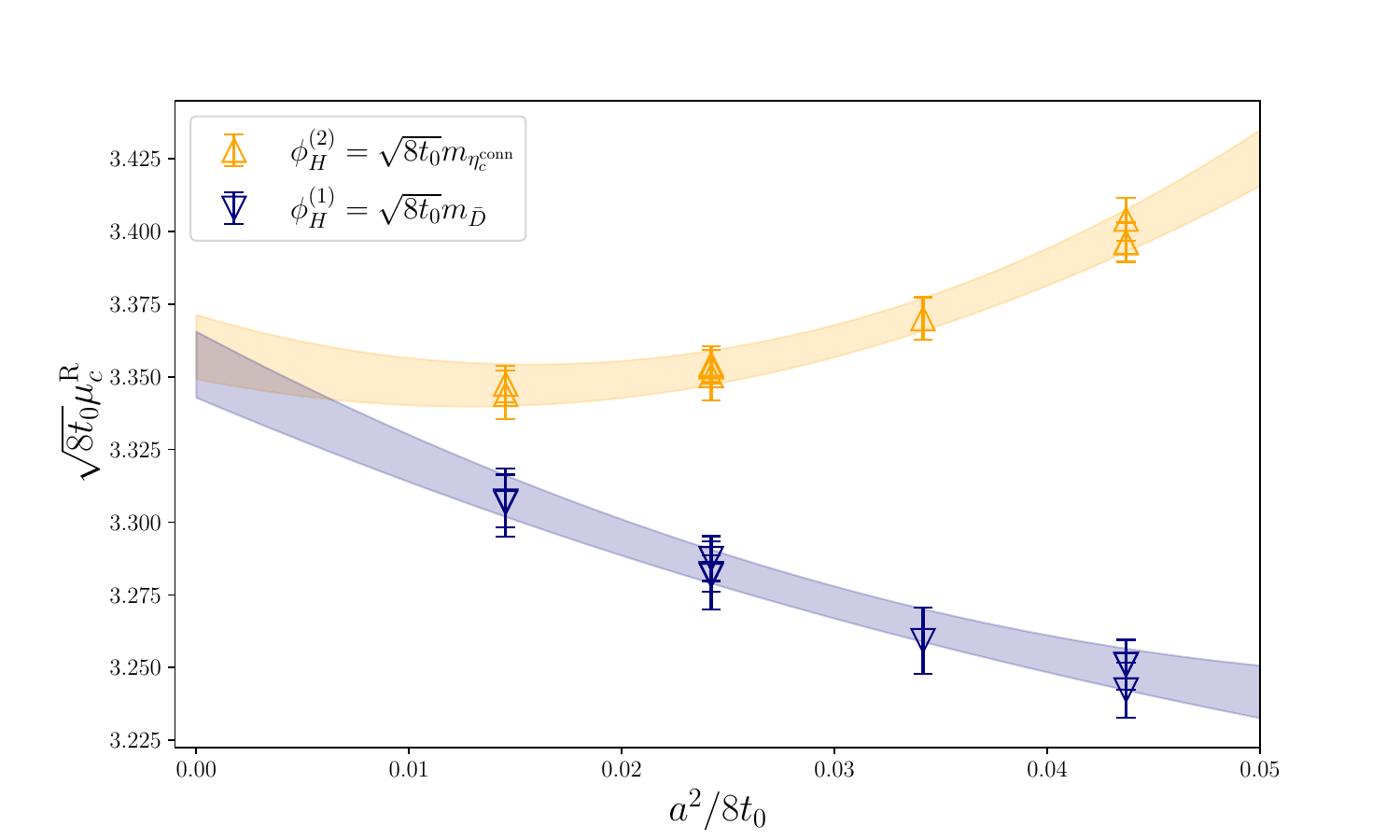}
	\caption{Comparison of the continuum limit approach for the two  charm matching 
	prescriptions. Shown are two of the fits with highest weights from the TIC, projected onto the lattice 
	spacing dimension. In yellow we show results for the $\eta_h^{\mathrm{conn}}$ matching condition, while  the blue 
	points illustrate  the flavour-averaged matching. Each data-point in this plot is projected to the 
	physical pion mass and the physical charm quark mass, in order to properly visualise the lattice 
	spacing dependence. }
	\label{fig:mc_continuum_limit}
\end{figure}

\begin{figure}
	\centering
	\includegraphics[scale=0.52]{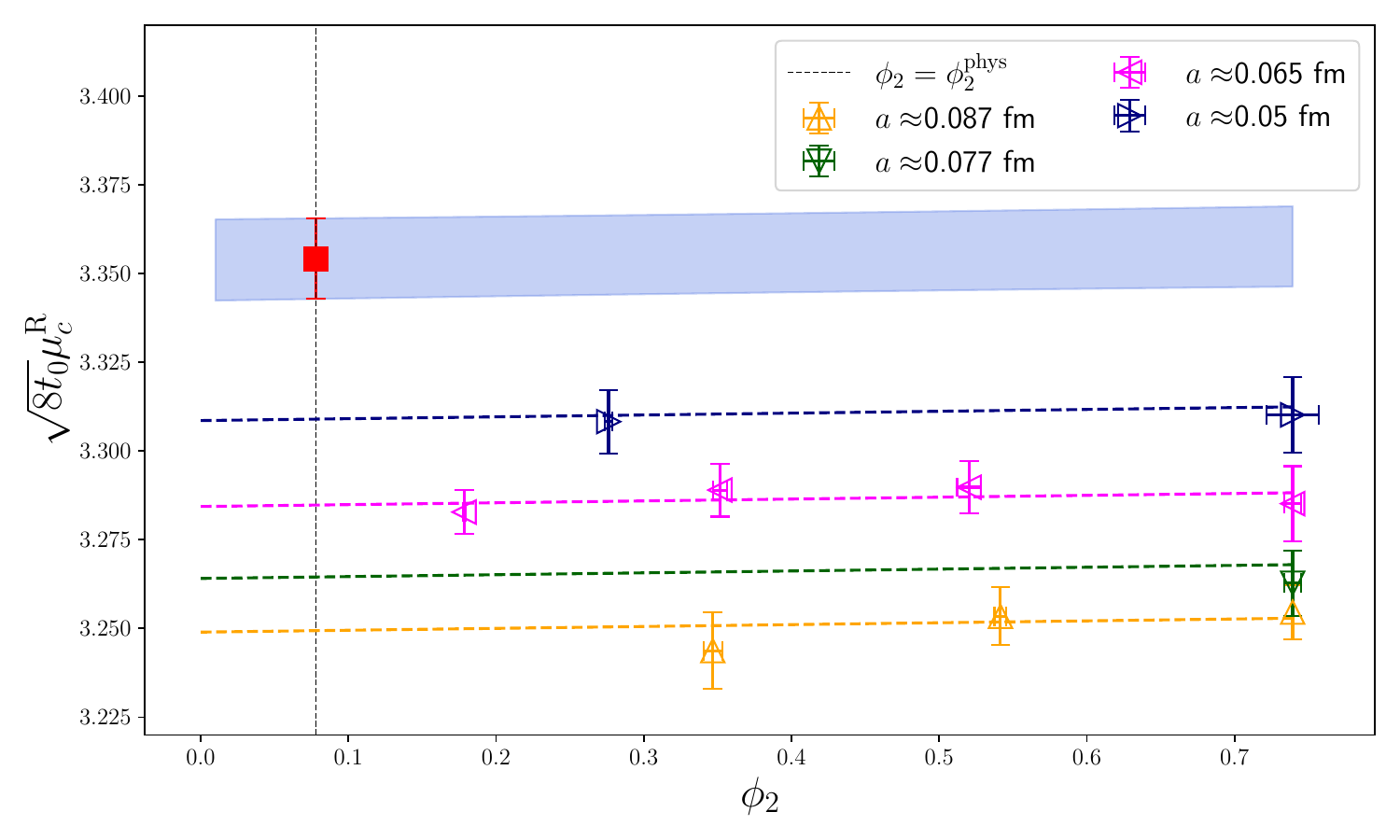}
	\caption{Pion mass dependence of the charm quark mass for one of the best  fits according to the TIC criteria. Results are shown for the flavour-averaged matching condition. Each point corresponds to the  value for a given ensemble, projected to the physical charm quark mass. The dashed lines represent the chiral trajectories at finite lattice spacing, while the blue shaded band is a projection to the continuum limit. The red point shows our final result extrapolated at the physical point in the continuum. }
	\label{fig:mc_pion_dependence}
\end{figure}

Figure~\ref{fig:mc_continuum_limit} illustrates typical fits for each of the matching conditions, chosen 
among those with higher weights according to the TIC prescription. The plot shows  the continuum limit behaviour of 
the charm quark mass in units of $\sqrt{8t_0}$. Results coming from the two matching strategies perfectly 
agree in the continuum, in spite of displaying a qualitatively different structure in cutoff effects.
We observe a scaling of the charm quark mass in reasonable
agreement with the $O(a^2)$ leading order, confirming the automatic $O(a)$-improvement of our setup;
nevertheless, we notice that given the current statistical accuracy, fits with  $O(a^4)$ terms are the 
preferred ones from the model average, since they allow to properly describe the curvature in our data. 
Note also the overall small size of scaling violations, which are at the few percent level.
Finally, Figure~\ref{fig:mc_pion_dependence} shows the pion  mass dependence of the charm quark mass. As 
expected, we observe a mild dependence of the charm mass on the light quark masses.
 
\subsection{Results for the charm quark mass}

The renormalised charm quark mass 
$\mu_c^{\rm\scriptscriptstyle R}$ can be obtained once we combine the results collected in Table~\ref{tab:mc_results_all_matching} with our determination of $\sqrt{t_0^{\mathrm{phys}}}$ in Eq.~(\ref{eq:t0phys}). As discussed at the beginning of this section, the knowledge of the renormalisation group running factors allows  to quote
results for the RGI and $\MSbar$ values of the charm quark mass.

After combining the results from our 128 fitting models through the model average procedure,
and using the running factor in \req{eq:rgi_running_factor}, we quote for the three-flavour theory
the value for the RGI quark mass
\begin{eqnarray}
  M_c^{\mathrm{RGI}}(\NF=3) &=& 1.485(8)(3)(14)[17]\ \mathrm{GeV}\,,
	\label{eq:rgi_charm_mass_result}
\end{eqnarray}
where the first error is statistical, including the uncertainty on  $t_0^{\mathrm{phys}}$,  the second accounts for the systematic uncertainty, derived from the model average, the third is the error contribution from the RGI running factor in Eq.~(\ref{eq:rgi_running_factor}), and the last error in brackets is the total uncertainty. 

\begin{figure}[t!]
	\centering
	\includegraphics[scale=0.7]{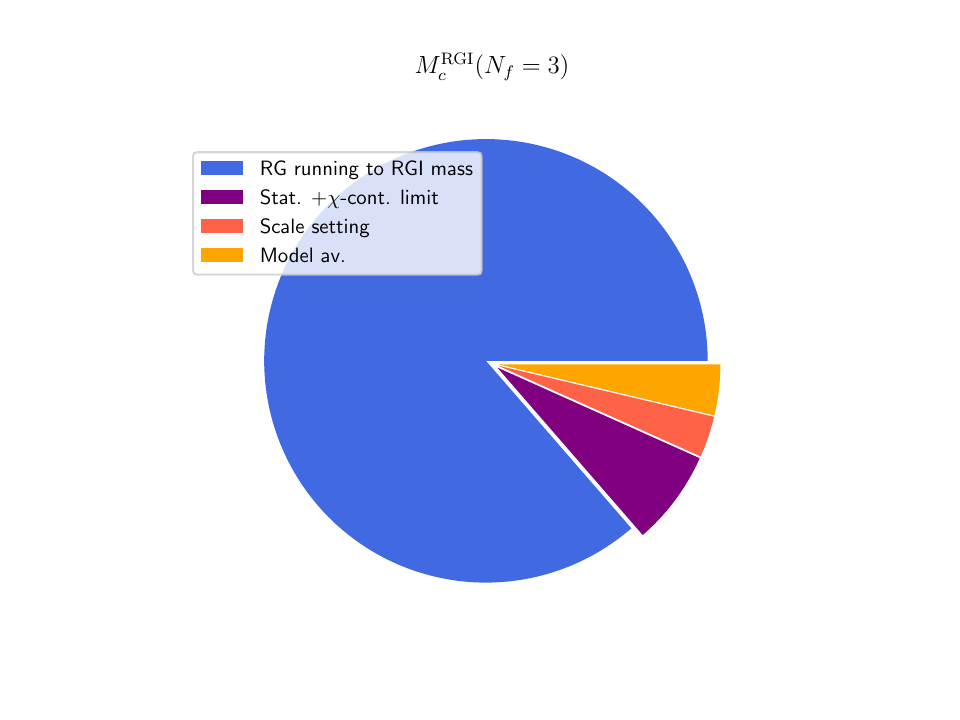}
	\caption{Relative contributions to the total variance of our final result for $M_c^{\mathrm{RGI}}$. The dominant piece comes from the error in the non-perturbative determination of the renormalisation group running factor to the RGI mass quoted in Eq.~(\ref{eq:rgi_running_factor}). The label statistical plus $\chi$-continuum limit stands for the error arising from the statistical accuracy of our data and the chiral-continuum extrapolation, while the scale setting piece comes from the physical value of the gradient flow scale $t_0^{\mathrm{phys}}$. Finally, the model average piece illustrates the systematic error arising from the set of models considered in this work.
          }
	\label{fig:mc_error_contributions}
\end{figure}

Figure~\ref{fig:mc_error_contributions} illustrates the relative contribution of various sources of error to the
uncertainty of our determination of $M_c^{\mathrm{RGI}}$. The dominant source of error comes from the 
renormalisation group running of Eq.~(\ref{eq:rgi_running_factor}), while the second most relevant 
contribution arises from the statistical error of  the correlation functions computed in each ensembles.  
The  error coming from  the uncertainty on $t_0^{\mathrm{phys}}$ based on our  scale setting  procedure~\cite{MA1}, as well as the 
systematic error from the model average  are subleading contributions. We therefore expect
that the 
inclusion in this charm quark mass analysis of further ensembles -- with finer lattice spacings and at physical pion masses --  will only have a significant impact if combined with improved determinations of the RGI running factor and the scale setting procedure.

In order to quote results in the $\MSbar$ scheme, we use five-loop perturbation theory for the quark
mass anomalous dimension~\cite{Baikov:2014qja,Luthe:2016xec,Baikov:2017ujl} and the beta function~\cite{Baikov:2016tgj,Herzog:2017ohr,Luthe:2017ttc}.
The matching between the $\NF=3$ and $\NF=4$ theories uses the four-loop decoupling effects~\cite{Liu:2015fxa}
incorporated into the RunDec package~\cite{Chetyrkin:2000yt,Schmidt:2012az,Herren:2017osy}. Renormalisation group equations are solved using as input the value 
$\Lambda^{(3)}_{\overline{\mathrm{MS}}} = 341(12)\ \mathrm{MeV}$ from~\cite{Bruno:2017gxd}. The correlation arising from the fact that a common subset of gauge field configuration ensembles were employed in the computation of $\Lambda^{(3)}_{\overline{\mathrm{MS}}}$ and the non-perturbative running factor in Eq.~(\ref{eq:rgi_running_factor}) is taken into account. We thus arrive to the following results for the RGI and $\MSbar$-scheme charm quark masses in the 4-flavour theory,
\begin{eqnarray}
  M_c^{\mathrm{RGI}}(\NF=4) &=& 1.546(8)(3)(14)(4)_\Lambda(3)_{\rm trunc.}[17] \ \mathrm{GeV}\,,\\
  \overline{m}_c(\mu=3\ \mathrm{GeV}, \NF=4) &=& 1.006(5)(2)(9)(6)_\Lambda(3)_{\rm trunc.}[13] \ \mathrm{GeV}\,,
	\\
	\overline{m}_c(\mu=\overline{m}_c, \NF=4) &=& 1.296(5)(2)(8)(11)_\Lambda(5)_{\rm trunc.}[16] \ \mathrm{GeV}\,,
\end{eqnarray}
where the first and second errors arise from the statistical and systematic errors, respectively, in the value of $M_c^{\mathrm{RGI}}(\NF=3)$ in Eq.~(\ref{eq:rgi_charm_mass_result}), the third error is due to the non-perturbative running factor in Eq.~(\ref{eq:rgi_running_factor}), the fourth error is related to the uncertainty in $\Lambda^{(3)}_{\overline{\mathrm{MS}}}$, the fifth error is an estimate of the truncation uncertainty from the deviation between the 5-loop and 4-loop results, and the last error in brackets is the total error. We observe that at the lower renormalisation scale, $\mu=\overline{m}_c$, the scale invariant $\MSbar$ charm mass, $\overline{m}_c(\mu=\overline{m}_c, \NF=4)$, receives a large contribution to its error from the uncertainty of $\Lambda^{(3)}_{\overline{\mathrm{MS}}}$ and from the truncation error. These specific sources of uncertainty are less prominent in the RGI mass, $M_c^{\mathrm{RGI}}(\NF=4)$.

\begin{figure}[t!]
	\centering
	\hspace{-0mm}
	\includegraphics[scale=0.47]{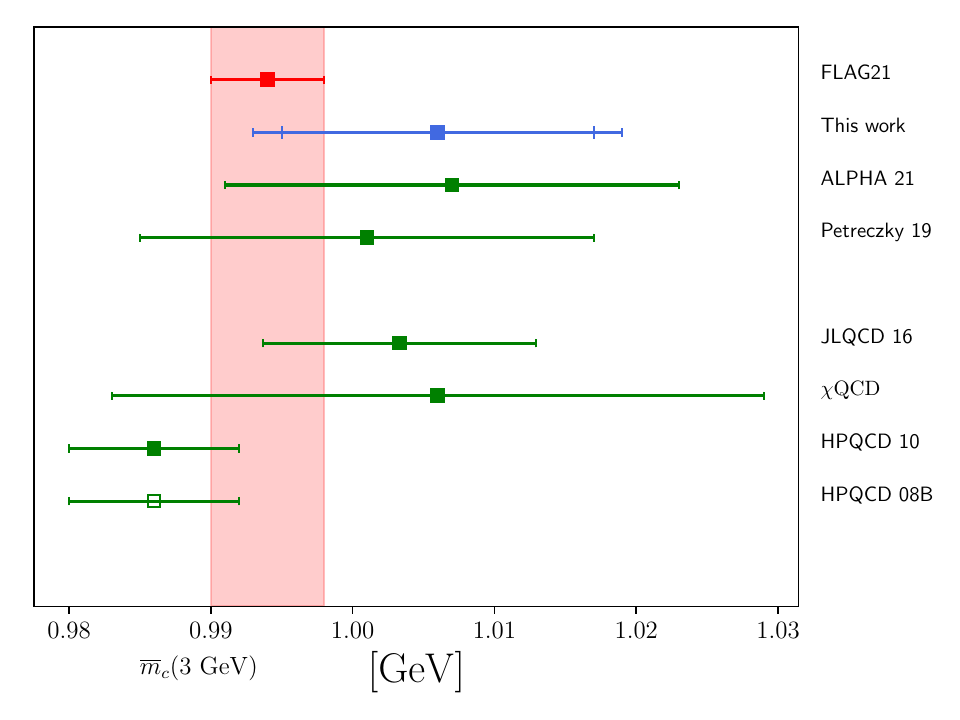}
	\includegraphics[scale=0.47]{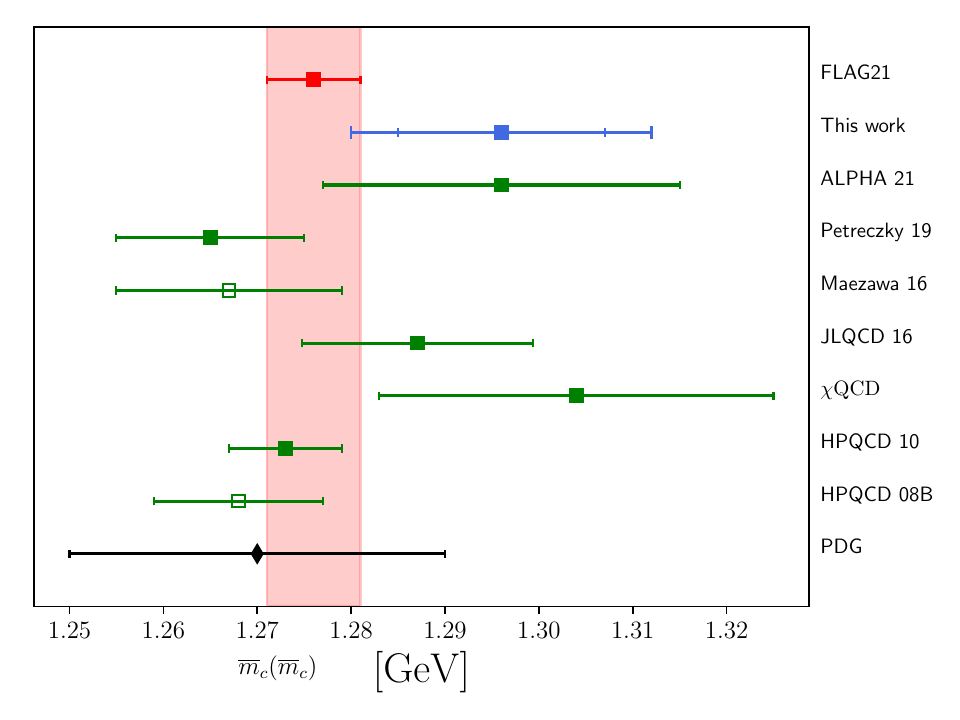}
	\caption{Comparison of our charm quark mass determinations in the $\MSbar$ scheme with the FLAG average~\cite{FlavourLatticeAveragingGroupFLAG:2021npn} and the results from other lattice QCD calculations based on $\NF=2+1$ dynamical simulations. In our results, shown in blue, we indicate both the total uncertainty and the error when excluding the uncertainty arising from $\Lambda^{(3)}_{\overline{\mathrm{MS}}}$. \textit{Left}: comparison for the  $\overline{m}_c(\mu=3\ \mathrm{GeV}, \NF=4)$. \textit{Right}: comparison for $\overline{m}_c(\mu=\overline{m}_c, \NF=4)$.  Starting from the bottom, results are taken from: PDG \cite{ParticleDataGroup:2022pth}, HPQCD 08B \cite{HPQCD:2008kxl}, HPQCD 10 \cite{McNeile:2010ji}, $\chi$QCD \cite{Yang:2014sea}, JLQCD 16 \cite{Nakayama:2016atf}, Maezawa 16 \cite{Maezawa:2016vgv}, Petreczky 19 \cite{Petreczky:2019ozv}, ALPHA 21 \cite{Heitger:2021apz}.
       }
	\label{fig:mc_comparison}
\end{figure}

In Figure~\ref{fig:mc_comparison} we compare our determinations of the charm quark mass in the $\MSbar$ scheme with the results from other lattice QCD calculations also based on $\NF=2+1$ dynamical simulations and with the corresponding FLAG average~\cite{FlavourLatticeAveragingGroupFLAG:2021npn}. We observe in particular a good agreement with the results from \cite{Heitger:2021apz} which are also based on CLS ensembles but employ Wilson fermions in the valence sector.

\section{Determination of decay constants of charmed mesons}
\label{sec:fDs}

\subsection{Computation of decay constants}

Along with the charm quark mass, in this paper we present a first computation of the $D_{(s)}$
meson decay constants within our setup. In the absence of electromagnetic interactions, the
decay constant fully determines the leptonic decay amplitude of flavoured pseudoscalar mesons,
and is given by the matrix element of the axial current as 
\begin{equation}
	\big| \langle 0 | A_0^{qr} | P^{qr}(\mathbf{p=0})\rangle\big| = \frac{f_{qr} m_{\rm\scriptscriptstyle PS}}{\sqrt{2m_{\rm\scriptscriptstyle PS}L^3}},
	\label{eq:decay_constant_definition}
\end{equation}
where the state $|P^{qr}\rangle$ is the ground state for a pseudoscalar meson with flavour content $qr$,
and $m_{\rm\scriptscriptstyle PS}$ its mass.
The factor $1/\sqrt{2m_{\rm\scriptscriptstyle PS}L^3}$ comes from the usual relativistic normalisation
of one-particle states in finite volume.

\begin{figure}
	\centering
	\includegraphics[scale=0.5]{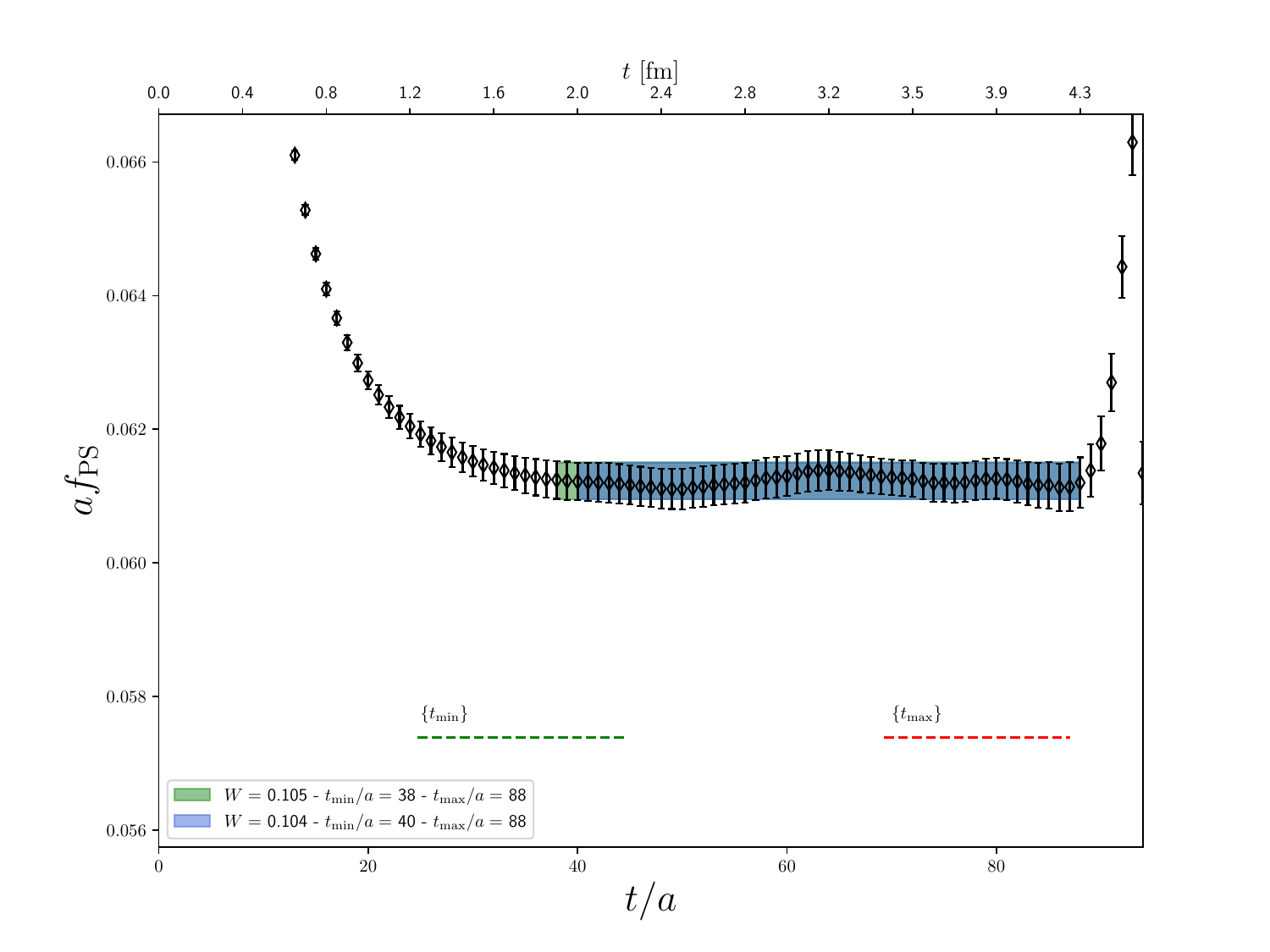}
	\includegraphics[scale=0.5]{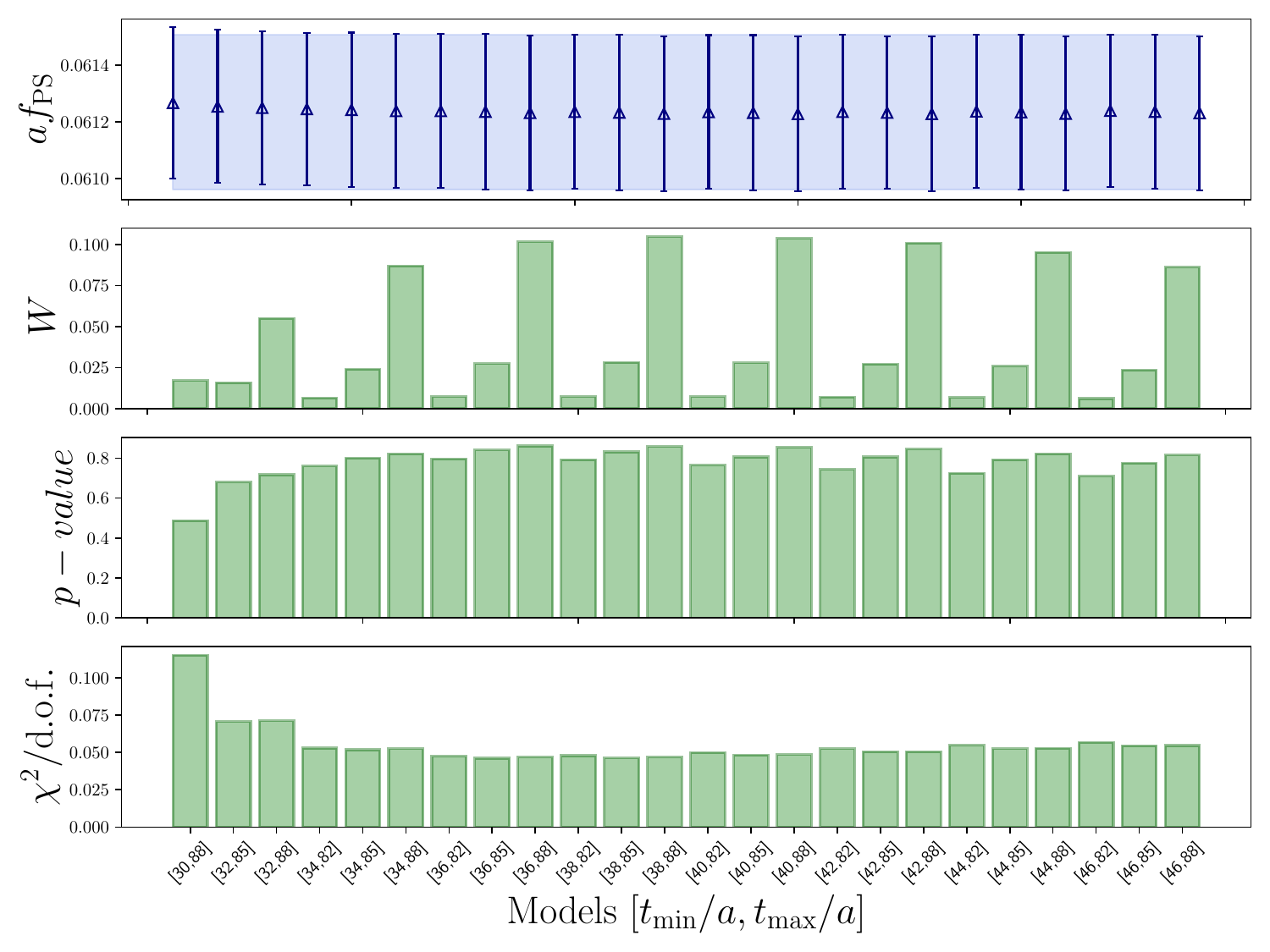}
	\caption{Illustration of the extraction of the heavy-light pseudoscalar decay constants, after applying a GEVP analysis, for ensemble J303. \textit{Top}: plateau for the heavy-light pseudoscalar decay constant for the two fit intervals with higher weights in the model average. \textit{Bottom}: summary of results from  different fit ranges together with weights $W$, p-values and $\chi^2/d.o.f.$. The shaded blue band represents the model average result. }
	\label{fig:decay_plateau} 
\end{figure}

With Wilson fermions, the computation of the above matrix elements requires the finite current
normalisation factor $\ZA$ and, if $O(a)$ effects are to be subtracted, a number of improvement
coefficients. With our fully twisted valence sector this is completely bypassed: when $qr$ belong
in a twisted quark doublet --- i.e., have different signs in the twisted mass matrix in \req{eq:mval} ---
the physical axial current, expressed in twisted quark variables, becomes a vector current,
and the Ward identity in \req{eq:vector_ward_id} allows to obtain it from the pseudoscalar
two-point function.
The resulting expression of the correctly normalised pseudoscalar decay constant reads
\begin{equation}
	f_{PS} = \sqrt{\frac{2L^3}{m_{PS}^3}} (\mu_q+\mu_r) | \langle 0 | P^{qr} | P^{qr}(\mathbf{p=0})\rangle |
	\label{eq:renormalised_decay_constant}.
\end{equation}
We will extract the matrix element $ \langle 0 | P^{qr} | P^{qr}(\mathbf{p=0})\rangle $  from the 
normalised eigenvector $v_n(t,t_0)$ of the GEVP  according to \req{eq:effective_matrix_element}. In 
order to extract the large time plateau where excited state contributions are suppressed we perform 
several fits to constant behaviour by varying the fit ranges,
and we assign a weight to each fit by means of the TIC prescription as
described in App.~\ref{app:TIC}. The results  for the ground state 
matrix element are then extracted through the model average given by \req{eq:model_average}.
In Figure  \ref{fig:decay_plateau} we show a representative plateau for a heavy-light decay constant, 
together with a summary of the model average with different fit intervals.

\subsection{Chiral-continuum fits and results for $f_{D_{(s)}}$}

The chiral-continuum fits for the $D_{(s)}$ meson decay constants are performed similarly to the
ones for the charm quark mass. By exploiting Chiral Perturbation Theory with heavy quarks 
\cite{Grinstein:1992qt, Goity:1992tp} to construct appropriate fit functions, we 
extract the physical point observables trough a global fit of the $f_D$ and $f_{D_s}$ decays,
and estimate the systematic effects by applying the model average procedure based on the TIC.

The quantities we fit to are combinations of meson masses and decay constants of the form
\begin{equation}
  \Phi_{D_{(s)}} = (8t_0)^{3/4}f_{D_{(s)}} \sqrt{m_{D_{(s)}}},
  \label{eq:defphiD}
\end{equation}
for which a Heavy Quark Effective Theory (HQET) scaling law in powers of the inverse
heavy quark mass exists.
The general continuum heavy and light quark mass dependence can be expressed as the product of the individual contributions to arrive at the generic expression 
\begin{equation}
	\Phi_{D_{(s)}} = \Phi_{\chi} \left[
	1 + \delta\Phi_{\chi\mathrm{PT}}^{D_{(s)}}
	\right]
	\left[
	1 + \delta\Phi_a^{D_{(s)}}
	\right]\,.
	\label{eq:fds_different_pieces}
\end{equation}
Here $\Phi_\chi$ governs the heavy-quark mass dependence while  $\delta\Phi_{\chi\mathrm{PT}}^{D_{(s)}}$ controls the light quark behaviour as approaching the physical point. Finally the lattice spacing dependence describing cut-off effects is regulated by $\delta\Phi_a^{D_{(s)}}$. In the following, we analyse these terms independently to arrive at a final expression for the $\Phi_{D_{(s)}}$ approach to the physical point.

The continuum heavy-quark mass dependence, $\Phi_\chi$, admits an expression in HQET of the form
\begin{align}
	\Phi_{\chi} = C_{\mathrm{HQET}}(m_h)\, \Phi_0 \left[ 1 + p_h^{(1)} \frac{1}{\phi_H} + p_h^{(2)} \left( \frac{1}{\phi_H} \right)^2 + \dots \right]
	\,,
	\label{eq:phichi}
\end{align}
where $\phi_H=\sqrt{8t_0}m_H$ monitors the heavy quark mass dependence with $m_H$ being the flavour-
average $m_{\bar{H}}$ or the $\eta_h^{\mathrm{conn}}$ pseudoscalar meson masses.
In general, this expression is not expected to have high accuracy in the charm mass region,
due to it being at the limit of applicability of HQET. Furthermore, perturbative values
for the matching factor $C_{\mathrm{HQET}}(m_h)$ have notoriously poor convergence behaviour.\footnote{This
is readily observed in the expression for the coefficient in the $\MSbar$ scheme~\cite{Manohar:2000dt,
	Ji:1991pr},  
\begin{equation} 
	C_{\mathrm{HQET}}(m_h) = \left[\alpha_s(m_h)\right]^{{\gamma_0}/{2\beta_0}} \left[1 + \frac{\alpha_s(m_h)}{4\pi}
	\left(-\frac{8}{3} + \frac{\gamma_1}{2\beta_0} - \frac{\gamma_0\beta_1}{2\beta_0^2} \right) + {\mathrm{O}}(\alpha_s^2) \right]\,,
	\label{eq:Wilson-coefficient}
\end{equation}
where, for QCD, $\gamma_0 = -4$, $\gamma_1 = -254/9 - 56\pi^2/27 + 20 \NF/9$, while
the perturbative coefficients of the $\beta$ function have their usual values
$\beta_0 = (11 - 2\NF/3)$ and $\beta_1 = (102 - 28\NF/3)$.
}
However, we are {\em not} interested in modelling the heavy quark mass dependence in a wide
region of masses --- we rather want to interpolate to the charm point from the nearby values
of the heavy masses we compute at. Therefore, we will simply take an expression with
the same functional form for the $m_h$ power corrections, and a constant overall coefficient,
as a convenient ansatz for the interpolation part of our fits. In HQET terms, this amounts
to neglecting the small logarithmic dependence on $m_h$ in a short interval of values.

The light quark mass dependence term, following Heavy Meson $\chi$PT (HM$\chi$PT) considerations,
reads~\cite{Goity:1992tp, Bazavov:2017lyh}
\begin{equation}
	\begin{split}
		\delta \Phi_{\chi {\mathrm{PT}}}^{D} =& - \frac{1+3g^2}{64\pi^2 \phi_f^2} \left[ 3 \mathcal{L}_{\pi} + 2 \mathcal{L}_{K} + \frac{1}{3} \mathcal{L}_{\eta} \right] +
		\frac{4  \phi_2 }{\phi_f^2} \left( p_\chi^{(0)} +  p_\chi^{(2)} \frac{\phi_2}{\phi_f^2} + \frac{p_\chi^{(4)}}{\phi_H} \right),
		\\
		\delta \Phi_{\chi {\mathrm{PT}}}^{D_s} =& - \frac{1+3g^2}{64\pi^2 \phi_f^2} \left[ 4 \mathcal{L}_{K} + \frac{4}{3}  \mathcal{L}_{\eta} \right] +
		\frac{8 \left( \phi_4 - \phi_2 \right)} {\phi_f^2} \left( p_\chi^{(0)} +  p_\chi^{(2)} \frac{\phi_2}{\phi_f^2} + \frac{p_\chi^{(4)}}{\phi_H} \right),
		\label{eq:deltaphichis} 
	\end{split}
\end{equation}
where $p_\chi^{(0,1\,\dots)}$ are fit parameters and $g^2$ is the
$H^\ast H \pi$ coupling in the static and chiral limits, here treated as a free fit parameter alongside
$p_\chi^{(i)}$.  In \req{eq:deltaphichis} we introduced  the notation for the chiral logarithm corrections 
\begin{eqnarray}
	\mathcal{L}_\pi &=& \phi_2 \log(\phi_2), 
	\\
	\mathcal{L}_K &=& \bigg(\phi_4 - \frac{1}{2}\phi_2\bigg)\log(\phi_4 - \frac{1}{2}\phi_2), 
	\\
	\mathcal{L}_\eta &= &\bigg(\frac{4}{3}\phi_4 - \phi_2\bigg)\log(\frac{4}{3}\phi_4 - \phi_2).
\end{eqnarray}
Here $\phi_2$ and $\phi_4$ are the usual hadronic combinations introduced  in \req{eq:phi2_and_phi4},
which control the light and strange quark mass dependence. When working at NLO in the chiral expansion, 
the term
$\phi_f$ appearing in \req{eq:deltaphichis}, which introduces the $\chi$PT scale,
is here replaced by the continuum
physical value  of $\sqrt{8 t_0} f_{\pi K}$, as determined from our setup \cite{MA1} at full twist, with $f_{\pi K}$ given by\footnote{We remind the reader that $f_{\pi K}$ is the quantity used to extract the physical scale $t_0^{\mathrm{phys}}$ in our setup.}
\begin{equation}
	f_{\pi K} = \frac{2}{3} \left(
	f_K + \frac{1}{2}f_\pi
	\right).
\end{equation}
Finally, with similar arguments to the one discussed in the case of the charm quark mass,
the lattice spacing dependence $\delta\Phi_a^{D_{(s)}}$ for the observables $\Phi_{D_{(s)}}$ can be 
parameterised as 
\begin{equation}
	\begin{split}
		\delta \Phi_{a}^{D} &= \frac{a^2}{8t_0} \left[ p_a^{(0)} +  \phi_2 \left( p_a^{(1)} + p_a^{(3)} \phi_H^2 \right) +  p_a^{(2)} \phi_H^2   \right] + {\mathrm{O}}(a^4)
		,
		\\
		\delta \Phi_{a}^{D_s} &= \frac{a^2}{8t_0} \left[ p_a^{(0)} + 2 \left( \phi_4 - \phi_2 \right) \left( p_a^{(1)} + p_a^{(3)} \phi_H^2 \right) +  p_a^{(2)} \phi_H^2   \right]  + {\mathrm{O}}(a^4),
		\, \label{eq:phias}
	\end{split}
\end{equation}
where $p_a^{(0,1,2,\dots)}$ are fit parameters. 

To summarise, for the continuum quark mass dependence  of $\Phi_D$ and $\Phi_{D_s}$
we adopt the expressions
\begin{equation}\small
	\begin{split}
		\Phi_D(0,\phi_2, \phi_H)  &= p_0 + \frac{4p_1}{\phi_f^2}\phi_2 + \frac{p_2}{\phi_H}
		-\frac{1+3g^2}{64\pi\phi_f^2}\bigg(
		3\mathcal{L}_\pi + 2 \mathcal{L}_K + \frac{1}{3}\mathcal{L}_\eta  
		\bigg) +
		\frac{4  \phi_2 }{\phi_f^2} \left( p_\chi^{(0)} +  p_\chi^{(2)} \frac{\phi_2}{\phi_f^2} + \frac{p_\chi^{(4)}}{\phi_H} \right),
		\\
		\Phi_{D_s}(0,\phi_2, \phi_H)  &= p_0 + \frac{8p_1(\phi_4-\phi_2)}{\phi_f^2} + \frac{p_2}{\phi_H}
		-\frac{1+3g^2}{64\pi\phi_f^2}\bigg(
		4 \mathcal{L}_K + \frac{4}{3}\mathcal{L}_\eta\bigg)\\
		&\qquad~
	+\frac{8 \left( \phi_4 - \phi_2 \right)} {\phi_f^2} \left( p_\chi^{(0)} +  p_\chi^{(2)} \frac{\phi_2}{\phi_f^2} + \frac{p_\chi^{(4)}}{\phi_H} \right),
	\end{split}
\end{equation}\normalsize
obtained by combining the light and heavy quark dependencies $\delta\Phi_{\chi \mathrm{PT}}$ and $\Phi_\chi$, 
respectively.  
Following \req{eq:fds_different_pieces}, this then leads to the final ansatz for $\Phi_{D_{(s)}}$
of the form
\begin{equation}
	\Phi_{D_{(s)}}(a,\phi_2, \phi_H) = 
	\Phi_{D_{(s)}}(0,\phi_2, \phi_H) \left[ 1   + \delta\Phi_a^{D_{(s)}}\right].
	\label{eq:fds_combined_fit}
\end{equation}
Since many fit parameters are shared between $\Phi_D$ and $\Phi_{D_s}$, we opt for a global fit for 
determining the two quantities. Moreover, at the symmetric point, i.e.,  for those ensembles with 
degenerate light and strange quark masses $\mu_l=\mu_s$, the two decay constant coincide, and
 $\Phi_D=\Phi_{D_s}$. Therefore, a global fit also helps to constrain the parameters at the symmetric 
 point.

Similarly to the case of the charm quark mass, we consider several specific forms of the fit ansatz,
by setting some combination of fit parameters to zero. We furthermore again match the charm scale using
the two different procedures described in Sec.~\ref{sec:charm_basics}. The result is a total
of 57 different models  for each matching condition,
and we use our TIC criterion to extract a systematic uncertainty associated to the variation
within the full set of fits. In this work, our current approach deliberately excludes fits involving cuts in $\beta$ or pion masses, as with the current subset of ensembles they are significantly penalised by the TIC. As we look ahead to future updates with the complete set of ensembles  we will incorporate cuts in the data within our analysis.

\begin{figure}
	\centering
	\includegraphics[scale=0.50]{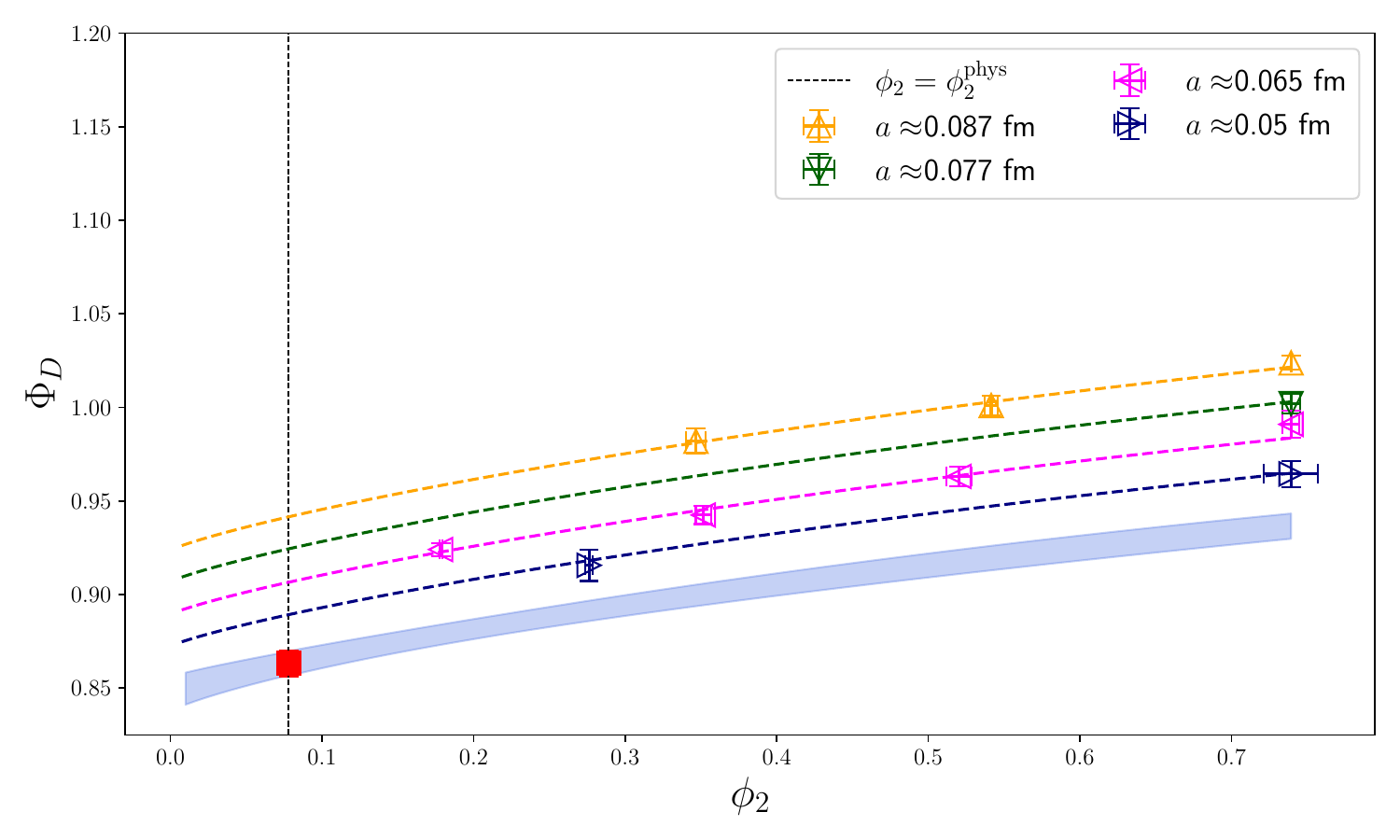}
	\includegraphics[scale=0.50]{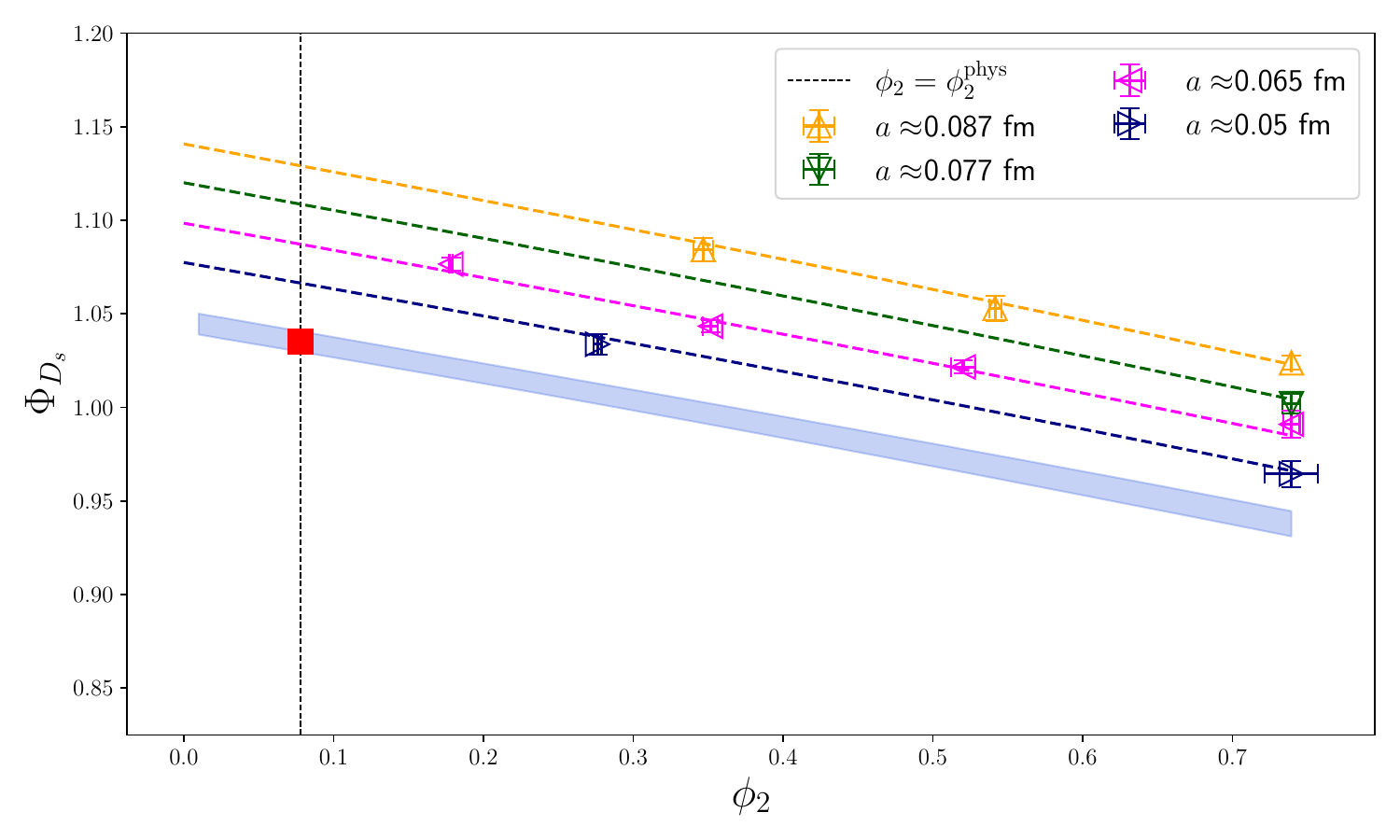}
	\caption{Chiral behaviour of the best fits according to the TIC criteria applied to  $\Phi_D$ (\textit{top}) and $\Phi_{D_s}$ (\textit{bottom}). Each point is projected to the physical charm quark mass, and results are shown for the flavour-averaged matching condition $\phi_H^{(1)}$. Dashed lines refer to the mass dependence at finite  values of the lattice spacing, while the blue band represents the projection to the continuum limit. Finally, the red square symbols indicate the physical point results.}
	\label{fig:chiral_fits_fds}
\end{figure}

\begin{figure}
	\centering
	\includegraphics[scale=0.5]{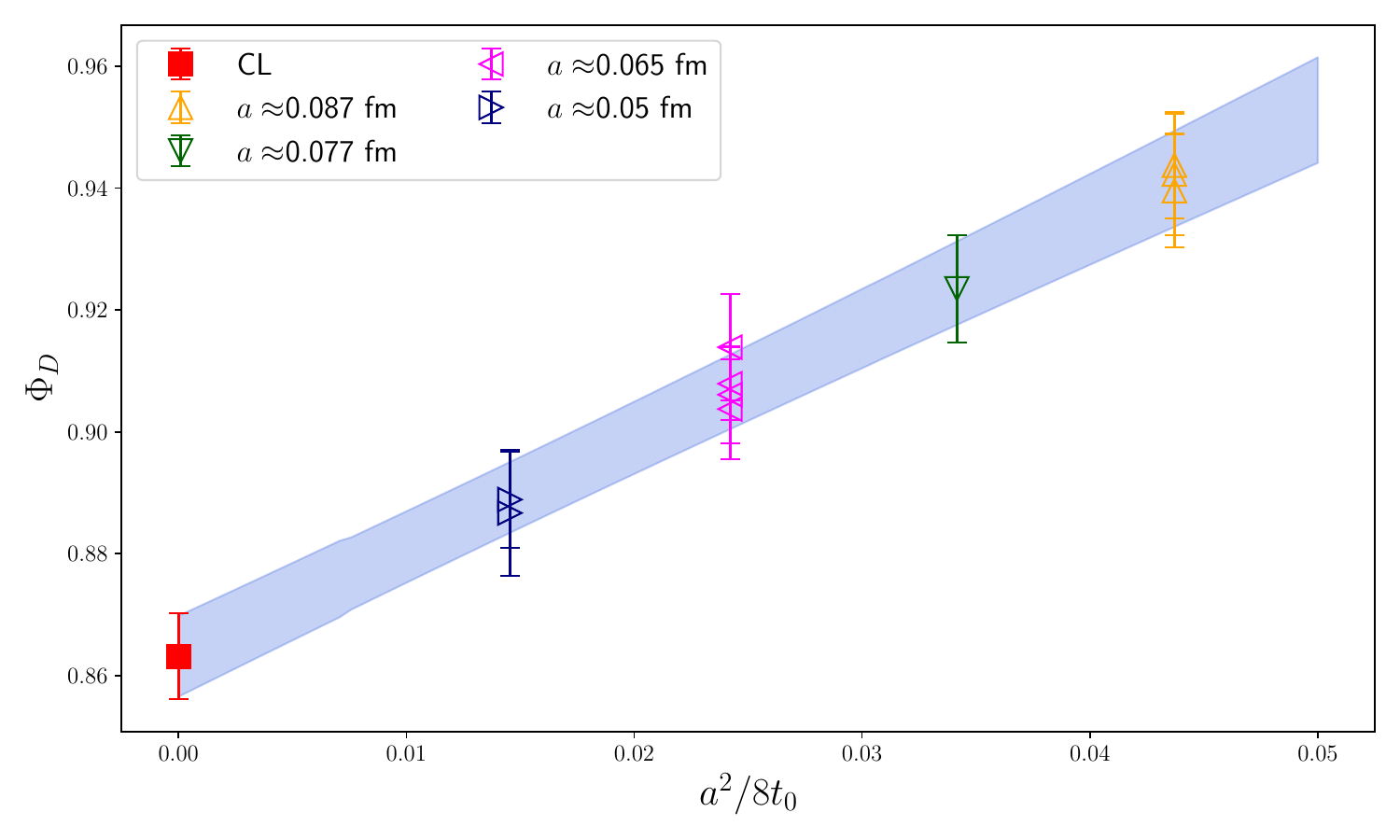}
	\includegraphics[scale=0.5]{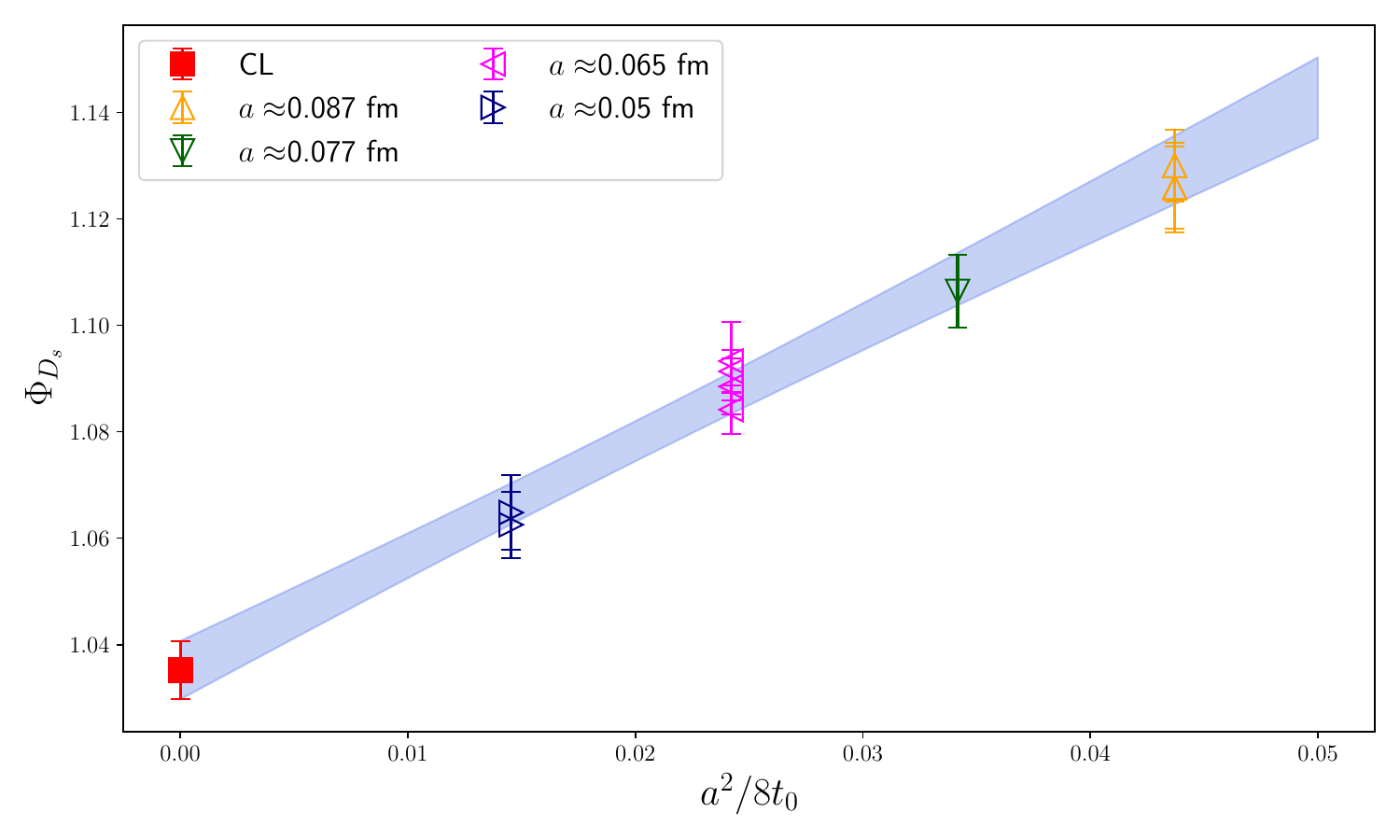}
	\caption{Continuum limit extrapolation of the best fits according to the TIC criteria applied to  $\Phi_D$ (\textit{top}) and $\Phi_{D_s}$ (\textit{bottom}).  Results are shown for the flavour-averaged matching condition $\phi_H^{(1)}$. The blue band represents the projection to the physical $\phi_2 = \phi_2^{\mathrm{phys}}$ and $\phi_H = \phi_H^{\mathrm{phys}}$, while the red square symbols denote the results in the continuum.}
	\label{fig:continuum_fits_fds}
\end{figure}

In Figure~\ref{fig:chiral_fits_fds} we show the chiral extrapolations for $f_D$ and $f_{D_s}$
with larger weights in the model average.  From our chiral-continuum extrapolations of $\Phi_D$ and $\Phi_{D_s}$, we observe a mild 
dependence on the  choice of the $\phi_H$ used to match the charm scale. Therefore, in the Figures we 
illustrate the flavour-averaged matching condition only. We also notice that $\Phi_D$ shows some 
curvature in $\phi_2$ arising from the chiral logs, while $\Phi_{D_s}$ presents a more linear behaviour 
while approaching the physical point. Figure~\ref{fig:continuum_fits_fds} shows an illustration
of the scaling towards the continuum limit of $\Phi_D$ and $\Phi_{D_{s}}$. We observe that the continuum 
approach is very well described by leading cutoff effects of $O(a^2)$, as expected for our valence action when it is tuned to maximal twist.

\begin{figure}[t!]
\begin{center}
\begin{minipage}{.40\linewidth}
\includegraphics[width=\linewidth]{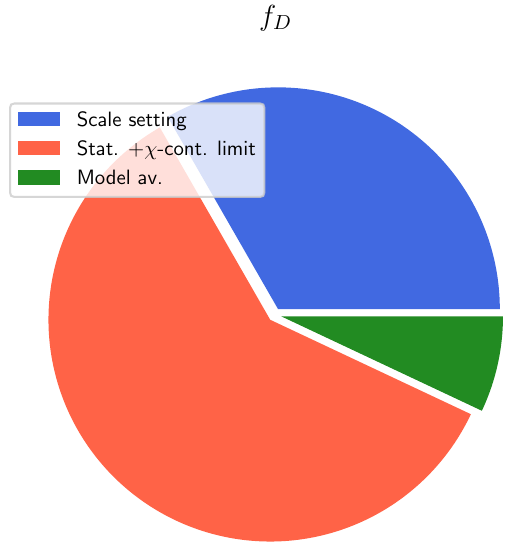}
\end{minipage}
\hspace{10mm}
\begin{minipage}{.39\linewidth}
\includegraphics[width=\linewidth]{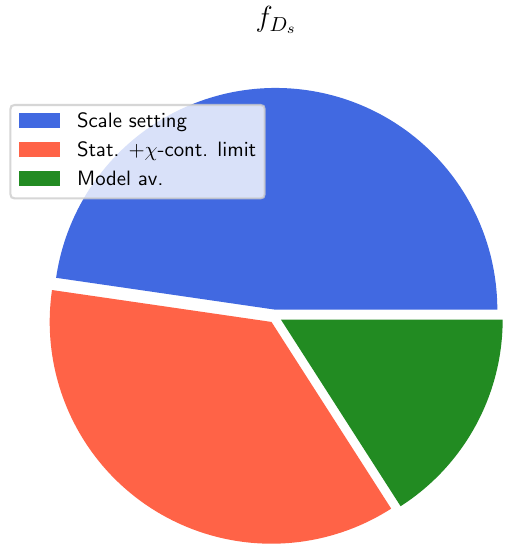}
\end{minipage}
\end{center}
\vspace{-5mm}
	\caption{Relative contributions to the total error of our determinations of $f_D$ (\textit{left}) and $f_{D_s}$ (\textit{right}). The label statistical plus $\chi$-continuum limit represents the error arising from the statistical accuracy of our data and the chiral-continuum extrapolations. The scale setting label denotes the error coming from the physical value $t_0^{\mathrm{phys}}$ as determined within our setup \cite{MA1}, while the model average represents the systematic error arising from the model variation according to the TIC procedure.	}
	\label{fig:fds_error_sources}
\end{figure}

\begin{figure}[t!]
	\centering
	\includegraphics[scale=0.70]{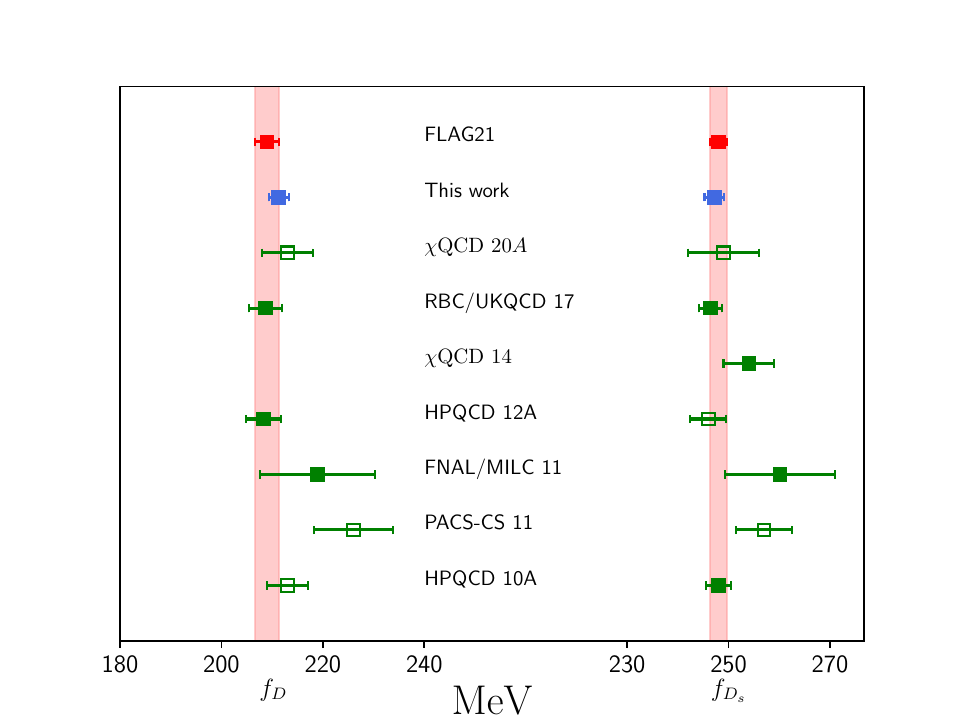}
	\caption{Comparison of our results for $f_D$ and $f_{D_s}$  with those from lattice QCD collaborations based on simulations with $\NF=2+1$ dynamical flavours as well as with FLAG21 averages~\cite{FlavourLatticeAveragingGroupFLAG:2021npn}.
          Only data points with filled symbols contribute to  the FLAG averages. Starting from the bottom, results are taken from: HPQCD 10 \cite{Davies:2010ip}, PACS-CS 11 \cite{PACS-CS:2011ngu}, FNAL/MILC 11 \cite{FermilabLattice:2011njy}, HPQCD 12A \cite{Na:2012iu}, $\chi$QCD 14 \cite{Yang:2014sea}, RBC/UKQCD 17 \cite{Boyle:2017jwu},  $\chi$QCD 20A \cite{Chen:2020qma}.
          }
	\label{fig:fds_comparison}
\end{figure}

In Table~\ref{tab:dec_res_all_matching} we show our determinations of $\Phi_D$
and $\Phi_{D_s}$ for each of the two procedures to match the charm scale, as well
as the result from their combination. Using this combination we arrive at the following results for the the $D_{(s)}$ meson decay constants,
\begin{eqnarray}
	f_D &=& 211.3(1.9)(0.6) \ \mathrm{MeV},
	\\
	f_{D_s} &=& 247.0(1.9)(0.7) \ \mathrm{MeV},
\end{eqnarray}
where the first error is statistical and the second the systematic uncertainty from the model average.
The error budget for the $D_{(s)}$ decay constants is dominated by the statistical uncertainty of 
correlators and the  error on chiral-continuum extrapolations. Therefore, we expect that a future addition of other
ensembles with finer lattice spacing and physical pion masses will contribute to significantly reduce the uncertainty of our current 
determination. The different contributions to the variance of $D_{(s)}$ meson decay constants are 
shown in Figure~\ref{fig:fds_error_sources}. Finally, in  Figure~\ref{fig:fds_comparison} we show a comparison between our results and other $\NF=2+1$ lattice QCD determinations.
\begin{table}[t!]
	\begin{center}	
		\begin{tabular}{c ||  c c  c    }
			\hline
			&  $\phi_{H}^{(1)}$ & $\phi_{H}^{(2)} $  &  \text{combined} \\ [0.5ex]
			\hline\hline
			$\Phi_D$ &  0.8624(78)(7) & 0.8583(75)(8) &   0.8606(76)(21)   \\ [0.5ex]
			$\Phi_{D_s}$ & 1.0352(61)(9) & 1.0295(60)(11) &  1.0328(60)(30) 
		\end{tabular}
		\caption{Model average results for the observables $\Phi_D$ and $\Phi_{D_s}$ --- defined in Eq.~(\ref{eq:defphiD}) ---  which are related to the $f_D$ and $f_{D_s}$ decay constants, respectively, for
		the two different matching quantities $\phi_H^{(i)}$. The last column reports the result of the combination of these two matching conditions. The first error is statistical while the second is the estimate of systematic uncertainty arising from the model averaging procedure. }
		\label{tab:dec_res_all_matching}
	\end{center}
\end{table}

\subsection{Direct determination of $f_{D_s}/f_D$}

In addition to the determination of $f_D$ and $f_{D_s}$, we investigate the direct determination
of the ratio $f_{D_s}/f_D$ from a dedicated fit. This allows for a consistency check, since
the ratio is dimensionless and thus does not require normalisation with a reference scale
such as $\sqrt{8t_0}$. One particular consequence is thus that this approach is only
indirectly subject to the uncertainty of the lattice scale setting. Another advantage
is that the ratio is exactly~1 by construction when $m_s=m_l$, i.e., the symmetric
point of our $\phi_4={\rm constant}$ trajectory, which is part of our line of constant
physics. We can thus perform a fit that is highly constrained in the unphysical masses
region, although at the price of reducing the total number of ensembles entering in the study of the approach to the physical point.

A first set of fit ansaetze is derived from the HM$\chi$PT expressions considered above
for $\Phi_{D_{(s)}}$. The generic form is
\begin{equation}
	\frac{\Phi_{D_s}}{\Phi_D} = \left[
	1 + \left(
	\delta\Phi_{\chi\mathrm{PT}}^{D_s} - \delta\Phi_{\chi\mathrm{PT}}^{D}
	\right)
	\right]
	\left[
	1 + \left(
	\delta\Phi_{a}^{D_s} - \delta\Phi_{a}^{D_s}
	\right)
	\right].
	\label{eq:ratio_fds_expansion}
\end{equation}
Here $\delta\Phi_{\chi\mathrm{PT}}^{D_{(s)}}$  introduced in Eq.~(\ref{eq:deltaphichis}) labels the light quark mass dependence of the ratio, while $\delta\Phi_a^{D_{(s)}}$ from Eq.~(\ref{eq:phias}) controls the continuum approach. It is worth noticing that at leading order the physical dependence on $\phi_H$, and also the lattice spacing dependence related to $\phi_H$, cancel out when expanding the ratio.
Collecting all the terms entering in Eq.~(\ref{eq:ratio_fds_expansion}) from the previous section,
we end up with 
\begin{equation}
	\begin{split}
		\frac{\Phi_{D_s}}{\Phi_D} =&
		\left[  1 - \frac{1+3g^2}{64\pi^2 \phi_f^2} \left[ 2 \mathcal{L}_{K} + \mathcal{L}_{\eta} - 3 \mathcal{L}_{\pi} \right] 
		+ \frac{4 \left( 2\phi_4 - 3 \phi_2 \right)}{\phi_f^2}  \left( p_\chi^{(0)} +  p_\chi^{(2)} \frac{\phi_2}{\phi_f^2} + \frac{p_\chi^{(4)}}{\phi_H} \right)  \right]  
		\\
		&\times \left[ 1+ \frac{a^2}{8t_0} \left( 2 \phi_4 - 3 \phi_2 \right) \left( p_a^{(1)} + p_a^{(3)} \phi_H^2 \right)  \right]. 
		\label{eq:ratiophi}
	\end{split}
\end{equation}
In this expression we consider all the possible combinations of non-vanishing fit parameters,
and perform our TIC-weighted model average among the different functional forms tested to
quote a systematic uncertainty.  

Given that various terms cancel in the  HM$\chi$PT expressions, we will further explore the systematic uncertainties by considering  also functional forms based on a Taylor expansion of $\Phi_{D_{(s)}}$. The generic
expression then reads
\begin{align}
	\Phi_{D_{(s)}}= \left( \Phi_{D_{(s)}}\right)_{\chi} \left[ 1 + \delta \Phi_{{h,\mathrm{Taylor}}} \right] \left[ 1 + \delta \Phi_{{m,\mathrm{Taylor}}}^{D_{(s)}} \right] \left[ 1 + \delta \Phi_a^{D_{(s)}}  \right]
	\,,
	\label{eq:phiqcontT}
\end{align}
where $ \left( \Phi_{D_{(s)}}\right)_{\chi}$ is the value in
the chiral limit and at the physical value of the heavy-quark mass.
In this expansion, the heavy and light mass dependence terms read
\begin{equation}
	\begin{split}
		\delta \Phi_{{h,\mathrm{Taylor}}} &=  p_h^{(0)} \left( \frac{1}{\phi_H} - \frac{1}{\phi_H^{\mathrm{phys}}} \right) + p_h^{(1)} \left( \frac{1}{\phi_H} - \frac{1}{\phi_H^{\mathrm{phys}}} \right)^2 ,
		\\
		\delta \Phi_{{m,\mathrm{Taylor}}}^{D} &=  p_m^{(0)} \phi_4 + \phi_2 \left[ p_m^{(1)}  +  p_m^{(2)} \phi_2 + p_m^{(3)} \left( \frac{1}{\phi_H} - \frac{1}{\phi_H^{\mathrm{phys}}} \right)  \right] ,
		\\
		\delta \Phi_{{m,\mathrm{Taylor}}}^{D_s} &=  p_m^{(0)} \phi_4 + 2 (\phi_4 - \phi_2)  \left[ p_m^{(1)}  + p_m^{(2)} \phi_2 + p_m^{(3)} \left( \frac{1}{\phi_H} - \frac{1}{\phi_H^{\mathrm{phys}}} \right)  \right].
		\label{eq:deltaphiTs} 
	\end{split}
\end{equation}
The lattice spacing dependence  $\delta \Phi_{a}^{D_{(s)}}$ can be parameterised in a similar fashion to that in \req{eq:phias}.
Combining these expressions into a functional form for the ratio of decay constants one then has
\begin{align}
	\frac{\Phi_{D_s}}{\Phi_D}   =& \left[  1 +  \left( 2 \phi_4 - 3 \phi_2 \right)  \left[ p_m^{(1)}  +  p_m^{(2)} \phi_2 + p_m^{(3)} \left( \frac{1}{\phi_H} - \frac{1}{\phi_H^{\mathrm{phys}}} \right) \right]  \right]  
	\nonumber \\
	& \times \left[ 1+ \frac{a^2}{8t_0}  \left( 2 \phi_4 - 3 \phi_2 \right) \left( p_a^{(1)} + p_a^{(3)} \phi_H^2 \right)  \right] \,.
	\label{eq:ratiophiT}
\end{align}

Then, in order to arrive at a final determination of $f_{D_s}/f_D$ we perform a model average among all the HM$\chi$PT and Taylor functional forms for the two different matching conditions simultaneously. In Table~\ref{tab:ratio_res_all_matching} we report our results for the
ratio of decay constants from the model average separately for each charm matching
condition, as well as their combination. Also for the ratio we observe good agreement for the two different $\phi_H^{(i)}$ tested in this work. 
Finally, for the  result combining the two matching conditions, we quote 
\begin{equation}
	\frac{f_{D_s}}{f_D} = 1.177(15)(5),
\end{equation}
where  the first error is  statistical and the second is the systematic uncertainty based on  the model average procedure. 

\begin{figure}[!htb]
	\centering
	\includegraphics[scale=0.50]{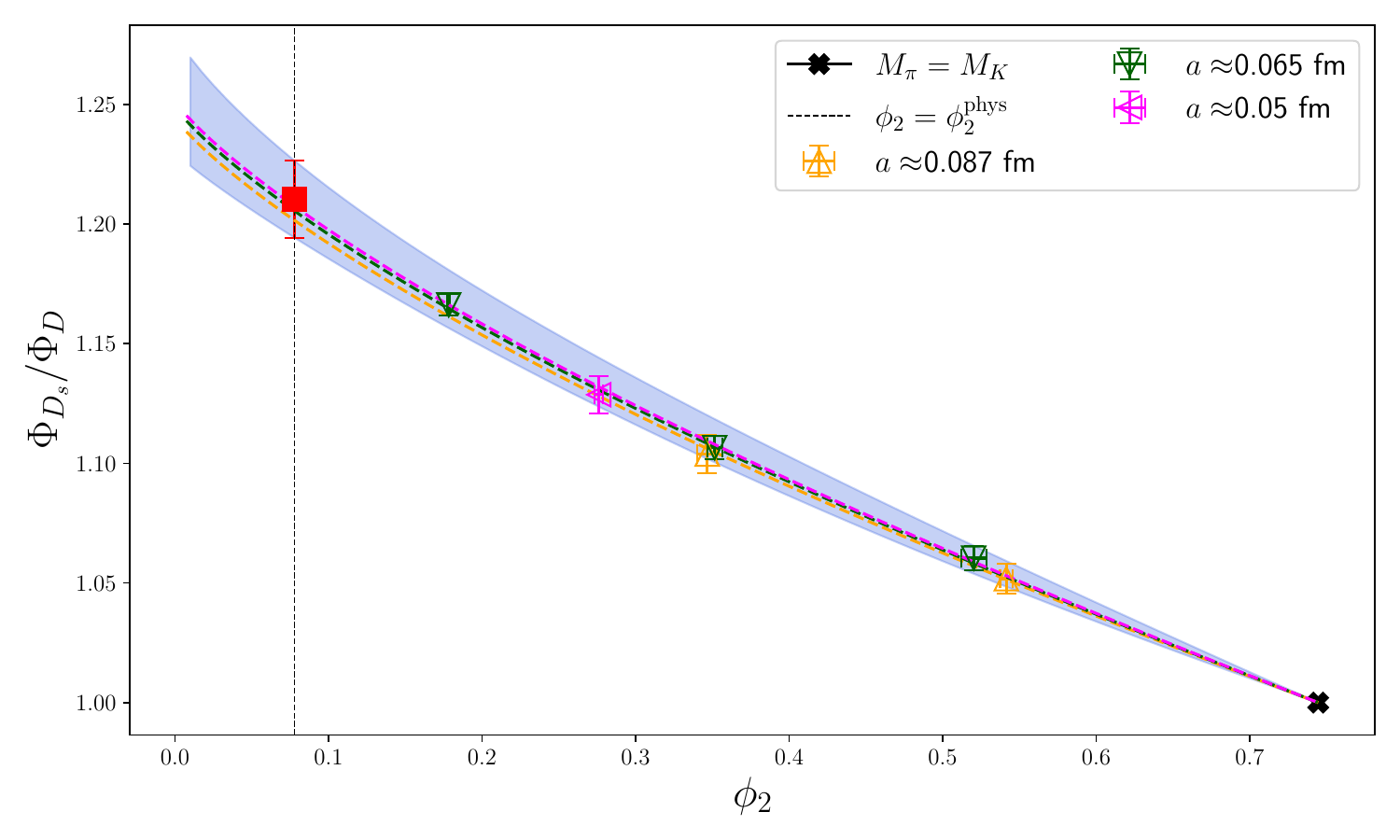}
	\includegraphics[scale=0.50]{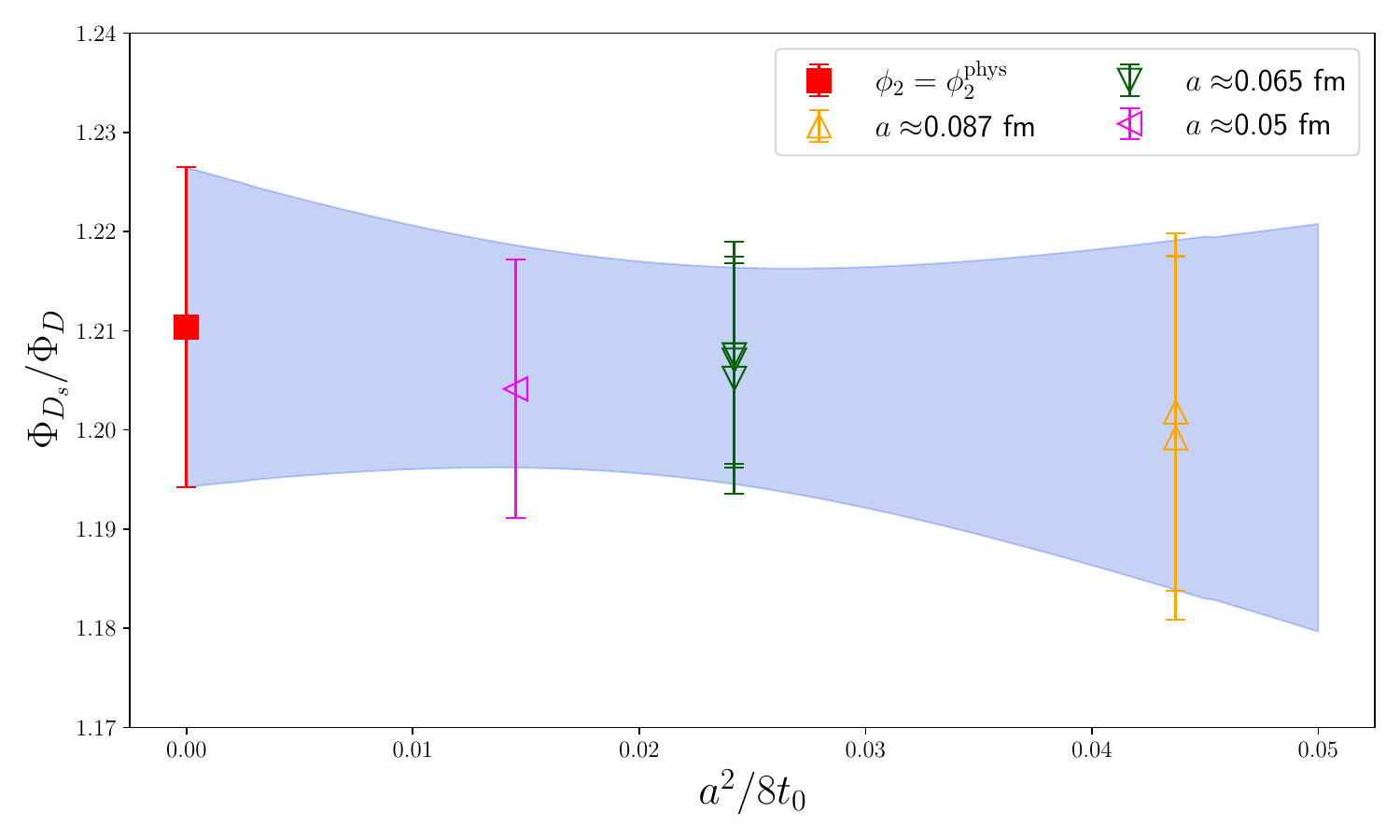}
	\caption{Illustration of the chiral-continuum extrapolation of the ratio $\Phi_{D_s}/\Phi_D$ for the HM$\chi$PT model with highest TIC value. Results are shown for the flavour-averaged matching condition. \textit{Top}:  Chiral approach to the physical point. The dashed lines illustrate the chiral trajectories at finite lattice spacing, while the blue shaded band is a projection of the continuum fit. The red square symbol represents the physical result in the continuum. The black cross symbol corresponds to the symmetric point. Data points at finite lattice spacing are projected to the physical charm quark mass. \textit{Bottom}: Lattice spacing dependence of  $\Phi_{D_s}/\Phi_D$. The red square symbol indicates the continuum result, while the blue shaded band shows the fitted functional dependence on the lattice spacing.  Points at finite lattice spacing are projected to the physical values of $\phi_2$ and $\phi_H$.}
	\label{fig:fds_ratio}
\end{figure}

In Figure~\ref{fig:fds_ratio} we show the HM$\chi$PT chiral-continuum fit of the 
$\Phi_{D_s}/\Phi_D$ ratio with highest weight in the model averaging procedure. In particular the plot on the left shows the chiral approach to the physical point, 
while the plot on the right represents the lattice spacing dependence.
The observed dependence on $\phi_2$ shows only a mild curvature arising from the chiral logs, while cutoff effects appear to be highly suppressed at the current level of statistical precision of our data.

\begin{figure}[!t]
	\centering
	\includegraphics[scale=0.4]{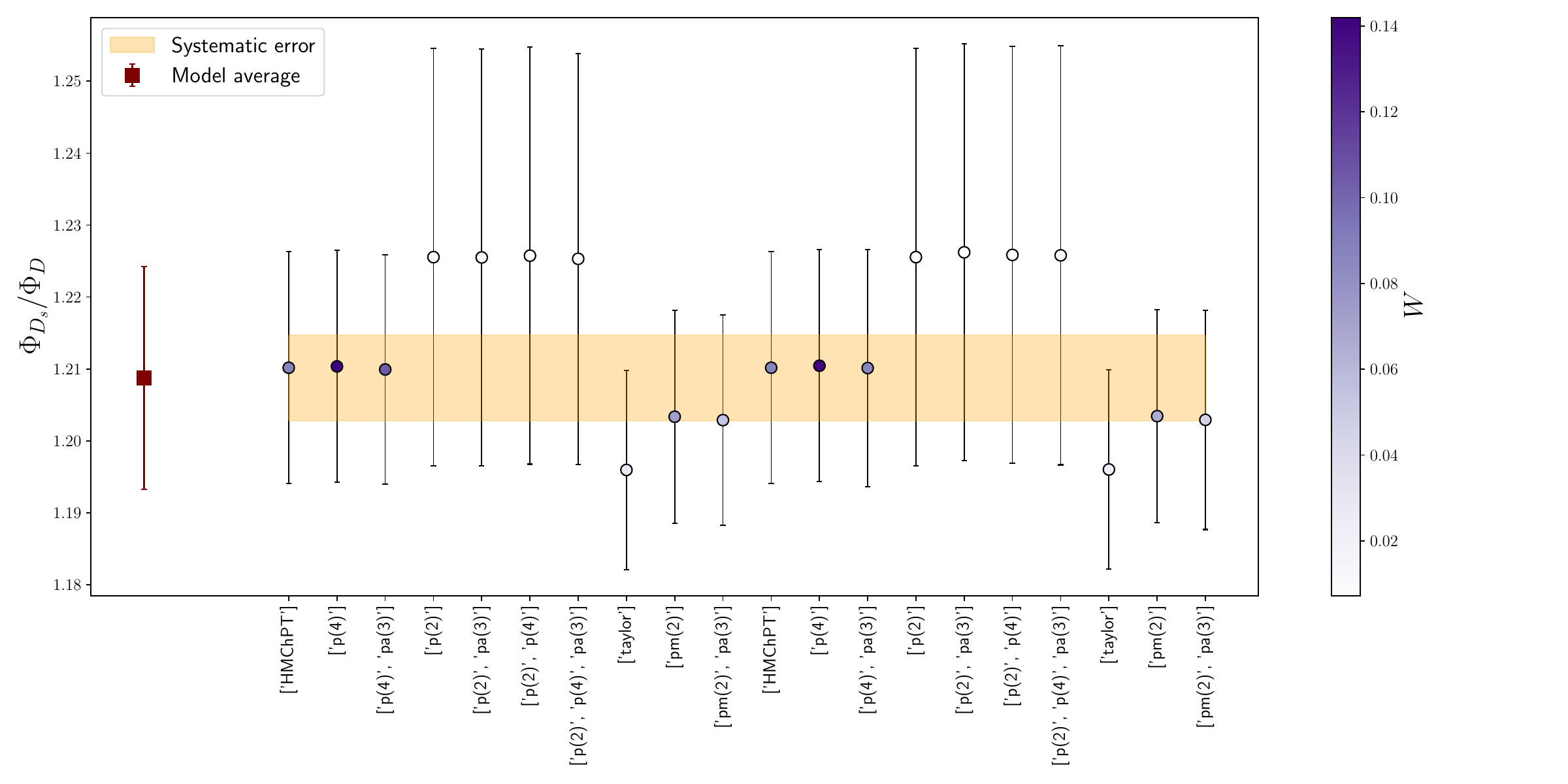}
	\caption{Summary of the model average procedure for the ratio $\Phi_{D_s}/\Phi_D$ based on the combination of the two matching conditions, $\phi_{H}^{(1)}$ and $\phi_{H}^{(2)}$. Each circular symbol  represents the result of a specific functional form, and the opacity is associated to the normalised weight $W$ of the model based on its TIC value. The yellow band represents the systematic uncertainty arising from the set of tested models, while the left-most red point is our final  averaged result. The labels of the 20 models specified in the horizontal axis are related to the terms characterising the dependencies on the mass and lattice spacing in the following way:  \texttt{`HMChPT'}  stands for the expression in Eq.~(\ref{eq:ratiophi}) where only the leading terms depending on the fit parameters $p_\chi^{(0)}$ and $p_a^{(1)}$ are considered . Similarly,  \texttt{`taylor'} refers to Eq.~(\ref{eq:ratiophi}) where only the terms depending on the fit parameters $p_m^{(1)}$ and $p_a^{(1)}$ are kept. The labels \texttt{`p(2)'} and \texttt{`p(4)'} correspond to the addition of the higher order terms depending on the parameters $p_\chi^{(2)}$ and   $p_\chi^{(4)}$ in Eq.~(\ref{eq:ratiophi}), respectively, while  \texttt{`pm(2)'} denotes the addition of $p_m^{(2)}$ from Eq.~(\ref{eq:ratiophiT}). Finally,  \texttt{`p(3)'}  denotes the inclusion of the fit parameter $p_a^{(3)}$ parameterising  higher order  lattice spacing dependence appearing in both the HM$\chi$PT and Taylor functional forms in Eq.~(\ref{eq:ratiophi}) and Eq.~(\ref{eq:ratiophiT}).
        } 
	\label{fig:fds_ratio_model_av}
\end{figure}

 Figure~\ref{fig:fds_ratio_model_av}
shows a summary of the model average procedure for the ratio $\Phi_{D_s}/\Phi_D$, displaying the fit results for the two 
matching conditions together with the associated weights, for the HM$\chi$PT and Taylor functional forms.

\begin{figure}[!h]
\begin{center}
\begin{minipage}{.43\linewidth}
\includegraphics[width=\linewidth]{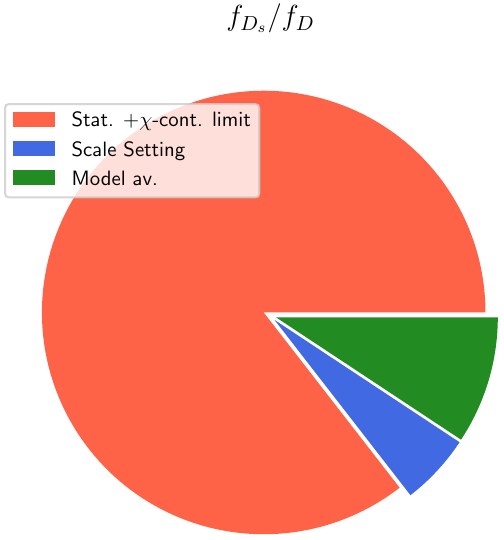}
\end{minipage}
\hspace{15mm}
\begin{minipage}{.4\linewidth}
\includegraphics[width=\linewidth]{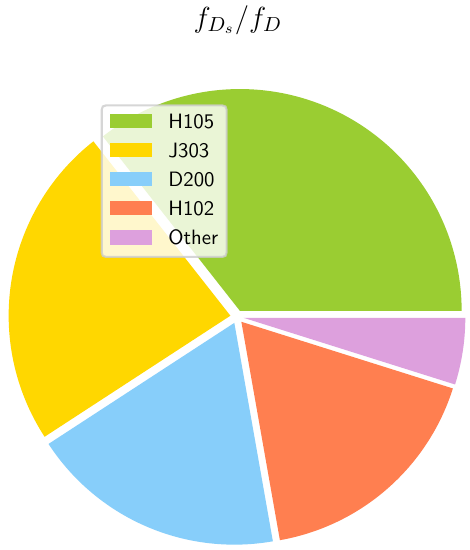}
\end{minipage}
\end{center}
\vspace{-5mm}
	\caption{\textit{Left}: Relative contributions to the total error on the determination of the ratio $f_{D_s}/f_D$. The label statistical plus $\chi$-continuum limit represents the error arising from the statistical accuracy of our data and the chiral-continuum extrapolation. The scale setting label denotes the error coming from the physical value $t_0^{\mathrm{phys}}$, while the model average represents the systematic error arising from the model variation according to the TIC procedure. \textit{Right}: Details of the relative contributions to the statistical and chiral-continuum extrapolation error arising from specific gauge field configuration ensembles. 
          }
	    \label{fig:fds_ratio_error}
\end{figure}

\begin{figure}[!h]
	\centering
	\includegraphics[scale=0.7]{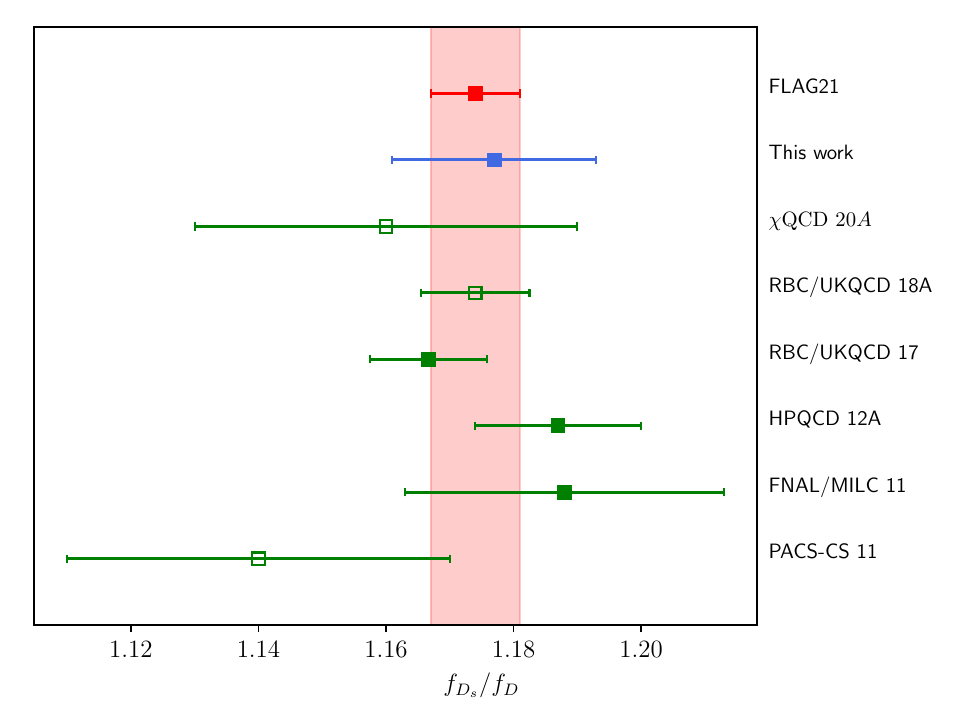}
	\caption{Comparison of our determination of $f_{D_s}/f_D$ with those of the other lattice QCD collaborations based on $\NF=2+1$ dynamical simulations as well as with the FLAG average~\cite{FlavourLatticeAveragingGroupFLAG:2021npn}. Only the results with filled symbols contribute to this average. Starting from the bottom, results are taken from:  PACS-CS 11 \cite{PACS-CS:2011ngu}, FNAL/MILC 11  \cite{FermilabLattice:2011njy}, HPQCD 12A \cite{Na:2012iu},  RBC/UKQCD 17 \cite{Boyle:2017jwu}, RBC/UKQCD 18A \cite{Boyle:2018knm},  $\chi$QCD 20A \cite{Chen:2020qma}. }
	\label{fig:fds_over_fd_comparison}
\end{figure}

\begin{table}[t!]
	\begin{center}	
		\begin{tabular}{c ||  c c  c}
			\hline
			 &  $\phi_{H}^{(1)}$ & $\phi_{H}^{(2)} $   & combined \\ [0.5ex]
			\hline\hline
			$f_{D_s}/f_D$   &  1.177(15)(6)& 1.178(15)(6) &  1.177(15)(5)
		\end{tabular}
		\caption{Results of the model average for $f_{D_s}/f_D$ for the two charm-quark matching conditions. The last column reports the combined result. The first error is statistical while the second is the systematic uncertainty arising from the model variation procedure. }
		\label{tab:ratio_res_all_matching}
	\end{center}
\end{table}
  In Figure \ref{fig:fds_ratio_error}  we show the major error sources contributing to our final determination of the ratio, where we notice that the major contribution is given by the statistical and chiral-continuum error. Finally, in Figure \ref{fig:fds_over_fd_comparison} we show a comparison between our result for $f_{D_s}/f_D$, the FLAG21 average and results from other collaborations.

\section{Conclusions and outlook}
\label{sec:conc}

In this work we have described our first computations of physical observables in the charm sector using the
Wilson fermion mixed-action setup described in greater detail in~\cite{MA1}. Emphasis
is put in setting up our methodology and exhibiting the characteristics of the framework. Our results for the charm quark mass and the $D_{(s)}$ meson decay constants are based
on a subset of CLS ensembles, yet they already sport a level of precision similar to that of several state-of-the-art results. We quote the values
\begin{gather}
\begin{split}
	M_c^{\mathrm{RGI}}(\NF=3) &= 1.485(8)(3)(14)[17]\ \mathrm{GeV},\\
	f_D &= 211.3(1.9)(0.6)[2.0] \ \mathrm{MeV},\\
	f_{D_s} &= 247.0(1.9)(0.7)[2.1] \ \mathrm{MeV},
	\\
	f_{D_s}/ f_D & = 1.177(15)(5)[16],
\end{split}
\end{gather}
as our main results. For the RGI charm quark mass in the 3-flavour theory, $M_c^{\mathrm{RGI}}(\NF=3)$, the first uncertainty is statistical, the second corresponds to the systematic error arising from the model selection, the third arises from the  RGI running factor in Eq.~(\ref{eq:rgi_running_factor}), and the last one in brackets is the total error. For the decay constants $f_D$, $f_{D_s}$ and their ratio $f_{D_s}/f_D$, the first error is statistical and the second is the systematic uncertainty from the model averaging, and the total error is given in brackets.

We foresee that these results could be improved in the future by means of a more extensive analysis
including additional CLS
ensembles with a finer value of the lattice spacing and physical pion mass simulations.
This is expected to have a significant impact in reducing the statistical uncertainty of
the decay constants. The error on the charm quark mass, on the other hand, is dominated
by the uncertainty induced by the non-perturbative renormalisation group running and thus work on that front would be required to improve the precision significantly.

In a related line of work, we are also applying our framework to the computation of
semileptonic form factors for charmed meson decay, for which preliminary results have
already been presented in~\cite{Frison:2019doh,Frison:2023axn}. Together with the computations illustrated in this
paper, they show how a comprehensive programme of precision heavy-flavour physics can be
pursued in the framework of Wilson fermion regularisations, reaching an excellent compromise
between the latter's advantages from the point of view of field-theoretical control
and the aim of high-precision computations.

\section*{Acknowledgements}

We are grateful to our colleagues in the Coordinated Lattice Simulations (CLS) initiative for the generation of the gauge field configuration
ensembles employed in this study.
Discussions with, and input from, our colleagues Mattia Bruno, Fabian Joswig, Simon Kuberski,
Alberto Ramos and Stefan Schaefer is gratefully acknowledged.
We acknowledge PRACE for awarding us access to
MareNostrum at Barcelona Supercomputing Center (BSC), Spain and to HAWK at GCS@HLRS, Germany.
The authors thankfully acknowledge the computer resources at MareNostrum and the technical support provided by Barcelona Supercomputing Center (FI-2020-3-0026).
We thank CESGA for granting access to Finis Terrae II.
This work is partially supported
by grants PGC2018-094857-B-I00 and PID2021-127526NB-I00, funded by MCIN/AEI/10.13039/501100011033 and by
“ERDF A way of making Europe”, and by the Spanish Research Agency (Agencia Estatal
de Investigaci\'on) through grants IFT Centro de Excelencia Severo Ochoa SEV-2016-0597
and No CEX2020-001007-S, funded by MCIN/AEI/10.13039/501100011033. We also acknowledge support from the project H2020-MSCAITN-2018-813942 (EuroPLEx), under grant agreementNo.~813942, and the
EU Horizon 2020 research and innovation programme, STRONG-2020 project, under grant
agreement No 824093.

\begin{appendix}
\section{GEVP implementation}\label{app:gevp}

In this work, ground state meson masses and matrix elements  are extracted from a generalised eigenvalue problem (GEVP) variational method following \cite{Blossier:2009kd}.
The GEVP has the form described in Section~\ref{sec:charm_basics}, cf. \req{eq:gevp_sec3}
and the discussion that follows.

\begin{figure}[!t]
	\centering
	\includegraphics[scale=0.9]{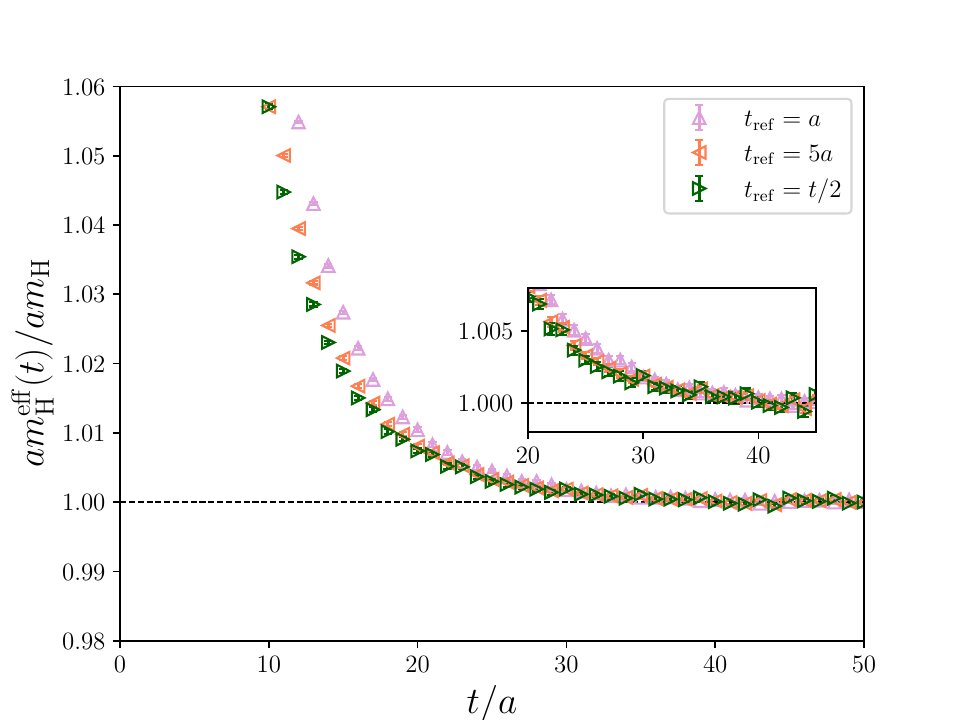}
	\caption{Illustration of the  ground-state effective masses determined from a GEVP analysis with three different ways of setting the value of $t_{\mathrm{ref}}$ for the ensemble J303. The effective masses are normalised by the central value of the mass extracted from conservative plateau choices.
	  The parameter $t_{\mathrm{ref}}$ is either kept fixed, $t_{\mathrm{ref}}/a=1,5$, or varied by setting $t_{\mathrm{ref}}=t/2$  in such a way that the condition  $t_{\mathrm{ref}} \geq t/2$ is fulfilled.
        }
	\label{fig:different_tnot}
\end{figure}

Considering only the first $N$ state contributions in the spectral expansion, we can extract their effective energies from the eigenvalues $\lambda_n(t,t_{\mathrm{ref}})$ as 
\begin{equation}\label{eq:eff_en_gevp}
	aE_n^{\mathrm{eff}}(t,t_{\mathrm{ref}}) \equiv \log\bigg(\frac{\lambda_n(t,t_{\mathrm{ref}})}{\lambda_n(t+a,t_{\mathrm{ref}})}\bigg) = aE_n +\mbox{O}(e^{-(E_{N+1}-E_n)t}).
\end{equation}
Here the asymptotic behaviour $\mbox{O}(e^{-(E_{N+1}-E_n)t})$ is ensured exclusively in the regime $t_{\mathrm{ref}} \geq t/2$. Whenever $t_{\mathrm{ref}}$ is kept fixed the first unresolved excited state is the $n+1$, and the asymptotic scaling behaves as $\mbox{O}(e^{-(E_{n+1}-E_n)t})$, therefore providing shorter plateaus.  In Fig. \ref{fig:different_tnot} we show a comparison  of low-lying heavy-heavy pseudoscalar  states
 as extracted from the GEVP with different values of $t_{\mathrm{ref}}$. In general, we observe a similar behaviour  when comparing different values of $t_{\mathrm{ref}}$, with a slightly better convergence when the condition $t_{\mathrm{ref}}\geq t/2$ is fulfilled. In this work we therefore stick to this choice for plateau extraction by setting $t_{\mathrm{ref}}=t/2$.
As explained in the main text,
in order to assess the systematic uncertainty associated with the extraction of the ground state signal from a plateau behaviour in the effective energies, we perform numerous fits by varying the time ranges of the 
fitting interval, and apply the model averaging procedure described in Appendix~\ref{app:TIC} ---
cf. the illustration in Fig.~\ref{fig:meff_plateau}.

\begin{figure}[!t]
	\centering
	\includegraphics[scale=0.9]{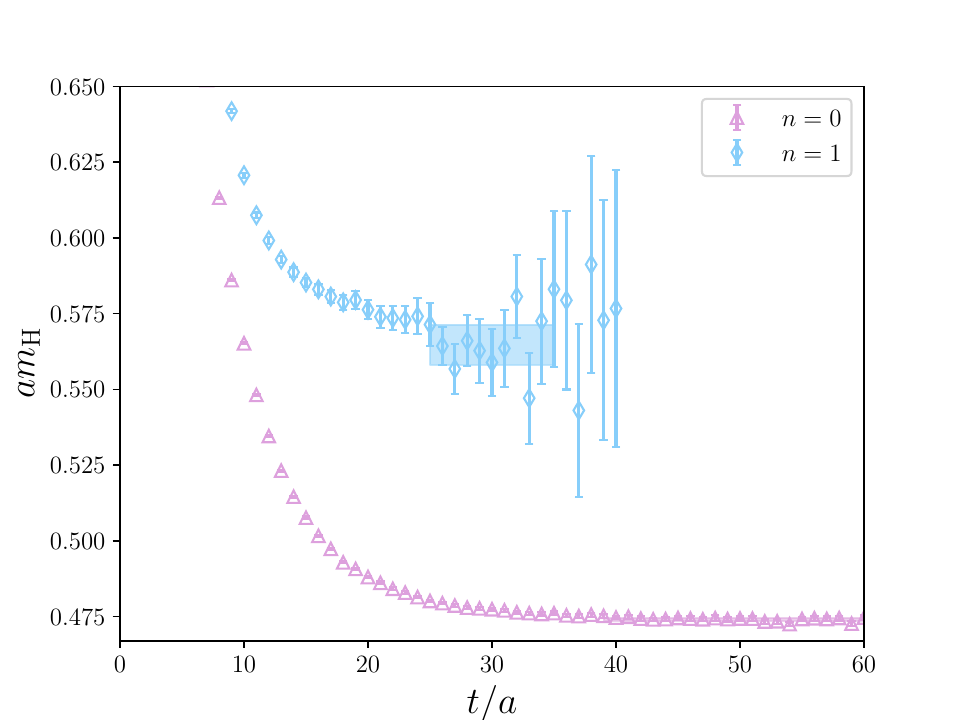}
	\caption{Illustration of the ground state and first excited state for a heavy-light pseudoscalar meson mass as extracted from the GEVP for the ensemble J303. We use $t_{\mathrm{ref}}=t/2$ such that the condition $t_{\mathrm{ref}} \geq t/2$ is fulfilled. The shaded bands correspond to the plateaus of choice.
        }
	\label{fig:different_gevp_sizes}
\end{figure}

As additional cross-checks and stability tests we also computed the first excited state from the GEVP.  A comparison of the ground state and first excited state as  is given in Fig. \ref{fig:different_gevp_sizes} together with the plateaus of choice.  As we are only interested in ground state,  we choose to stick to the $2\times 2$  matrix formulation of the GEVP.

In addition to the meson spectrum, in this work we  also extract the matrix element $ \langle 0 | P^{qr} | P^{qr}(\mathbf{p=0})\rangle $ from the GEVP analysis by considering the normalised eigenvector $v_n(t,t_{\mathrm{ref}})$ in Eq.~(\ref{eq:gevp_sec3}),
where we remind that $| P^{qr}(\mathbf{p=0})\rangle$ stands for a ground state.  Namely, we define for each state $n$ the number \cite{Blossier:2009kd}
\begin{equation}
	R_n = \left( v_n(t,t_{\mathrm{ref}}),C_{\rm\scriptscriptstyle P}) (t)v_n(t,t_{\mathrm{ref}})\right)^{-1/2} e^{E_nt/2},
	\label{eq:gevp_effective_operator}
\end{equation}
where $(\cdot, \cdot)$ is the usual scalar product and $C_{\rm\scriptscriptstyle P}$ is the GEVP matrix from Eq.~(\ref{eq:gevp_matrix}).  Then, the ground state matrix element is given by
\begin{equation}
	p_0^{\mathrm{eff}}(t,t_{\mathrm{ref}}) = (v_0(t,t_{\mathrm{ref}}), C_{\rm\scriptscriptstyle P,0})) R_0, \qquad (C_{\rm\scriptscriptstyle P,0})_k = (C_{\rm\scriptscriptstyle P})_{k0} 
	\label{eq:effective_matrix_element}
\end{equation}
The large distance behaviour of the effective matrix element is governed by 
\begin{equation}
	p_0^{\mathrm{eff}}(t,t_{\mathrm{ref}}) = p_0 + \mbox{O}(e^{-(E_{N+1}-E_0) t_{\mathrm{ref}}}), \qquad p_0 =  \langle 0 | P^{qr} | P^{qr}(\mathbf{p=0})\rangle,
\end{equation}
in the regime where the condition $t_{\mathrm{ref}}\geq t/2$ is satisfied. We perform constant fits in a number of  time intervals and use the model averaging procedure
in  Appendix~\ref{app:TIC} to estimate the systematic uncertainty due to excited-state contamination.
In Figure \ref{fig:decay_plateau} we show a representative plateau for a heavy-light decay constant, together with a summary of the model average with different fit intervals.

\section{Model averaging procedure}
\label{app:TIC}

In this work, the systematic uncertainties are estimated from a model averaging procedure discussed
in detail in~\cite{MA1}. Here we collect the main ideas and point to the relevant background
references.

As is often the case in lattice QCD calculations, in this study we deal with fits to
highly correlated data. The dichotomy thus arises between trying correlated $\chi^2$ fits,
which typically leads to numerical instabilities and potential biases in statistical estimators, or keeping
an uncorrelated $\chi^2$, which is however not a suitable
quantity to assess the goodness-of-fit. To overcome this situation, we follow
an approach introduced in \cite{Bruno:2022mfy} based on the expectation value of the $\chi^2$, denoted $\chi^2_{\mathrm{exp}}$, and its corresponding p-value,
which does allow to quantify the goodness-of-fit in a controlled manner.
Furthermore, we make use of the Takeuchi Information Criteria (TIC) proposed in \cite{Frison:2023lwb} to assign a weight to each model, which then allows to perform a weighted model average to arrive at a final result for the systematic uncertainty~\cite{Jay:2020jkz}. Specifically, the value of the TIC assigned to each fitting model is
\begin{equation}
	\mathrm{TIC} = \chi^2 - 2\chi^2_{\mathrm{exp}}\,.
\end{equation}
To each model $m$ in the complete set, consisting of $M$ models, we assign a normalised weight $W_m$ defined as follows
\begin{equation}
	W_m \propto \exp\bigg(
	-\frac{1}{2}\mathrm{TIC}_m
	\bigg), \qquad \sum_{m=1}^{M}W_m =1\,.
	\label{eq:weight_ic}
\end{equation}
The result of the model average for an observable $\cO$ that has been determined for each of the models is then given by 
\begin{equation}
	\langle \cO \rangle = \sum_{m=1}^{M} W_m \langle \cO \rangle_m\,.
	\label{eq:model_average}
\end{equation}
Finally, to estimate the systematic uncertainty arising from the model variation we employ the weighted variance defined as follows
\begin{equation}
	\sigma^2_{\cO} = \sum_{m=1}^{M} \bigg(W_m \langle \cO\rangle_m^2\bigg) 
	- 
	\bigg(
	\sum_{m=1}^{M}
	W_m \langle \cO\rangle_m
	\bigg)^2\,.
	\label{eq:weighted_variance}
\end{equation}

\end{appendix}

\bibliographystyle{JHEPg6l10}
\bibliography{biblio}

\end{document}